\documentclass[12pt,twoside,a4paper]{book}

\usepackage[dvips]{graphicx}
\usepackage{amssymb}

\begin{document}

\frontmatter

\begin{titlepage}

\begin{center}
\Large{University of Padua \\
Department of Physics ``Galileo Galilei''}
\end{center}

\vspace{1.5cm}

\begin{center}
\large{Ph.D. Thesis}
\end{center}

\vspace{1cm}

\begin{center}
\huge{A BACK-REACTION APPROACH \\
TO DARK ENERGY}
\end{center}

\vspace{3cm}

\begin{center}
\large{\textbf{Ph.D. candidate}: Valerio Marra\\
\textbf{Advisors}: Prof. Sabino Matarrese\\
$\quad \quad \quad \quad \; \; \,$Prof. Antonio Masiero}
\end{center}

\vspace{5cm}

\begin{center}
March 14, 2008
\end{center}

\end{titlepage}

\def \lta {\mathrel{\vcenter  {\hbox{$<$}\nointerlineskip\hbox{$\sim$}}}}
\def \gta {\mathrel{\vcenter {\hbox{$>$}\nointerlineskip\hbox{$\sim$}}}}
\def\a{\alpha}
\def\BF{B_F(\phi)}
\def\D{\Delta}
\def\beqra{\begin{eqnarray}}
\def\eeqra{\end{eqnarray}}
\def\beq{\begin{equation}}
\def\eeq{\end{equation}}

\def\L{\Lambda}
\def\p{\partial}
\def\half{{1 \over 2}}

\font\bigastfont=cmr10 scaled \magstep 3
\def\bdot{\hbox{\bigastfont .}}

\newcommand{\Frac}[2]{\frac
{\begin{array}{@{}c@{}}\strut#1\strut\end{array}}
{\begin{array}{@{}c@{}}\strut#2\strut\end{array}}}

\begin{quotation}

\flushright{
\textit{A Fabiana}
}

\end{quotation}

\tableofcontents

\chapter{Introduction}

This thesis is mainly about how to set up and carry out the idea of back-reaction in a physically meaningful way.

\ 

Most, if not all, observations are consistent with the cosmic concordance model
according to which, today, one-fourth of the mass-energy of the universe is
clustered and dominated by cold dark matter.  The remaining three-quarters is
uniform and dominated by a fluid with a  negative pressure (dark energy, or
$\Lambda$). 

While the standard $\Lambda$CDM model seems capable of accounting for the
observations, it does have the feature that approximately 95\% of the
mass-energy of the present universe is unknown. We are either presented with
the opportunity of discovering the nature of dark matter and dark energy, or
nature might be different than described by the $\Lambda$CDM model. Regardless,
until such time as dark matter and dark energy are completely understood, it is
useful to look for alternative cosmological models that fit the data.  

One non-standard possibility is that there are large effects on the {\it
observed} expansion rate (and hence on other observables) due to the
back-reaction of inhomogeneities in the universe.  The basic idea is that
all evidence for dark energy comes from the observational determinations of the
expansion history of the universe.  Anything that affects the observed expansion
history of the universe alters the determination of the parameters of dark
energy; in the extreme it may remove the need for dark energy.

The ``safe'' consequence of the success of the concordance model is that the
isotropic and homogeneous $\Lambda$CDM model is a good  {\it phenomenological}
fit to the real inhomogeneous universe. And this is, in some sense, a
verification of the cosmological principle:  the inhomogeneous universe can be
described by means of an isotropic and  homogeneous solution.  However, this
does not imply that a primary source of dark energy  exists, but only that it
exists as far as the  phenomenological fit is concerned. For example, it is not
straightforward that the universe is accelerating. If dark energy does not
exist at a fundamental level, its presence in the concordance model would tell
us that the pure-matter inhomogeneous model has been renormalized, from the
phenomenological point of view  (luminosity-distance and redshift of photons),
into a homogeneous $\Lambda$CDM model.

There are two ways to approach the problem of the back-reaction.\\
The first point of view is strictly observational and focuses directly on the past light cone, on
the effects of large-scale nonlinear inhomogeneities on observables such as
the  luminosity-distance--redshift relation.\\
The second point of view tries, instead, to interpret the inhomogeneous universe by means of an effective model.\\
These two different points of view actually share the same idea of smoothing out
the inhomogeneities. The differences are in the way this process is carried out.
In the approach which uses an effective model the averaging is explicit, while
in the observational approach the averaging is implicit: inhomogeneities are indeed
``integrated out'' in the luminosity-distance--redshift relation.

This duality will shape the structure of the present thesis: we will present the theoretical backgrounds
and work out illustrative models for both the approaches.

After an outline of the evidences for dark energy, Chapter \ref{deintro} will introduce the candidates we are interested in: the cosmological constant, the quintessence and the back-reaction.
This will give us the opportunity to give our point of view about the dark energy problem
and to justify the idea of the back-reaction as a possible alternative explanation for dark energy.

Chapter \ref{brset} will be about the dual point of view we just talked about.
In line with Ellis's work, we will first introduce the smoothing process in general relativistic cosmology
and then we will try to work it out theoretically by setting the problem
in a physically meaningful way.

In Chapters \ref{pool} and \ref{s-c} we will attempt to work out the theoretical concepts introduced by means of concrete models, even if it is difficult to analyze a realistic inhomogeneous model of the universe in an exact way. In order to understand the physics of the back-reaction, therefore, we have looked at exact general relativistic solutions where some symmetries will enable us to carry out the calculations. We have chosen the spherically symmetric dust Lema\^{i}tre-Tolman-Bondi (LTB) solution as our starting point.

There is increasing attention to the LTB model because it was shown that it is possible to fit the observed luminosity-distance--redshift relation by adjusting the LTB free functions. To achieve this result, however, it is necessary to place the observer at the center of a rather big underdensity.
Even though we will build nothing more than toy models, we will try to avoid this fine-tuning.
The ultimate goal is indeed to find a realistic dust model that can explain observations (like the
luminosity-distance--redshift relation) without a need for dark energy. The
ultimate aim being to have an exactly solvable realistic inhomogeneous
model. The models we will build are but a first small step in this pragmatic and necessary
direction.

In Chapter \ref{pool} we will study the average scheme developed by Buchert building
a ``homogeneous'' universe model where the parameters are chosen in order not to single out the center.
We will study both the curved and the flat case.

Then, in Chapter \ref{s-c}, we will build a Swiss-cheese model with the observer in
the cheese looking through a series of holes.
The cheese consists of the usual Friedmann-Robertson-Walker (FRW)
solution and the holes are constructed out of a LTB
solution.
We will study this model under both points of view.
First, we will focus on the effects of large-scale non-linear inhomogeneities on
observables such as the luminosity-distance--redshift relation.
Then we will try to apply the fitting scheme developed by Ellis and Stoeger:
we propose a fitting procedure that is intermediate between the fitting approach  and
the averaging one: a fit with respect to light-cone averages.

In Appendix \ref{alphaevo} we will present  a general model for the cosmological evolution of the fine structure constant driven by a typical Quintessence scenario.  
We consider a coupling, between the Quintessence scalar and  the electromagnetic kinetic term, given by a general function. 
We study the dependence of the cosmological variation of the fine structure constant upon the functional form of the general function chosen and discuss the constraints imposed by the data.
We find that different cosmological histories for the fine structure constant are possible within the avaliable constraints.
Appendices \ref{ltbm} and \ref{sewing} are, instead, about LTB models.

For a discussion of the results see the Conclusions.

\cleardoublepage

\mainmatter

\chapter{Dark Energy} \label{deintro}

In this chapter we will introduce the Dark-Energy problem, at first its experimental evidences and then the possible explanations in which we are interested.
It will help us to introduce the Back-Reaction problem in the next chapter.

\section{Evidences}

The fact that matter energy density does not dominate the universe has been one of the most stunning discoveries in the last few years in cosmology.

Most, if not all, observations are consistent with the cosmic concordance model
according to which, today, one-fourth of the mass-energy of the universe is
clustered and dominated by cold dark matter.  The remaining three-quarters is
uniform and dominated by a fluid with a negative pressure which we call dark energy. 

\begin{figure}[htbp]
\begin{center}
\includegraphics[width=14cm]{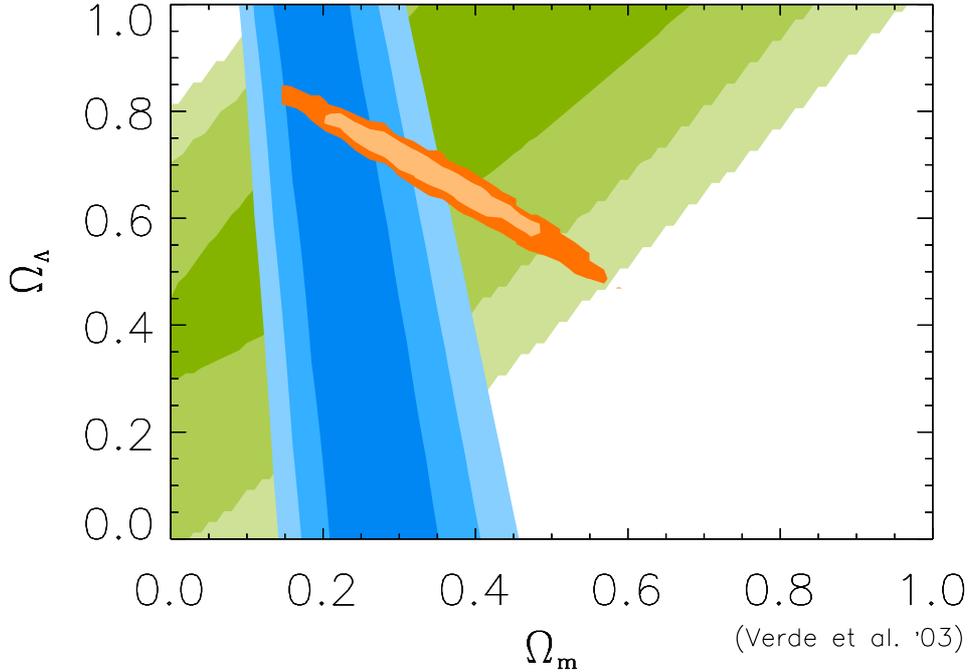}
\end{center}
\caption{\small \slshape Experimental constraints on the plane $\Omega_{M}$-$\Omega_{DE}$ from the 2dF galaxy survey (blue), CMB (orange) and Supernovae Ia (green).}
\label{verde03}
\end{figure}

The three main experimental evidences for dark energy are summarized in Fig.~\ref{verde03}.\\
The blue area is about the results of the 2dF galaxy survey which gives $\Omega_{M} \simeq 0.3$. 
This measure is independent of $\Omega_{DE}$ and so it is vertical\footnote{The blue area is not actually vertical, but slightly tilted. This because the 2dF galaxy survey is mainly at $z=0.14$. In order to evolve the data till $z=0$ a cosmological model (in particular a $\Lambda$CDM model) has been used and therefore a dependence on dark energy introduced.} in the plane $\Omega_{M}$-$\Omega_{DE}$.\\
The orange area represents the constraints from the WMAP observations of the CMB anisotropies.
They give $1 \simeq \Omega_{TOT}=\Omega_{M}+\Omega_{DE}$ and so the orange area is aligned to the line for $(0,1)$ and $(1,0)$.
\newpage
Finally, the green area shows the Supernovae Ia measurements about the deceleration parameter which, in the $\Lambda$CDM model, is given by $q= \Omega_{M}/2-\Omega_{DE}$. Therefore the green area is almost perpendicular to the orange one.

The fact that these constrains cross perpendicularly and consistently gave the name {\it cosmic concordance model} to a $\Lambda$CDM with:
\begin{eqnarray}
	\Omega_{M} &\simeq& 0.25 \cr
	\Omega_{DE} &\simeq& 0.75 \cr
	w_{DE} &\simeq& -1
\end{eqnarray}
%

\clearpage
\section{Candidates}

In this section we will take into account three possible explanations for dark energy: the cosmological constant, a cosmological scalar field and inhomogeneous universe models.
In this thesis we are really interested only in the last possibility.
We will describe the first two because, beside being interesting in themselves, they will help us understand the problem and will turn out to be useful to describe inhomogeneous models effectively.
They are indeed based on the cosmological principle of homogeneity and isotropy.

For a more comprehensive examination of possible approaches and different points of view to dark energy see \cite{Nobbenhuis:2006yf}.

\subsection{The Cosmological Constant}

Trying to explain dark energy through the cosmological constant $\Lambda$, that is by using the concordance model, has the advantage of simplicity and agreement with experimental data.

The cosmological constant appears as a free parameter in the Einstein's equations:
\begin{equation}
R_{\mu \nu}-{1 \over 2} \, g_{\mu \nu} \, R -\Lambda \, g_{\mu \nu}= 8 \pi \, T_{\mu \nu}
\end{equation}
where we are using the signature $(-,+,+,+)$ and geometric units, $c=1=G$.

The freedom about the value of $\Lambda$ is cause of problems.
In spite of being treated as the single problem of the value of $\Lambda$,  from a conceptual point of view they are actually three distinct problems\footnote{For a similar approach see \cite{Quartin:2008px}.}.
This distinction will help us understand which ones are more pertinent to the cosmological problem and which are actually the advantages in using alternative explanations for dark energy.

For a wider overview of cosmology in the presence of the cosmological constant we refer to \cite{weinberg-lambda, carroll:lambda0, Carroll:2000fy, Peebles:2002gy, Padmanabhan:2002ji}.

\begin{figure}[htbp]
\begin{center}
\includegraphics[width=12cm]{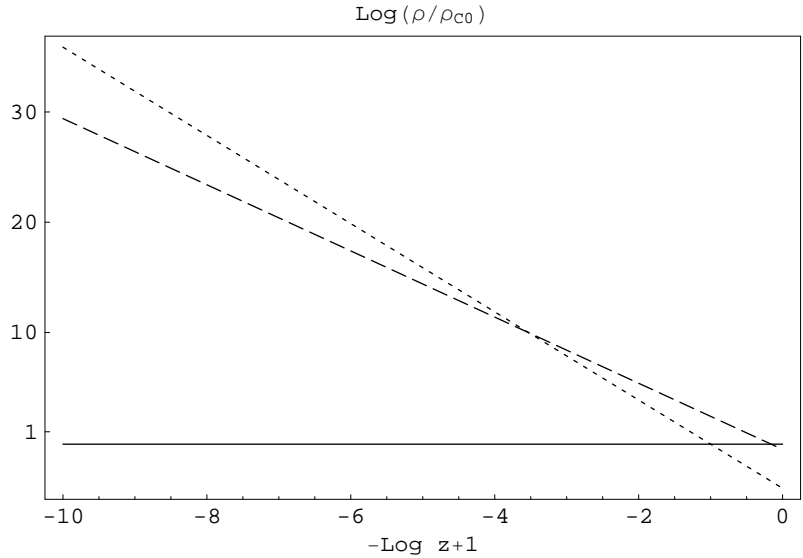}
\end{center}
\caption{\small \slshape Evolution of the energy density of radiation (short-dashed line), matter (long-dashed line) and cosmological constant (solid line) with respect to the redshift. All energy densities are expressed in units of the present critical energy density  $\rho_{C 0}$.}
\label{cosmoconst}
\end{figure}

\subsubsection{Initial conditions problem}
The first problem is about the initial conditions that we have to give to the concordance model: it would be expectable to have comparable values of the initial energy densities. However, as you can see from Fig.~\ref{cosmoconst}, matter and radiation initial energy-density values are much bigger than the cosmological constant one. The concordance model implies therefore a sizeable fine tuning of the initial conditions.

\subsubsection{Coincidence problem}
The second issue is about the values of the present-day energy density of matter and cosmological constant. As you can see from Fig.~\ref{cosmoconst}, they evolve differently and therefore generally, during the evolution of the universe, we will expect them to have different values. However their values happen to be of the same order at present time. This fact needs a sizeable fine tuning to happen.

\subsubsection{Origin problem}
Finally, problems show up if we ask where the cosmological constant comes from.\\
Classically, it is a free parameter given as an initial condition, $\rho_{\Lambda}$.\\
Quantum mechanically, we have the contribution from the ground state of the energy-momentum tensor which does not need to be zero:
\begin{equation}
\langle 0 |T_{\mu \nu}|0 \rangle \equiv T_{\mu \nu}^{GS} = \rho_{GS} \, g_{\mu \nu}
\end{equation}
$\rho_{GS}$ has the same equation of state of $\rho_{\Lambda}$ and therefore we can combine them into the vacuum energy density $\rho_{V}$:
\begin{equation} \label{vacu}
\rho_{V}=\rho_{\Lambda}+\rho_{GS}
\end{equation}
The problem is that any quantum-mechanic estimation of $\rho_{V}$ largely exceeds the experimental value of
$\rho_{DE} \sim (10^{-3} \mbox{ eV})^{4}$. Typical estimations are around $\rho_{V} \sim (10^{27} \mbox{ eV})^{4}$, that is, $30$ orders of magnitude bigger.

\subsection{Quintessence models} \label{quiquoqua}

Quintessence or cosmological scalar-field models will be, as effective models, a useful tool in our analysis of a inhomogeneous universe.

The idea is to let the cosmological constant be a dynamical quantity. The Lagrangian of a quintessence field is, therefore:
\begin{equation} \label{lphi}
{\cal L}_{\phi} = \frac{1}{2} \partial^{\mu}\phi \, \partial_{\mu}\phi - V(\phi)
\end{equation}
from which, assuming spatial homogeneity, we obtain:
\begin{equation} \label{rho-p}
\rho_{\phi}={1 \over 2} \dot{\phi}^{2} + V(\phi)
\quad \mbox{ e } \quad
p_{\phi}={1 \over 2} \dot{\phi}^{2} - V(\phi)
\end{equation}
The equation of state is:
\begin{equation} \label{wphi}
w_{\phi}={p_{\phi} \over \rho_{\phi}}={\dot{\phi}^{2}/2 - V(\phi) \over \dot{\phi}^{2} /2 + V(\phi)}
\end{equation}
Quintessence models are extensively studied (see for example \cite{Copeland:2006wr}) mainly because of their flexibility. In particular there exist classes of potentials which exhibit attractor solutions independent of initial conditions \cite{Steinhardt:1999nw}.

\begin{figure}[!htbp]
\begin{flushright}
\includegraphics[width=16cm]{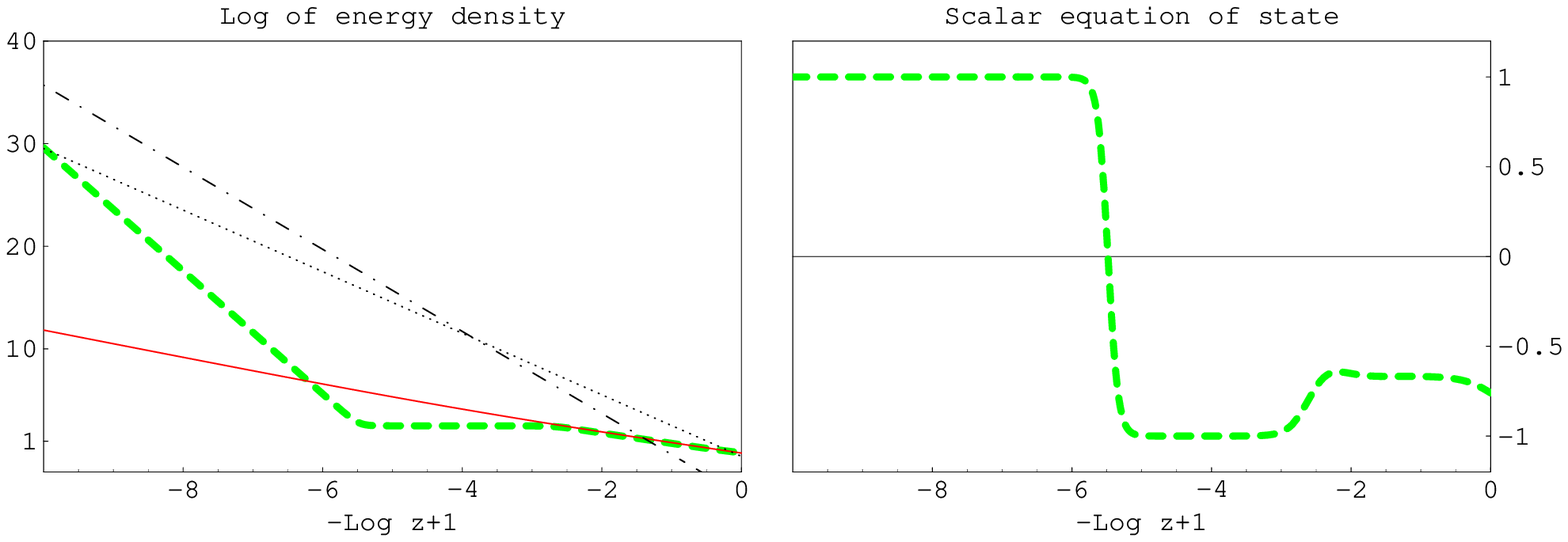}
\caption{\small \slshape Evolution of the energy densities (left) and scalar equation of state (right) for a  quintessence model with potential $V=M^{5} \phi^{-1}$ and  initial conditions $\rho_{\phi}^{in}/\rho_{C 0}=10^{30}$ at $z=10^{10}$.
The dot-dashed line represents the energy density of radiation, the dotted line the energy density of matter, the green dashed line the energy density of quintessence and the red solid line the attractor. All of the energy densities are expressed in units of the present critical energy density  $\rho_{C 0}$. From \cite{Marra:2005yt}.}
\label{quinti2}
\end{flushright}
\end{figure}

For illustrative purposes, in Fig.~\ref{quinti2} we show the evolution of the energy density and of the equation of state for a scalar potential $V=M^{5} \phi^{-1}$ and initial conditions $\rho_{\phi}^{in}/\rho_{C 0}=10^{30}$ at $z=10^{10}$. $M$ gives the energy scale of the potential and has to be fine-tuned in order to have the right present-day quintessence energy density.

As you can see from Fig.~\ref{quinti2}, at early times the kinetic energy is dominant over the potential energy: during this period, called kination and characterized by $w_{\phi}\simeq 1$, the scalar field rapidly rolls down along the potential. Then the friction performed by the Hubble term\footnote{The continuity equation for a quintessence field can be written as $\dot{\rho}_{\phi}=-3 H \dot{\phi}^{2}$.} slows down the run of the scalar till its freeze: during this period, called slow-roll and characterized by $w_{\phi}\simeq -1$, the potential energy dominates and the scalar mimics a cosmological constant. Finally the scalar reaches the attractor and follows a dynamics characterized by its features. In the illustrative case shown, the attractor is characterized by $w_{\phi}\simeq -0.67$. See Appendix~\ref{alphaevo} for more details.

\subsubsection{Pros}
A first advantage of the example shown is that it solves the initial conditions problem: we can change of many order of magnitude the chosen value of $\rho_{\phi}^{in}/\rho_{C 0}=10^{30}$ at $z=10^{10}$ and still the late time dynamics will be the same: only the moment at which the scalar reaches the attractor will change. This is a real improvement in comparison with the cosmological constant.

In the second place, a dynamical scalar field may have chances to solve the coincidence problem. Indeed, if it couples to the matter energy density, it could be possible to explain why, today, they are of the same order of magnitude.
However, quintessence models generally feature a sizeable fine-tuning about the mass parameter of the potential, similarly to what happens with the cosmological constant.

\subsubsection{Cons}
Beside the problem of the origin of quintessence, its first drawback is about its mass which has to be ultra-light:
\begin{equation} \label{mphi}
m_{\phi_{0}} \sim H_{0} \sim 10^{-33} \mbox{ eV}
\end{equation}
Quantum-mechanics direct or indirect corrections will easily bring $m_{\phi_{0}}$ toward much bigger values.

Other problems come from the possible coupling of the quintessence field with the other terms in the matter-radiation Lagrangian and from the fact that such an ultra-light field has long-range interactions: $\lambda \sim m_{\phi_{0}}^{-1} \sim H_{0}^{-1}$. Variations of fundamental constants and violations of the equivalence principle are  therefore expected. For more details see Appendix~\ref{alphaevo}.

\subsubsection{Effective theory?}
In the case quintessence models are used as effective models, none of the drawback examined above will be applicable.
They indeed are about the quantum mechanics implications a primary quintessence field brings to light.

As for this thesis, the most appealing feature of quintessence is its dynamical nature which can be linked to the evolution of the universe. In particular, a coupling of an effective quintessence field to inhomogeneities could solve the coincidence problem.
We will discuss this point in the next chapter.

\clearpage
\subsection{Inhomogeneous models - a point of view} \label{poview}

This overview of advantages and drawbacks of $\Lambda$CDM and quintessence models gives us some hints to better set the dark energy problem. We will present now the point of view that will characterize this thesis.
It is clarifying to write:
\begin{equation} \label{core}
\rho_{V} \equiv (\rho_{\Lambda} \; \, +) \; \, \rho_{GS} \ggg \rho_{DE} \sim \rho_{M}
\end{equation}
On the left-hand side, the two terms $\rho_{\Lambda}$ and $\rho_{GS}$ are really different and could be inappropriate to mix them. The latter comes from quantum mechanics, the former comes from general relativity. $\rho_{\Lambda}$ is likely connected to cosmology and therefore to $\rho_{DE}$ while for $\rho_{GS}$ it could not be the case. Indeed typical estimations for $\rho_{GS}$ are far away from the cosmological value.\\
Therefore it could be possible that we should be only concerned about $\rho_{\Lambda}$ the role of which has to be understood in order to have a satisfactory general-relativist cosmological model.

On the right-hand side, the similarity in density values of the dark energy and matter components suggests a direct or indirect connection: the dark energy problem might be a cosmological one.

If we accept this reasoning and remember that the homogeneous $\Lambda$CDM model is a good fit to the real inhomogeneous universe, we could think that the connection is by means of the smoothing of the inhomogeneities: {\it the back-reaction of inhomogeneities makes an inhomogeneous matter FRW model appear as a $\Lambda$CDM model.}\\
One immediate benefit is to turn the coincidence problem into a hint in favor of the back-reaction: only recently  have the cosmic structures evolved enough to have a sufficiently strong back-reaction of inhomogeneities which could affect observables.

The issue now is what we mean by back-reaction of inhomogeneities.
There are broadly speaking two distinct approaches as we will now discuss.
One is focused on theoretical aspects while the other on observations.\\
Both the approaches will, however, share the idea of smoothing out inhomogeneities.
The duality in the interpretation of this concept will characterize this thesis.

\clearpage
\subsubsection{Observational side - $\rho_{DE}$} \label{dake}

The ``safe'' consequence of the success of the concordance model is that the isotropic and homogeneous $\Lambda$CDM model is a good {\it observational} fit to the real inhomogeneous universe.
And this is, in some sense, a verification of the cosmological principle: 
the inhomogeneous universe can be described by means of an isotropic and 
homogeneous solution.

However this does not imply that a primary source of dark energy 
exists, but only that it exists effectively as far as the 
observational fit is concerned.
For example it is not straightforward that the universe, as Supernovae Ia measurements seem to tell us, is globally 
accelerating.
If dark energy does not exist at a fundamental level, 
its evidence coming from the 
concordance model would tell us that the purely-matter inhomogeneous
model has been renormalized, from the observational point of view 
(luminosity and redshift of photons), into a homogeneous $\Lambda$CDM model.
Moreover, the very homogeneous nature of dark energy seems a clue about its effective nature.

There could indeed be the possibility that there are large effects on the {\it observed} expansion rate due to the back-reaction of inhomogeneities in the universe (see, e.g., Ref.~\cite{Kolb:2005da} and refs. therein). 
The basic idea is that all evidence for dark energy comes from observational determination of the expansion history of the universe. 
Anything that affects the observed expansion history of the universe alters the
determination of the parameters of dark energy; in the extreme it may remove
the need for dark energy.\\

Summarizing, this approach is tied to our past light cone: it will focus on the effects 
of large-scale non-linear inhomogeneities on observables such as the 
luminosity-distance--redshift relation.
Even though there is no explicit averaging in here, the $d_{L}(z)$ is a about the luminosity and redshift of photons that travelled through inhomogeneities: their effects are therefore ``integrated out'', averaged in an implicit way.

\clearpage
\subsubsection{Theoretical side - $\rho_{\Lambda}$} \label{lambda}

We said that $\rho_{\Lambda}$ is likely connected to cosmology and that, in any case, we have to understand its meaning in order to have a satisfactory general-relativistic cosmological model.

It could be that $\rho_{\Lambda}$ is not directly related to what we measure as $\rho_{DE}$.
The latter comes from the fitting of the real inhomogeneous universe by means of a $\Lambda$CDM, while the former is about a free parameter of Einstein's equations.

As we will discuss in the next chapter, $\rho_{\Lambda}$ could be the result of a smoothing process \cite{ellis-1984, Buchert:2007ik}. This is an important issue in general relativity because, generally, every measurement involves some form of smoothing or averaging.

Following this point of view, $\rho_{\Lambda}$ could be a product of the back-reaction of inhomogeneities and have its freedom fixed by a measuring process. Inhomogeneities introduce a scale in the otherwise scale-free general relativity.
We think this is a crucial step in understanding how General Relativity effectively works in a lumpy universe.\\

As we will see, the issue is to understand how to carry out this smoothing meaningfully.
In particular if it is possible to connect $\rho_{\Lambda}$ to $\rho_{DE}$.
Within the previous approach we were only concern about the luminosity-distance--relation, while here we want, in addition to that, a phenomenological model that fits observations, in other words, we want a description by means of a mean field.

\chapter{The Back-Reaction Problem} \label{brset}

In this chapter, after setting the problem in the first section, we are going to explore the dual point of view sketched in Sect.~\ref{poview}.

The second section will suggest two possible approaches to the smoothing process.
The approach based on Buchert's equations will be carried out in Chapter \ref{pool}, while the one based on the light-cone fitting will be studied in Chapter \ref{s-c}.

The third section will be, instead, about the strictly observational point of view which will be worked out in Chapter \ref{s-c}.\\

\section{Foundations: the smoothing process}

We will here set the smoothing process.
We think that this is a crucial step in understanding how General Relativity effectively works in a lumpy universe.

This section will lay the foundations of the problem without actually telling how to pragmatically search for a back-reaction effect.
The following two sections will be about finding out how to achieve the smoothing process.
To set the problem we will follow \cite{ellis-1984,ellis-1987}.

\begin{figure}[htb]
\begin{center}
\includegraphics[width=12cm]{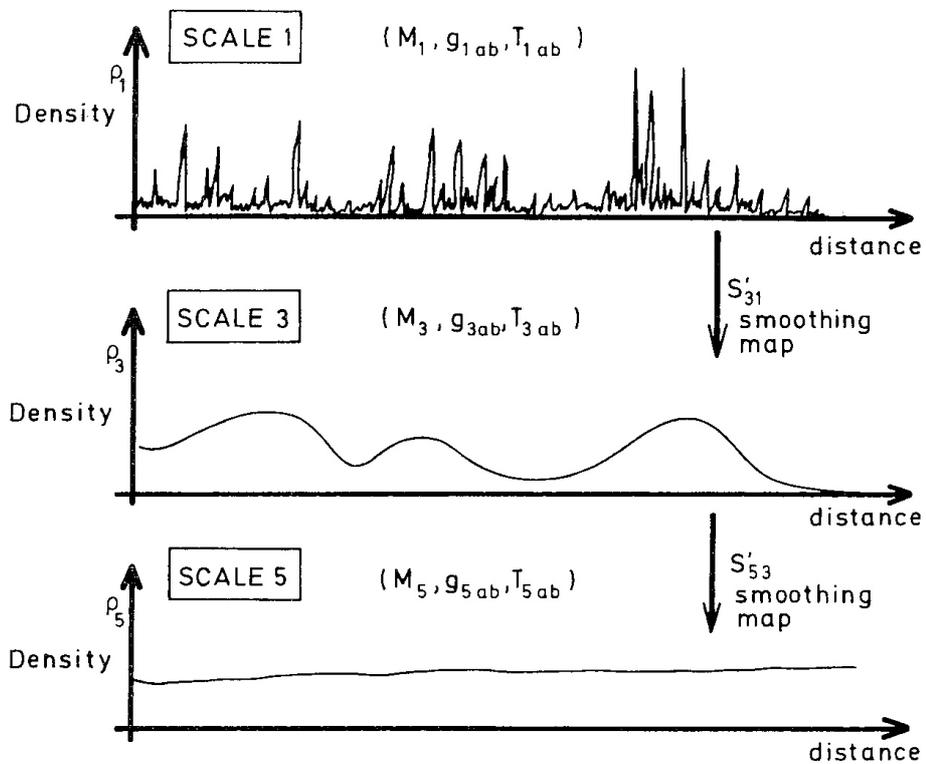}
\caption{\small \slshape Comparison of models of the same region of universe on three scales showing different amounts of detail.
Scale 1 represents all details down to stars. Scale 3 represents all details down to galaxies. Scale 5 represents only large scale features. From \cite{ellis-1984}.}
\label{ellis84}
\end{center}
\end{figure}

\clearpage
Let's look at fig. \ref{ellis84} as a starting point: it compares models of the same region of the universe on three scales showing different amounts of detail.
Scale 1 represents all details down to stars. Scale 3 represents all details down to galaxies. Scale 5 represents large scale features only.
The usual models of cosmology, like the concordance model, are shown at Scale 5.
It should be pointed out that these different matter tensors and metric tensors are intended to describe the same physical system and the same space-time, but at different scales of description.
Smoothing out inhomogeneities renormalizes the description of the universe.\\

\clearpage
General relativity tests confirm that Einstein's equations hold on Scale 1 which is the starting point in the flow chart below.

\begin{figure}[htb]
\begin{center}
\includegraphics[width=10cm]{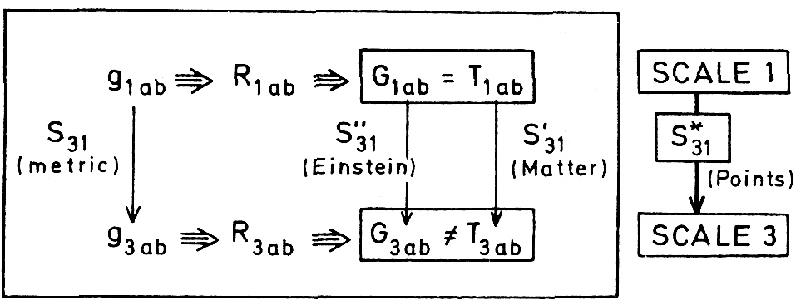}
\label{ellis84b}
\end{center}
\end{figure}
%
The flow chart summarizes tensors correspondence from Scale 1 to Scale 3.
The two models are related by:
\begin{itemize}

\item Maps $S^{*}$, which determines which points in the different underlying manifolds are related to each other by the smoothing procedure.

\item Maps $S$, which determines the metric tensor of the smoothed-out representation from the more detailed one.

\item Maps $S'$, which determines the matter tensor.

\item Maps $S''$, which determines the Einstein tensor.

\end{itemize}

Because of the non-linear nature of the fields equations, in general, the operations of smoothing will not commute with going to the field equations, that is:
\begin{equation}
S'' \neq S'
\qquad \mbox{ or } \qquad
{\langle G_{\mu \nu}(g_{\alpha \beta}) \rangle} \neq G_{\mu \nu}({\langle g_{\alpha \beta} \rangle})
\end{equation}
We can better set the problem defining a tensor $P_{3}$ representing the difference between the Einstein tensor $G_{3}$ defined from the smoothed-out metric $g_{3}$, and the smoothed-out matter tensor $T_{3}$. This correction will take care of the change of scale of description.
\begin{equation}
G_{3 \, \alpha \beta} \equiv R_{3 \, \alpha \beta} - \half  \, g_{3 \, \alpha \beta} \, R_{3}= 8\pi T_{3 \, \alpha \beta}+ 8 \pi P_{3 \, \alpha \beta}
\end{equation}
The tensor $P_{3}$ represents the effects of small-scale inhomogeneities in the universe on the dynamic behavior at the smoothed-out scale.

\subsubsection{Comments}

The implicit assumption in the usual approach is that at Scale 5 the term $P_{5}$ vanishes. Clearly this may not be true.
Moreover, $P$ does not need to obey the usual energy conditions. It is indeed an effective term. There would be no conceptual problems in having $w_{P}<-1$, for example.

It is difficult to refrain from considering the possibility of a connection between the cosmological constant and this correction term:
\begin{equation} \label{coconge}
P_{\alpha \beta}
\stackrel{?}{=}
\rho_{\Lambda} \, g_{\alpha \beta}
\qquad \mbox{that is} \qquad
w_{P}=-1
\end{equation}
$\rho_{\Lambda}$ could be indeed a product of the back-reaction of inhomogeneities and have its freedom fixed by the smoothing process which, generally, is introduced by every measurements.
Inhomogeneities introduce a scale in the otherwise scale-free general relativity.

Even if equation (\ref{coconge}) does not exactly hold, it still suggests the idea that the correction term $P$ may be the source for an effective quintessence model. In the next section we will show a concrete example about this.

\clearpage
\section{Smoothing out inhomogeneities - theoretical side}

In the last section we set the foundation of the problem of the back-reaction. Now we will consider the theoretical approach in smoothing out inhomogeneities, that is, we will look for a phenomenological model that fits observations, that is, for a description by means of a mean field.

The two approaches we are going to describe will give two conceptually different $\rho_{\Lambda}$-like terms. The problem is really to set the back-reaction problem in a physically meaningful way.

\subsection{Averaging the traces of Einstein's equations}
\label{avese}

We will describe now a way to average Einstein's equations based 
on averaging their traces~\cite{Buchert:2002ht,Buchert:1999er}.

\subsubsection{Equations}

The matter content be made of irrotational dust with four-velocity $u^{\mu}=(1,0,0,0)$, that is, the local observer is comoving with the energy flow of the fluid. Hence $T_{\mu \nu}= \rho \: u_{\mu} u_{\nu}$ and the Einstein equations are:
\begin{equation}
R_{\mu \nu}- \half g_{\mu \nu} R = 8 \pi \rho \, u_{\mu} u_{\nu}
\end{equation}
where $\rho$ is the matter energy density and $R$ the Ricci scalar.
We can work in the synchronous and comoving gauge with line element:
\begin{equation}
\label{linea}
ds^2=-dt^2 + g_{ij} \, dx^i dx^j
\end{equation}
where $t$ is cosmic time.
Greek indices run through 0...3, while Latin indices run through 1...3.

We will introduce the expansion tensor\footnote{The extrinsic curvature is the opposite} $\Theta^i_{\ j}$:
\begin{equation}
\Theta_{i j} \equiv h^{\alpha}_{\ i} \, h^{\beta}_{\ j} \, u_{\alpha ; \beta}
=u_{i ; j}
\quad \textrm{where} \quad
h^{\alpha}_{\ \beta}=g^{\alpha}_{\ \beta} + u^{\alpha} \, u_{\beta}
\end{equation}
from which we can define the volume expansion scalar $\theta$ and the traceless tensor $\sigma^i_{\ j}$, called shear:
\begin{equation}
\Theta^i_{\ j} \equiv {\theta \over 3} \, \delta^i_{\ j} + \sigma^i_{\ j}
\end{equation}

Now let us define the averaging operation which for a scalar field ${\cal F}$
is a covariant operation, given a foliation of spacetime:
\begin{equation} \label{m1}
\langle {\cal F} \rangle_D = {1 \over V_{D}} \int_D {\cal F} dV=
\frac{\int_D {\cal F} \sqrt{g}\,d^3\!x} {\int_D 
\sqrt{g}\,d^3\!x}
\end{equation}
where the average depends on content, shape and position of the comoving spatial domain of averaging, which is considered given.\\
We will introduce a dimensionless effective scale factor via the volume:
\begin{equation} \label{scala}
a_D(t) \equiv \left({V_D(t)\over V_{D_0}}\right)^{1/3}
\end{equation}
Thus, the averaged expansion rate may be written in terms of the scale factor\footnote{It follows from ${\dot J}= \theta J$ where $J=\sqrt{g}$}:
\begin{equation}
\langle\theta\rangle_D = {{\dot V}_D\over V_D} = 3 
{{\dot a}_D \over a_D} \equiv 3 H_{D}
\end{equation}
Dots denote derivatives with respect to cosmic time.

The smoothing procedure and the dynamical evolution does not commute as it is clear from this relation:
\begin{equation}
\label{commu}
{\langle {\cal F} \rangle}_D^{\bdot} - \langle \dot{{\cal F}} \rangle_D =
\langle {\cal F}\, \theta \rangle_D - \langle \theta\rangle_D\langle{\cal F}
\rangle_D
\end{equation}
Within this averaging scheme, this is the source of the backreaction effect.
Indeed, for ${\cal F}= \theta$ we will have:
\begin{equation}
{\langle \theta \rangle}_D^{\bdot} = {\langle \dot{\theta} \rangle}_D + {\langle \theta^{2} \rangle}_D - {\langle \theta\rangle}_D^{2} \geq {\langle \dot{\theta} \rangle}_D
\end{equation}
and therefore even if ${\langle \dot{\theta} \rangle}_D \leq 0$, it may happen that ${\langle \theta \rangle}_D^{\bdot} \geq 0$, that is the inhomogeneities allow the coarse-grained deceleration parameter
\begin{equation} \label{accece}
q_{D} \equiv - {3{\langle \theta \rangle}_D^{\bdot} +{\langle \theta\rangle}_D^{2} \over {\langle \theta\rangle}_D^{2}} \equiv - {\ddot{a}_{D} \over a_{D} H^{2}_{D}}
\end{equation}
to be negative in spite of its positive local value\footnote{It follows from the local Raychaudhuri's equation (\ref{ray}) that $q \equiv - ( 3 \dot{\theta} + \theta^2)/\theta^2= 6 (\sigma^2 + 2 \pi \rho)/ \theta^2 \geq 0$ and that ${\langle \dot{\theta} \rangle}_D \leq 0$}.

\subsubsection{Averaging} 
Now we can average the following local scalar equations:
\begin{eqnarray}
\label{cons}
{\bf II} + \half {\cal R} - 8 \pi \, \rho & = & 0 \qquad \textrm{energy constraint} \\
\label{ray}
\dot{{\bf I}} + {\bf I}^2 -2 \, {\bf II} + 4\pi \, \rho & = & 0 \qquad \textrm{Raychaudhuri's equation} \\
\dot\rho + {\bf I} \, \rho &= & 0 \qquad \textrm{continuity equation}
\end{eqnarray}
where two of the scalar invariants of the expansion tensor, namely its trace and the dispersion of its diagonal components, are:
\begin{eqnarray}
{\bf I} &\equiv& \Theta^{\ell}_{\,\;\ell} = \theta \cr
{\bf II} &\equiv& {1\over 2}\left(\theta^2 - \Theta^{\ell}_{\,\;k}\Theta^{k}_{\,\;\ell}\right)
= {1\over 3}\theta^2 - \sigma^2 \label{invari}
\end{eqnarray}
where $\sigma^2 \equiv \frac{1}{2}\sigma^i_{\ j} \sigma^j_{\ i}$. The ${\cal R}$ in (\ref{cons}), and in the following equations, is the trace of the spatial Ricci tensor\footnote{\  ${\cal R}_{i j}$ is the Ricci tensor of the metric $g_{ij}$ of the hypersurface and not $R_{i j}=h^{\alpha}_{\ i} \, h^{\beta}_{\ j} \, R_{\alpha \beta}$.} ${\cal R}_{i j}$.
Actually, the density is a scalar invariant as well, $T=g^{\mu \nu}T_{\mu \nu}= -\rho$.
The result of the averaging is:
\begin{eqnarray}
\label{mediate1}
\langle{\bf II}\rangle_D + \half {\langle {\cal R} \rangle}_D - 8\pi \langle \rho\rangle_D &=& 0\\
{\langle {\bf I} \rangle}_D^{\bdot} + {\langle {\bf I} \rangle}_D^2 - 2 \langle{\bf II}\rangle_D + 4\pi \langle\rho\rangle_D &=& 0\\
\label{mediate3}
\langle \rho \rangle^{\bdot}_D + {\langle {\bf I} \rangle}_D \langle\rho\rangle_D &=& 0
\end{eqnarray}
that is, provided we express the equations in terms of the invariants (\ref{invari}), the averaged quantities obey the same equations as the local ones in spite of the non-commutativity of the averaging procedure and the dynamical evolution, which is expressed by the commutation rule (\ref{commu}).\\
The reason for this nontrivial property is the special type of nonlinearities featured by the gravitational system, in particular the nonlinearity in $\theta$ contained in Raychaudhuri's equation.\\
Summarizing, the traces of Einstein's equations, expressed by means of the scalar invariants ${\bf I}$, ${\bf II}$ and $\rho$, are not  structurally affected by the averaging procedure: the corrections are in the renormalized invariants and not in their relations.
This result has some consonance with the idea of $\rho_{\Lambda}$ fixed by the smoothing process: not even in that case Einstein's equations are structurally changed.

The extra-terms produced by the averaging will appear if we rewrite the first two equations in a more familiar form:
\begin{eqnarray}
\label{cq1}
H^{2}_{D} &=& {8\pi \over 3} \langle \rho\rangle_D -{1 \over 6} {\langle {\cal R} \rangle}_D -{1 \over 6} Q_D\\
\label{cq2}
{{\ddot a}_D \over a_D} &=&  -{4\pi \over 3} \langle\rho\rangle_D + {1 \over 3} Q_{D}
\end{eqnarray}
where the (kinematical) back-reaction source term is:
\begin{equation}
Q_{D} \equiv 2 \langle{\bf II}\rangle_D - {2\over 3} \langle {\bf I}\rangle_D^2 \;=\;
{2\over 3} \left( \langle\theta^{2}\rangle_D - \langle\theta\rangle_D^{2} \right) - 2\langle\sigma^2\rangle_D
\end{equation}
The equations (\ref{cq1}-\ref{cq2}) can once more be rewritten in a dimensionless form:
\begin{eqnarray}
1 &=& \Omega_{M}^{D}+\Omega_{{\cal R}}^{D}+\Omega_{Q}^{D} \label{terz} \\
\label{dece}
q_{D} &=& \half \Omega_{M}^{D}+2 \Omega_{Q}^{D}
\end{eqnarray}
where we have defined the density parameters $\Omega^{D}_{M}=8 \pi {\langle \rho\rangle}_D / 3 H^{2}_{D}$, $\Omega^{D}_{Q}={- Q_{D}/6 H^{2}_{D}}$ and $\Omega^{D}_{{\cal R}}={- {\langle {\cal R} \rangle}_D/6 H^{2}_{D}}$ in the usual way.

The consistency of equations (\ref{cq1}-\ref{cq2}) requires that $Q_{D}$ and ${\langle {\cal R} \rangle}_D$ satisfy the integrability condition:
\begin{equation}
\label{condi}
\left(a_D^6 Q_{D} \right)^{\bdot} + a_D^4 \left(a_D^2 \langle {\cal R} \rangle_D\right)^{\bdot} \;=\;0
\end{equation}

\subsubsection{Effective back-reaction source}

Finally we can rewrite again (\ref{cq1}-\ref{cq2}) in term of the effective backreaction source:
\begin{eqnarray}
\label{br1}
H^{2}_{D} &=& {8\pi \over 3} \left( {\langle \rho\rangle}_D + \rho_{BR} \right) \quad \Longleftrightarrow \quad \Omega^{D}_{M}+ \Omega^{D}_{BR}=1 \\
\label{br2}
{{\ddot a}_D \over a_D} &=&  -{4\pi \over 3} \sum_{i}(\rho_{i}+3 p_{i})
\end{eqnarray}
where the effective backreaction source energy density and pressure are:
\begin{equation} \label{effe1}
\rho_{BR}=-{Q_{D} \over 16 \pi}-{{\langle {\cal R} \rangle}_D \over 16 \pi}
\qquad \qquad
p_{BR}=-{Q_{D} \over 16 \pi}+{1 \over 3}{{\langle {\cal R} \rangle}_D \over 16 \pi}
\end{equation}
The equation of state will be:
\begin{equation} \label{effe2}
w_{BR}={p_{BR} \over \rho_{BR}}=\frac{-Q_{D} + {\langle {\cal R} \rangle}_{D}/3}{-Q_{D} - {\langle {\cal R} \rangle}_D}
\end{equation}
We have therefore the non trivial result that the standard Friedmann equations are modified only by an extra source beside their dependence on the scale of averaging. Unfortunately the system of the equations (\ref{br1}-\ref{br2}) is not closed, the smoothing procedure has washed out too many details. To close the system it will be sufficient to have the equation of state: we will come back to this point later in this section.

These last equations look like the corresponding ones for quintessence, (\ref{lphi}-\ref{rho-p}), with kinetic energy $T_{\phi} \sim - Q_{D}$ and potential $V_{\phi} \sim - {\langle {\cal R} \rangle}_D$. For example, with zero potential (i.e. ${\langle {\cal R} \rangle}_D=0$) we have from (\ref{condi}) $T_{\phi} \sim 1/a_{D}^{-6}$ which is the usual kination with $w=1$.
Indeed a topic of research could be to analyze the problem with the quintessence formalism, trying to find potentials coupling $\phi$ to the inhomogeneities
or some other connected observables.
Because $\phi$ is an effective field, it does not suffer of the problems examined in Sect. \ref{quiquoqua}. It will be possible also to have $w<-1$. See \cite{Buchert:2006ya} for a work in this direction. 

\subsubsection{Acceleration in a dust universe} \label{acce}

The condition to have acceleration follows from the acceleration equation (\ref{cq2}) and (\ref{dece}):
\begin{equation}
- \Omega_{Q}^{D} >{ \Omega_{M}^{D} \over 4}
\qquad \textrm{or} \qquad
Q_{D}>4 \pi \langle\rho\rangle_D
\end{equation}
However, having significant backreaction is not enough: it is necessary that $Q_{D}$ and ${\langle {\cal R} \rangle}_D$ are strongly coupled, otherwise $Q_{D}$ will fall in kination and will become quickly negligible.
Therefore ${\langle {\cal R} \rangle}_D$ will evolve in a different way with respect to the usual curvature $k$: only if $Q_{D}$ and ${\langle {\cal R} \rangle}_D$ are decoupled it will be ${\langle {\cal R} \rangle}_D \sim 1/a_{D}^{2}$. So the CMB tells us only that the part of ${\langle {\cal R} \rangle}_D$ which evolves as $1/a_{D}^{2}$ is negligible.

\begin{figure}[htb]
\begin{center}
\includegraphics[width=13cm]{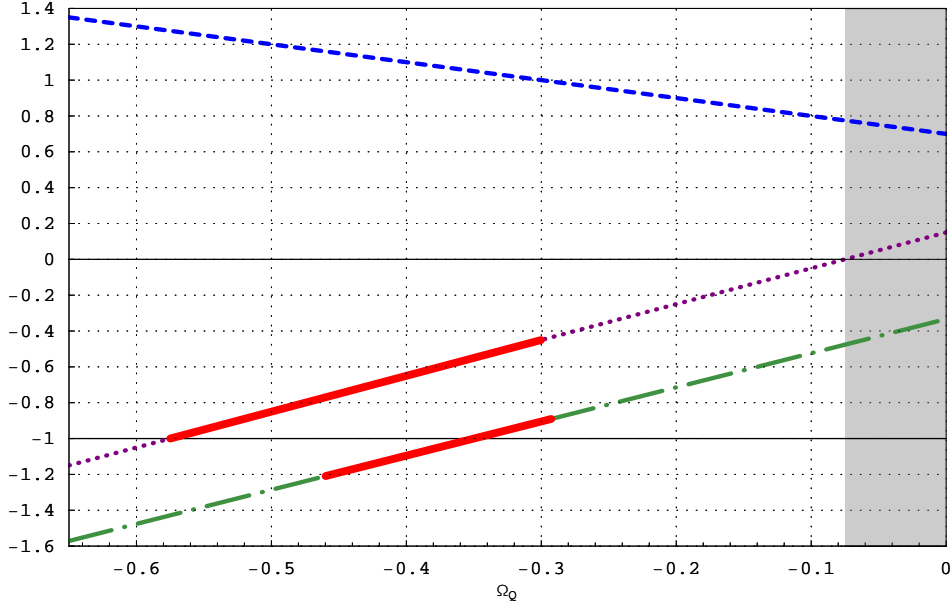}
\caption{\small \slshape We have plotted for $\Omega_{M}^{D}=0.3$ the value of the principal parameters as a function of $\Omega_{Q}^{D}$. The dashed line is $\Omega_{{\cal R}}^{D}$, the dotted line is $q_{D}$ and the dot-dashed line is $w_{BR}$. The measured values \cite{Riess:2004nr} are marked in solid red, while in the gray area there is no acceleration.}
\label{limitiq}
\end{center}
\end{figure}

Assuming $\Omega_{M}^{D}=0.3$ we have plotted in fig. \ref{limitiq} the value of the main parameters as a function of $\Omega_{Q}^{D}$ to see if they are in contrast with the present day data.
From the plot we see that the following values:
\begin{equation}
-0.45 \lesssim \Omega_{Q}^{D} \lesssim -0.3
\qquad \textrm{and} \qquad
1 \lesssim \Omega_{{\cal R}}^{D} \lesssim 1.2
\end{equation}
are in agreement with the data. We stressed again that the value we have found for $\Omega_{{\cal R}}^{D}$ is not ruled out by the CMB because the latter constrains only the part of $\Omega_{{\cal R}}^{D}$ which evolves as $1/a_{D}^{2}$. The volume $D$ refers to the scale at which the parameters are measured.

\subsubsection{A conjecture}

We can try to apply the conjecture that $P_{\alpha \beta}= \rho_{\Lambda} \, g_{\alpha \beta}$ to this formalism, that is:
\begin{equation} \label{congeq}
w_{BR}=-1
\end{equation}
We are now able to close the system (\ref{br1}-\ref{br2}):
\begin{equation} \label{conge2}
\Omega_{Q}^{D} =- {1 \over 3} \Omega_{{\cal R}}^{D} =\textrm{const}(a_{D})
\qquad \textrm{or} \qquad
Q_{D}=- {1 \over 3} {\langle {\cal R} \rangle}_D=\textrm{const}(a_{D})
\end{equation}
where the last equalities derive from the constance of $\rho_{BR}$ which is implied by (\ref{congeq}).
The solution (\ref{conge2}) satisfies the integrability condition (\ref{condi}).

\begin{figure}[htb]
\begin{center}
\includegraphics[width=13cm]{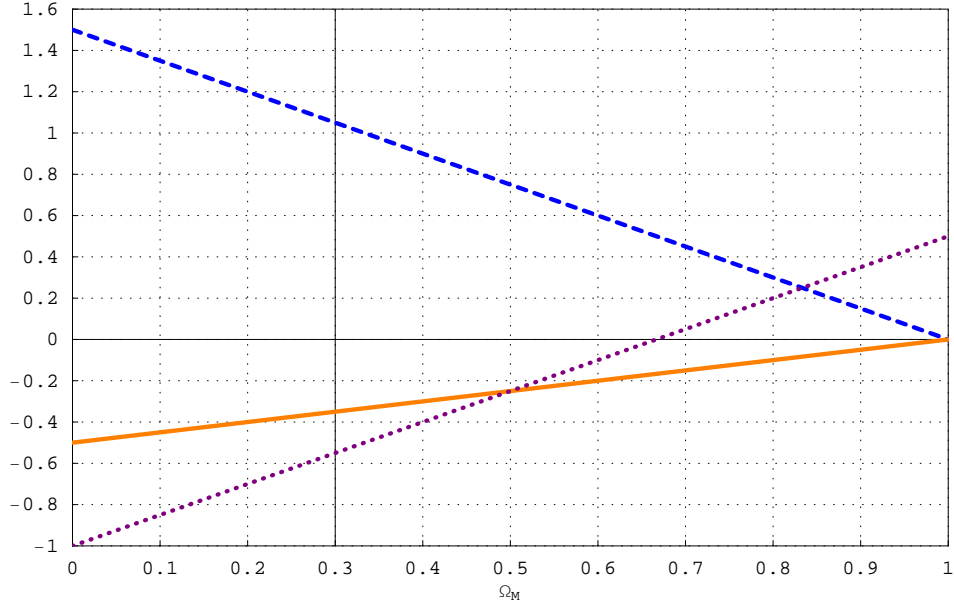}
\caption{\small \slshape Assuming $w_{BR}=-1$, we have plotted the principal parameters as a function of $\Omega_{M}^{D}$. The dashed line is $\Omega_{{\cal R}}^{D}$, the solid line is $\Omega_{Q}^{D}$ and the dotted line is $q_{D}$.}
\label{conge}
\end{center}
\end{figure}

It is possible to put a remedy to our ignorance about the dependence of the observable from $a_{D}$ expressing the scale as a function of the matter parameter: $a_{D}=a_{D}(\Omega_{M}^{D})$, so that to give $\Omega_{M}^{D}$ is as to give the domain of averaging.
We can then plot in fig. \ref{conge} this exact solution:
\begin{equation}
\Omega_{Q}^{D} = \half \Omega_{M}^{D} - \half
\qquad
\Omega_{{\cal R}}^{D}={3 \over 2}-{3 \over 2}\Omega_{M}^{D}
\qquad
q_{D}={3 \over 2}\Omega_{M}^{D}-1
\end{equation}

Summarizing:
\begin{eqnarray}
\Omega_{M}^{D}=0.3 \qquad &\Omega_{BR}^{D}&=0.7 \qquad w_{BR}=-1 \qquad q_{D}=-0.55 \nonumber \\
\textrm{where} \qquad &\Omega_{BR}^{D}&=\Omega_{Q}^{D}+\Omega_{{\cal R}}^{D} \\
\textrm{and} \qquad  &\Omega_{Q}^{D}&=-0.35 \qquad \Omega_{{\cal R}}^{D}=1.05 \nonumber
\end{eqnarray}
We are stressing again that the volume $D$ refers to the scale at which the parameters are measured.

\subsubsection{Bare and dressed cosmological parameters}

There is one more remark about the theoretical framework just described: the parameters of eq. (\ref{terz}), which is:
\begin{equation}
\Omega_{M}^{D}+\Omega_{{\cal R}}^{D}+\Omega_{Q}^{D}=1
\end{equation}
are not directly accessible to observations because the analogous FRW ones implicitly assume a constant curvature space \cite{Buchert:2002ht}.

Before making this comparison we need to average the inhomogeneous geometry into a constant curvature one.
The following sketch gives an idea of the process. It seems as if we have to correct only for volume effects when dressing the parameters.

\begin{figure}[htb]
\begin{center}
\includegraphics[width=8cm]{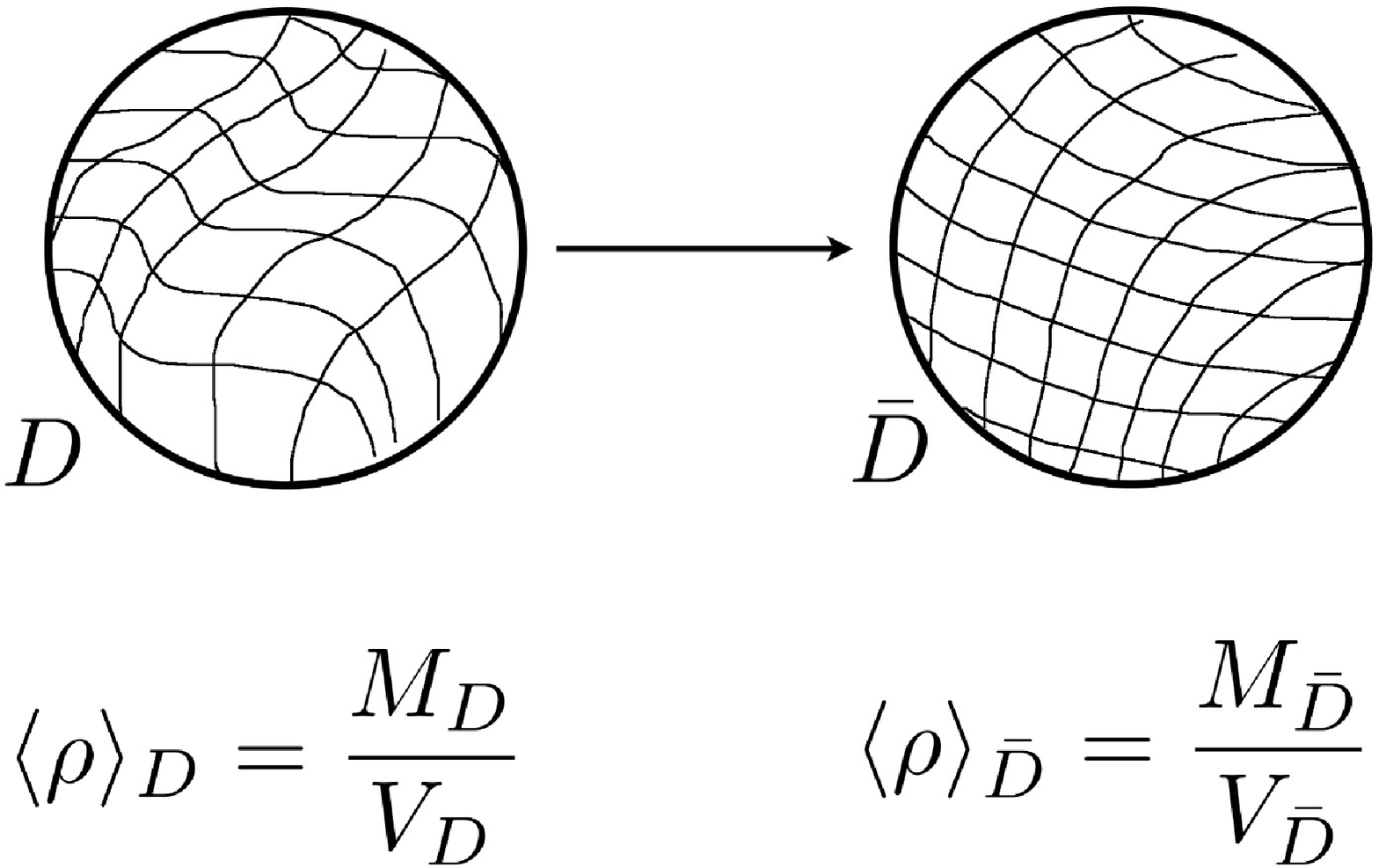}
\label{bare}
\end{center}
\end{figure}

However, averaging the curvature, we will find:
\begin{equation} \label{barec}
{\langle {\cal R} \rangle}_{\bar{D}}=\left ( {V_{\bar{D}} \over V_{D}} \right )^{-2/3} {\langle {\cal R} \rangle}_D  -  Q^{R}_{D}
\end{equation}
that is, beside volume effects, there is an extra correction named curvature back-reaction $Q^{R}_{D}$. Its analytical expression is similar to the kinematical back-reaction $Q_{D}$ but there are, instead of the scalar invariants of the extrinsic curvature, the ones of the intrinsic curvature.

From  eq. (\ref{barec}) we can understand the 
physical content of geometrical averaging. It makes transparent that, in the smoothed model, 
the averaged scalar curvature is `dressed' both by the volume effect 
and by the curvature backreaction effect itself. The volume effect is expected
precisely in the form occurring in (\ref{barec}), if we think of comparing two regions
of distinct volumes, but with the same matter content, in a constant
curvature space (remember that a constant curvature space 
is proportional to the inverse square of the radius of curvature, hence the volume-exponent 2/3).
Whereas the backreaction term encodes the deviation of
the averaged scalar curvature from a constant curvature model, e.g., a FLRW
space section.

Therefore the relation we have to use to compare results to experimental data is:
\begin{equation}
\bar{\Omega}_{M}^{\bar{D}}+\bar{\Omega}_{{\cal R}}^{\bar{D}}+\bar{\Omega}_{Q}^{\bar{D}}=1
\end{equation}
%

\subsection{The fitting problem} \label{fritto}

The other theoretical approach in smoothing out inhomogeneities we will consider is based on the fitting problem \cite{ellis-1987}.

There are broadly speaking, two distinct 
approaches, which have been applied to understand the large-scale structure 
of the universe.

\begin{figure}[h!]
\begin{center}
\includegraphics[width=10cm]{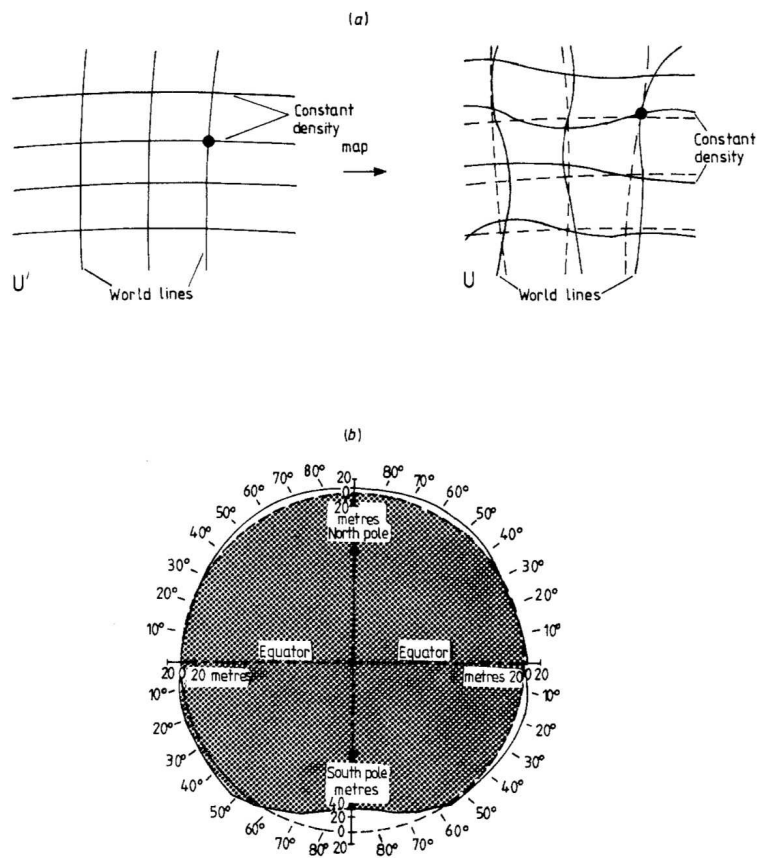}
\caption{\small \slshape (a) An exactly uniform and spherically symmetrical FRW universe $U'$ mapped into the lumpy universe $U$ so as to give the best possible fit. (b) An exactly spherical sphere fitted to the lumpy world to give the best fit possible. From \cite{ellis-1987}.}
\label{ellisstoeger}
\end{center}
\end{figure}

The standard approach is to make the assumption of spatial homogeneity 
and isotropy on a large enough scale. Then it follows that the universe 
is represented by a FRW model.\\
The main problem of this approach is that it simplifies the way the real 
lumpy universe should be averaged.
It does not deal with the correction term $P$ introduced in the previous section.\\
The concordance model, however, fits experimental data very well: the direct
consequence of its success is, indeed, that the 
isotropic and homogeneous $\Lambda$CDM model is a good {\it observational} 
fit to the real inhomogeneous universe.
And this is, in some sense, a verification of the cosmological principle of of spatial homogeneity 
and isotropy on a large enough scale: 
the inhomogeneous universe can be described by means of a isotropic and 
homogeneous solution.
However this does not imply that a primary dark energy component really 
exists, but only that it exists effectively as far as the observational fit is concerned.\\
If dark energy does not primary exist, its evidence coming from the 
concordance model would tell us that the purely-matter inhomogeneous model 
has been renormalized, from the observational point of view (luminosity 
and redshift of photons), into a homogeneous $\Lambda$CDM model.\\
So, following this reasoning, it is not true, as stated in the discussion about the previous section, that the standard approach assumes $P$ to vanish.
The standard approach simply takes the $P$ correction term as a primary source in the energy-momentum tensor.

The other approach is to make no a priori assumptions of global symmetry
and build up our universe model only on the basis of astronomical 
observations. The main problem with such an approach is the practical 
difficulty in implementing it.

An approach which is intermediate between the two outlined 
above is based on the fitting procedure. 
It asks the question about which is the FRW 
model which best fits our lumpy universe, Fig.~\ref{ellisstoeger} (a). This question will 
lead to a procedure that will make us better understand how to interpret the large-scale FRW solution.\\
The approach resembles that used in geodesy, where a perfect sphere is 
fitted to the ``pear-shaped'' earth; deviations of the real earth from the idealised model 
can then be measured and characterised, Fig.~\ref{ellisstoeger} (b).\\
The best-fit will be implemented along the past light cone. This because a 
meaningful fitting procedure should be related directly to astronomical 
observations.\\
The FRW model we have in mind to use in the fitting is a quintessence-like model with an effective source with varying 
equation of state. We will develop these ideas in Section~\ref{fitti}.

\clearpage
\subsection{Discussion}

Let's start discussing the approach about averaging the traces of Einstein's equations.
This scenario is theoretically consistent, but there are two main issues.
\subsubsection{Is $Q_{D}$ observationally relevant?}
We have to make it clear if this approach gives an observationally meaningful back-rection.
We have to understand how the $\rho_{\Lambda} \rightarrow \rho_{BR}$ that comes from this scenario is related to what we measure as $\rho_{DE}$.

A first important observation is that we should be mainly concerned about the luminosity-distance--redshift relation $d_{L}(z)$ and look at the effects inhomogeneities have on it. Therefore it seems approximate to average at constant cosmic time, we should instead average along the light cone.\\
To overcome this problem we will use the approach sketched in Sect. \ref{fritto}. As we will present in Chapter \ref{s-c}, we will use a fitting scheme based on light-cone averages.

The scenario developed by Buchert, beside giving precious hints toward the understanding of the back-reaction problem, can, however, be conceptually better understood if we regard the spatial average as an ensemble average \cite{Kolb:2005da}.\\
The inhomogeneities are seen as variables that take random values over different realizations of a volume D at a fixed location or over different locations of the volume D in a single realization (fair sample hypothesis).
In order to proper represent an observer in the probability ensemble or a randomly placed observer, we should average his measurements over the possible realizations.
Averaging at constant cosmic time over a volume of size comparable with present-day Hubble volume has, therefore, to be understood as averaging over the possible realization of the volume D inside the Hubble volume.
This averaging is necessary if there is a big variance, which is however not expected, in the cosmic realizations of the volume D and has to be intended from a statistical point of view.\\
However, this approach is not directly related to the averaged description the formalism developed by Buchert is about. It should be seen as a possible interpretation of the latter.

\subsubsection{Quantitative estimations}

Then, we have to check the strength and the properties of this effective $\rho_{BR}$ quantitatively in order to see if it can explain the concordance model and how far it is from the standard cosmological constant.

The following approaches have been mainly studied.
A first possibility is by means of the perturbation theory.
It has been found an instability in the perturbative expansion \cite{Kolb:2005da}.
Since the perturbation approach breaks down, it is not possible to predict on firm grounds that backreaction is responsible for the present-day acceleration of the universe. However, it is intriguing that such an instability shows up only recently in 
the evolution of the universe.

To overcome the problems coming from dealing perturbatively with non-linear perturbations, attention has been given to exact inhomogeneous solutions of Einstein's equations.\\
In particular the Lema\^itre-Tolman-Bondi (LTB) solution has been studied 
extensively in the literature  \cite{Alnes:2006pf, Biswas:2006ub, Apostolopoulos:2006eg, Alnes:2005rw, Celerier:1999hp,
Mansouri:2005rf, Vanderveld:2006rb, Rasanen:2004js, Tomita:2001gh, Chung:2006xh, Kai:2006ws}.
We will discuss this in Chapter~\ref{pool}.

\begin{figure}[htb]
\begin{center}
\includegraphics[width=15.5cm]{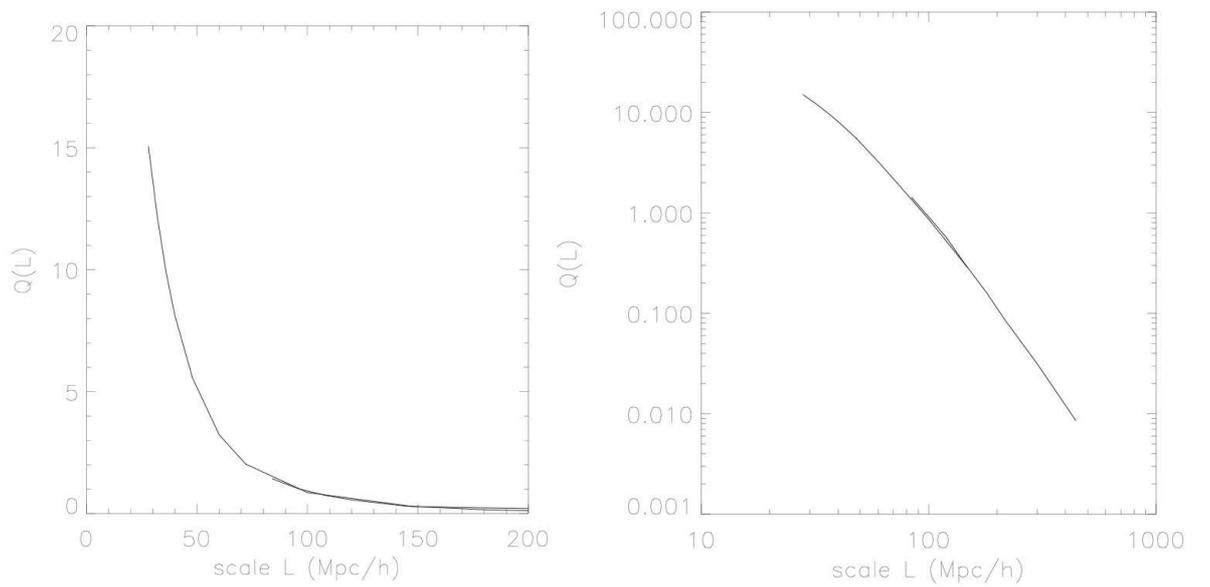}
\caption{\small \slshape The expected ``backreaction'' i.e., the 
quantity $|Q_{D}|/4 \pi \langle \rho\rangle_D$ 
against the scale $L \equiv a_{D} = V_{D}^{1/3}$
(measured in Mpc/h) in linear/linear (left panel) and log/log (right panel) 
format for a Standard--CDM model box of $1.8$Gpc/h in size.
This dimensionless quantity is still of the 
same order as the actual matter density on scales around $100$Mpc/h \cite{Buchert:1996tf}.} 
\label{nunume}
\end{center}
\end{figure}

Finally, numerical estimates were made in order to check quantitatively the importance of $Q_{D}$.
A plot of the absolute value of $Q_{D}$, normalized by the global 
mean density, against scale for a Standard--CDM is shown in Fig.~\ref{nunume}.
The absolute values are shown there because the calculations were done in the newtonian approximation with periodic boundary conditions and therefore $Q_{D}$  is, by assumption, zero: $Q_{D}$ can indeed be written as a  total divergence.
The absolute value is then an estimate of the expected ``backreaction'' on some scale.\\
From the plot we can see that the magnitude of the ``backreaction'' source term is of the same order 
as the mean density and higher on scales $\lambda < 100$Mpc/h for SCDM. It quickly drops to a $10\%$ effect
on scales of $\lambda \approx 200$Mpc/h.\\
This is however only a hint toward a proper general relativistic estimation. The newtonian approximation neglects indeed important effects like, for example, the coupling (\ref{condi}) between $Q_{D}$ and ${\langle {\cal R} \rangle}_D$ without which there cannnot be a sizeable back-reaction.

\clearpage
\section{Smoothing out inhomogeneities - observational side}

The other possible approach focuses directly on the luminosity-distance--redshift relation $d_{L}(z)$ and looks at the effects inhomogeneities have on it. 
To carry out this analysis of the back-reaction it is necessary to have an inhomogeneous model to work with, in particular it is desirable to have an exact inhomogeneous solution of Einstein's equations.

We remark that the $d_{L}(z)$ is a about the luminosity and redshift of photons that travelled through inhomogeneities and have their effects ``integrated out''. In other words, even if we are not averaging explicitly the universe, the luminosity-distance--redshift relation has an implicit meaning of average in itself.\\
This is the reason why this approach can be regarded as a possible answer to the smoothing proceses as well as the approches of the previous section are.

It has been shown that the LTB solution can be
used to fit the observed $d_{L}(z)$  without the need for dark energy (for
example in \cite{Alnes:2005rw, Alexander:2007xx}).  To achieve this result, however, it is necessary
to place the observer at the center of a rather big underdensity.
This is in contrast with the experimental verifications of the cosmological principle.\\
To overcome this  fine-tuning problem, in Chapter \ref{s-c} a Swiss-cheese model will be used where the observer is in the cheese and looks through Swiss-cheese holes constructed out of an LTB solution.

It is of great interest to compare this physically meaningful approach with the light-cone average fitting we will use.
Both are indeed tied to observations.

\chapter[A ``homogeneous'' LTB model]{The backreaction effect through a ``homogeneous'' dust LTB model}
\label{pool}

In this chapter we will try to build an inhomogeneous universe model searching for a back-reaction effect of sub-Hubble perturbations.
In particular we will aim at finding the average dynamics effects described in Sect.~\ref{avese}.

We will make use of the dust LTB model, described in Appendix~\ref{ltbm}.
It has been shown that the LTB solution can be
used to fit the observed $d_{L}(z)$  without the need of dark energy (for
example in \cite{Alnes:2005rw}).  To achieve this result, however, it is necessary
to place the observer at the center of a rather big underdensity.
This is in contrast with the experimental verifications of the cosmological principle.\\
To overcome this  fine-tuning problem we will build a model that does not single out the center as a preferred place.
To this end, we paid attention to choose appropriate densities and curvature profiles.

\begin{figure}[htb]
\begin{center}
\includegraphics[width=15cm]{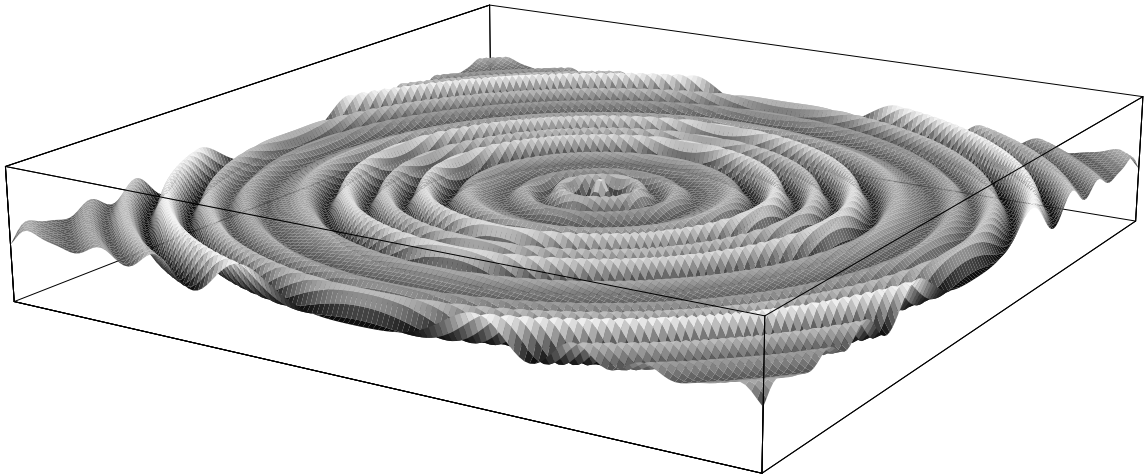}
\caption{\small \slshape An example of the density profile (\ref{seni}) we are going to use in the Lema\^{i}tre-Tolman-Bondi model.}
\label{densi}
\end{center}
\end{figure}

To be clear, the density profile we have in mind is:
\begin{equation} \label{seni}
\rho=\sum_{i}c_{i}\cos k_{i}r+\tilde{\rho}
\end{equation}
In fig.~\ref{densi} there is a sketch of that density profile.
%
%
This shape is intermediate between a cubic lattice without preferred positions and a standard LTB model in which the center is singled out by its different density and/or spatial curvature. It is a sort of exact ``spherical lattice'': even though the inhomogeneities of the density are centered on the center of symmetry, an observer moving around such a universe should not note sizeable differences between one place and the other.

\section{Flat case} \label{flafla}

We will now consider a flat LTB model. This analysis will be useful when dealing with more elaborate models with $E \neq 0$.

\subsection{Our model}

The density we will use has only inhomogeneities of one size:
\begin{equation} \label{densia}
\rho(r, \bar{t})=\rho_{M}+\rho_{M}\cos k \, r
\end{equation}
and does not have the weak singularity mentioned in \cite{Vanderveld:2006rb}.
This density profile is clearly nothing more than a toy-model: the present-day universe is far more inhomogeneous. However, it will have the main features in order to look for a back-reaction effect.\\
In (\ref{densia}) $k=2 \pi /\lambda$ and we will choose $\lambda=0.05$ which, with the units of Table \ref{units}, corresponds to $420$ Mpc.
The voids will be roughly 20 times smaller than $r_{BB}$, the comoving distance traveled by a photon since the big bang.

As said in Appendix~\ref{ltbm}, we fixed the gauge freedom by choosing $\bar{t}(r)=\bar{t}$ and $Y(r, \bar{t})=r$.

We will show how its density evolves in the next section when we will study the curved case. In the flat case the evolution is indeed reversed: structures are not forming, but spreading out. We will show better this in the Chapter~\ref{s-c}.

\subsection{Back-reaction effects}

We built this model in search of the back-reaction effects described in Sect.~\ref{avese}: we are interested in the quantities on the right hand side of eq. (\ref{cq1}), which in the units of Table~\ref{units} is:
\begin{equation}
H^{2}_{D} = {4 \over 9} \langle \rho\rangle_D -{1 \over 6} {\langle {\cal R} \rangle}_D -{1 \over 6} Q_D
\end{equation}
We have given the analytical expression of their constituents for a LTB model in Appendix~\ref{ltbm}.

In the flat case one finds out immediately from (\ref{riccis}) that $\langle {\cal R} \rangle_D=0$ and with a short calculation that $Q_{D}=0$, thus there is no backreaction at all i.e. the expansion rate of the LTB model and the one of the FRW model are the same.
It exactly holds:
\begin{equation}
q_{D}=1/2
\end{equation}
These results do not depend on the choice of the free functions of the LTB model.
This is not trivial: in the flat case the local equations are invariant under the averaging process: this is the cause of the absence of back-reaction.

However is not possible to conclude that there is no backreaction in the curved case.
In the flat case we have, from eq. (\ref{scala}), $a_{D}=Y$ and this can be seen, alternatively, as the reason why we found no back-reaction: the equation for $Y$ are already averaged as pointed out in Appendix~\ref{ltbm} and therefore there can not be any different average dynamics.
In the curved case, however, $a_{D} \neq Y$. The expansion is now:
\begin{equation}
\langle \theta \rangle_D = {1 \over V_{D}} \int_D \theta \, dV=
\frac{\int_{0}^{R}dr \; \theta \; Y^{2}Y' / W}
{\int_{0}^{R} dr \; Y^{2}Y' / W}
= {{\dot V}_D\over V_D} = 3 {{\dot a}_D \over a_D} = 3 H_{D}
\end{equation}
where $R$ marks the outer shell of the spherical domain $V_{D}$. It is not anymore true that $H_{D}=\dot{Y}/Y$.

\section{Curved case}
Introducing the curvature enriches the model with respect to the flat case. We need to introduce the curvature for two reasons: the first is to have a more natural evolution and the second is to have a non trivial domain dynamics.

\subsection{Our model}

Except for the non-zero curvature, we will use the same model studied in the flat case. 
As explained in Appendix~\ref{ltbm}, the LTB curvature $E$ is connected to the FRW curvature $k$ by $k=-2E/r^{2}$.
(It is not normalized to unity.)
If $E \sim \textrm{const}$, away from the observer there will be no clash with the CMB measurement of $k=0$.
However, the center will be singled out in contrast with our idea of ``homogeneous'' model.
To avoid this, when we will introduce the curvature, we will demand to have $E=0$ on the average.

We could think to use an oscillatory curvature, similarly to the behavior of the density. However there is too much freedom in the choice of $E$: it is, together with $\rho$ and $\bar{t}$, a free function of the model.\\
$E$, however, is fixed by the initial conditions: requesting $E=0$ forced the velocity of the shells to compensate the inhomogeneities to have a flat space.
Thus we can think that a more natural situation will happen if such a compensation does not take place.
To have a realistic evolution, we demand therefore that there are no initial peculiar
velocities at time $\bar{t}$, that is, to have an initial expansion $H$
independent of $r$. From Eq.\ (\ref{cucu}) this implies:
\begin{equation} 
\label{Er}
E(r)=\frac{1}{2} H_{FRW}^2(\bar{t}) \, r^{2}-\frac{1}{6\pi}\frac{M(r)}{r}
\end{equation}
The graph of $E(r)$ chosen in this way is shown in Fig.\ \ref{E}: as you can see we have exactly
an oscillating behavior without the arbitrariness of requesting it.  As seen from
the figure, the curvature $E(r)$ is small compared with unity. Indeed, in many
formulae $W=(1+ 2 E)^{1/2}\simeq 1+E$ appears, therefore one should compare $E$
with $1$. In spite of its smallness, the curvature will play a crucial role to
allow a realistic evolution of structures, as we will see in the next section.

\begin{figure}[htb]
\begin{center}
\includegraphics[width=16.2 cm]{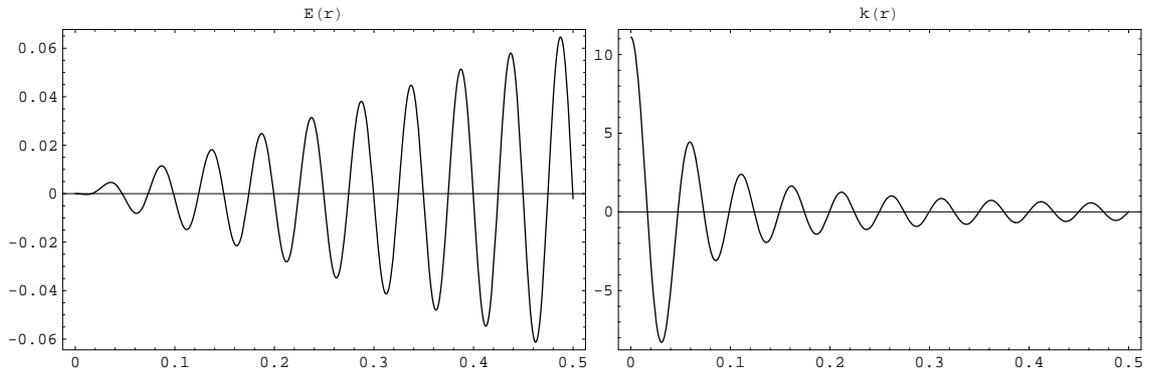}
\caption{\small \slshape Curvature $E(r)$ and $k(r)$ necessary for the initial conditions of no
peculiar velocities.}
\label{E}
\end{center}
\end{figure}

\subsection{The dynamics}

\begin{figure}[htb]
\begin{center}
\includegraphics[width=16 cm]{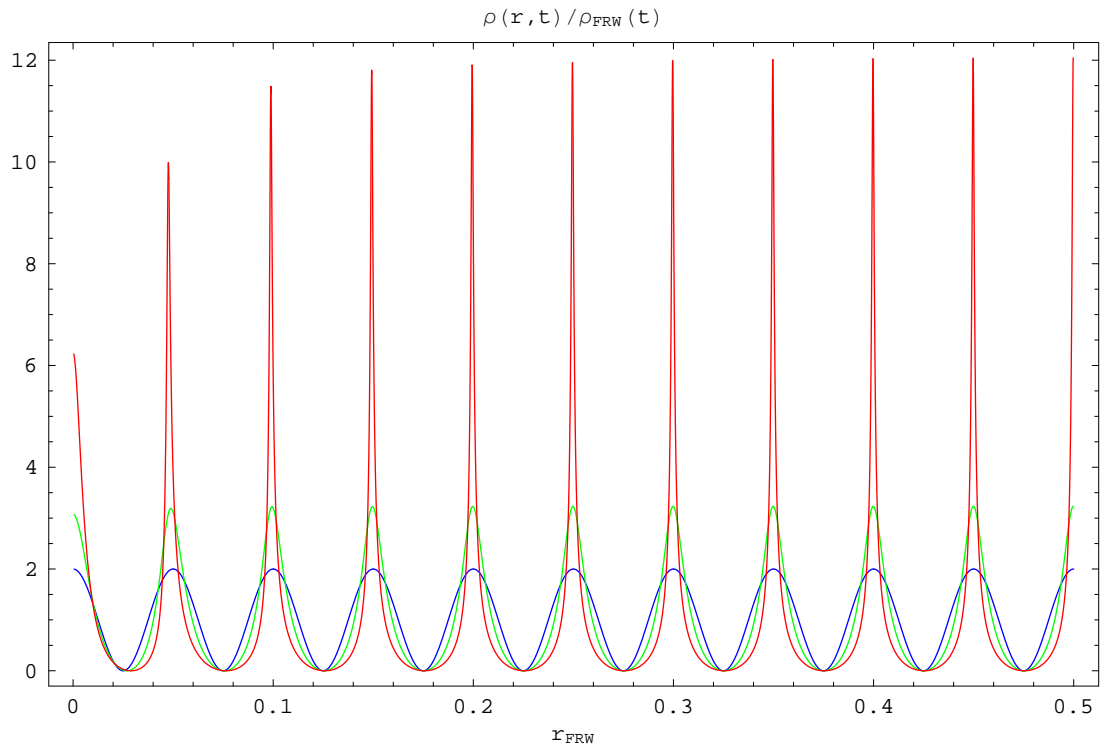}
\caption{\small \slshape Behavior of the
density profiles $\rho(r,t)$ with respect to $r_{FRW}=Y(r,t)/a(t)$, at times $t=\bar{t}=-0.8$ (blue), 
$t=-0.4$ (green) and $t=t_0=0$ (red).
The values of $\rho_{FRW}(t)$ are $1,\ 2.8,$ and $25$, for $t=0,\
-0.4,\ -0.8$,  respectively. }
\label{densifi2}
\end{center}
\end{figure}

In Fig.\ \ref{densifi2} we show the evolution of $\rho(r,t)$ for three times: 
$t=\bar{t}=-0.8$ (the Big Bang is at $t_{BB}=-1$), $t = -0.4$, and $t=0$ 
(corresponding to today).
As you can see, overdense regions start contracting and become thin shells (mimicking structures), while underdense
regions become larger (mimicking voids), and eventually occupy most of the
volume.

The densities are normalized with respect to the average density $\langle \rho\rangle_D \simeq \rho_{FRW}(t)$ where the latter is the density of the FRW model which is at the initial time $\rho_{FRW}(\bar{t})=\langle \rho(r,\bar{t})\rangle_D$.
The equalities are not exact because the oscillations in the density smooth out not exactly within a finite volume: this is due to the growing of the surface area with the respect to the radius.

Moreover, remember that $r$ is only a label for the shell whose Euclidean position at time
$t$ is $Y(r,t)$.  In Fig.\ \ref{densifi2} we have normalized $Y(r,t)$ using $r_{FRW}=Y(r,t)/a(t)$ where $a(t)$ is again the scale factor of the FRW model just introduced.

\clearpage
\subsection{Back-reaction effects}

We will now look again at the quantities on the right hand side of eq. (\ref{cq1}), in particular,  $Q_{D}$ and $\langle {\cal R} \rangle_D$.

We have found that $Q_{D}$ is negligible while this is not the case for $\langle {\cal R} \rangle_D$.
The reason is simple: the only difference with respect to the flat model in which both are zero is the curvature. While $E$ appears explicitly in $\langle {\cal R} \rangle_D$, in $Q_{D}$ it has only an indirect effect coming from the fact that now $a_{D} \neq Y$.

From the integrability condition (\ref{condi}), we can therefore deduce that, being $Q_{D}$ and $\langle {\cal R} \rangle_D$ decoupled, the latter will evolve like a standard curvature term: ${\langle {\cal R} \rangle}_D \sim 1/a_{D}^{2}$.
In other words, we will have $w_{BR}\simeq -1/3$.
We have verified this numerically: in Fig.~\ref{ricci} we plotted the time evolution of  $\langle {\cal R} \rangle_D$.

\begin{figure}[htb]
\begin{center}
\includegraphics[width=13 cm]{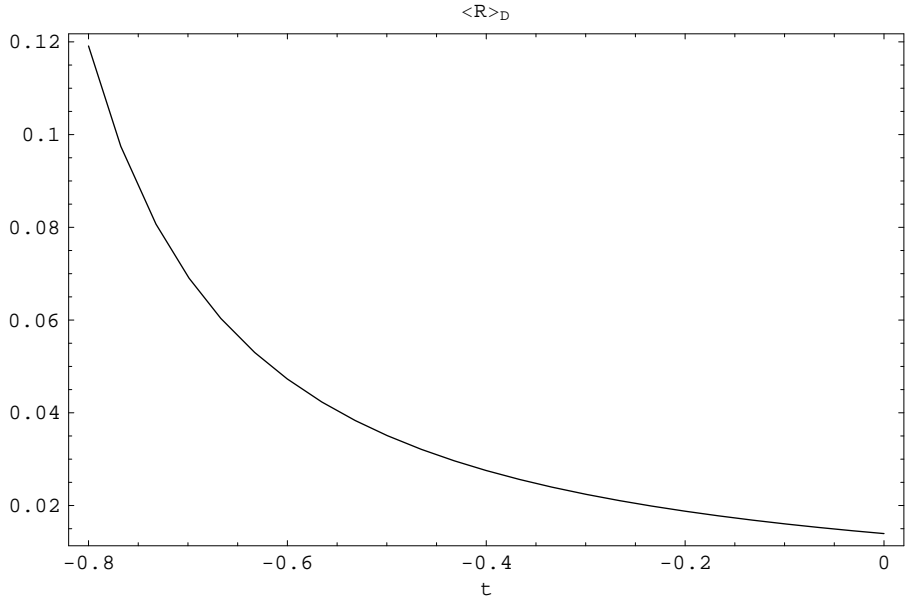}
\caption{\small \slshape Time evolution for $\langle {\cal R} \rangle_D$. It evolves as a FRW curvature term: ${\langle {\cal R} \rangle}_D \sim 1/a_{D}^{2}$.}
\label{ricci}
\end{center}
\end{figure}

\clearpage
\section{Discussion}

We have applied the averaging scheme of Sect.~\ref{avese} to a toy model that features the growing of structures.
We looked quantitatively at the importance of the back-reaction sources $Q_{D}$ and $\langle {\cal R} \rangle_D$.
While we found a (very small) contribution from the average curvature, no appreciable effects were coming from $Q_{D}$.

We trace this to the exact absence of back-reaction for the flat LTB model. The ultimate reason is that the equation automatically performs a ``euclidean'' average with respect to the local quantity $Y$: averaging something twice will not, of course, give sizeable effects. We believe that this is caused by spherical symmetry (Birkhoff theorem). \\
In the curved case, the curvature enters only indirectly in $Q_{D}$ and therefore we still do not have interesting domain effects.

From this study we learn that LTB models with observers at the center are not interesting to study the formalism developed by Buchert.
This could be, however, due to a poor use of them more than to their deficiency with respect to back-reaction studies in general.
Having already discussed the possibility that the questions put by Buchert's formalism are not the observationally-meaningful questions,
we will try to make a better use of the LTB model.
Having seen that spherical symmetry suppresses back-reaction effects, we will try to go beyond this limitation building a Swiss-cheese (toy) model for the universe in the next Chapter.

We will conclude acknowledging that recently and independently the reference \cite{Biswas:2006ub} developed a similar density profile, using, however, a different curvature: we have used an oscillating-around-zero curvature which does not single out the center of symmetry.

\chapter{Swiss Cheese}
\label{s-c}

In this chapter (see also  \cite{Marra:2007pm,Marra:2007gc}) we explore a cosmological toy model in order to attempt to
understand the role of large-scale non-linear cosmic inhomogeneities in the
interpretation of observable data. The model is based on a Swiss-cheese model,
where the cheese consists of the usual Friedmann-Robertson-Walker (FRW)
solution and the holes are constructed out of a Lema\^{i}tre-Tolman-Bondi (LTB)
solution.
We are focusing on a Swiss-cheese model because, even if it is made of spherical symmetric holes, it is not a spherical symmetric model as a whole.
It is a first step to go beyond spherical symmetry which will turn out to be the main limitation of LTB solutions.

This model will turn out to be well-suited to study the dual point of view sketched in Sect.~\ref{poview}.
See \cite{Biswas:2007gi, Brouzakis:2006dj, Brouzakis:2007zi} for other works in this direction.

\section{Our LTB model} \label{ourmodel}

We are going to study a Swiss-cheese model where the cheese consists of the usual
Friedmann--Robertson--Walker solution and the spherically symmetric holes
are constructed from a Lema\^{i}tre-Tolman-Bondi solution.  The
particular FRW solution we will choose is a matter-dominated, spatially-flat
solution, \textit{i.e.,} the Einstein--de Sitter (EdS) model.\\
In this section we will describe the LTB model parameters we have
chosen. We refer to Appendix~\ref{ltbm} for notation, units and an introduction to the LTB models.

First of all, for simplicity we will choose $\bar{t}(r)=\bar{t}$; \textit{i.e.,}
specifying the initial conditions for each shell at the same moment of time.

\begin{figure}[htb]
\begin{center}
\includegraphics[width=10cm]{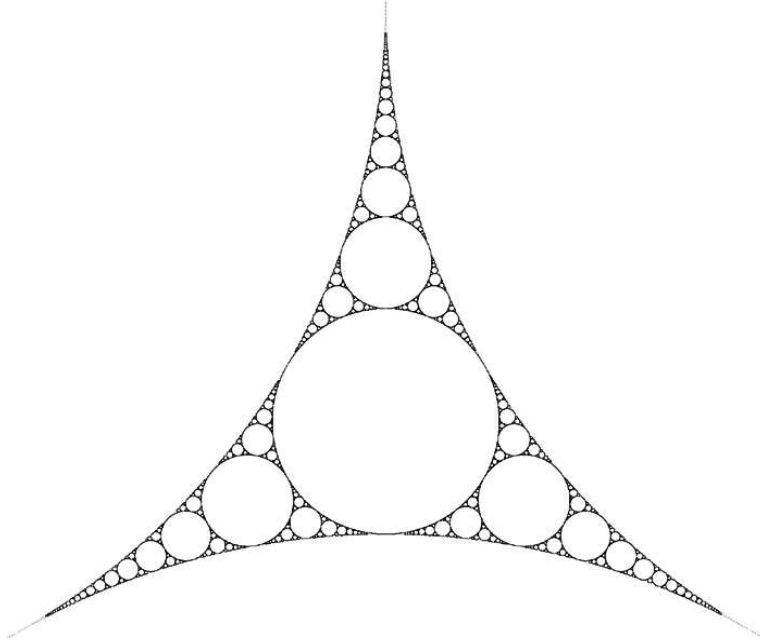}
\caption{\small \slshape The Apollonian Gasket.}
\label{apo}
\end{center}
\end{figure}

We will now choose $\rho(r,\bar{t})$ and $W(r)$ in order to match the flat FRW model
at the boundary of the hole: \textit{i.e.,} at the boundary of the hole
$\bar{\rho}$ has to match the FRW density and $W(r)$ has to go to unity. 
A physical picture is that, given a FRW sphere, all the matter in the inner
region is pushed to the border of the sphere while the quantity of matter
inside the sphere does not change.
With the density chosen in this way, an observer outside the hole will not feel
the presence of the hole as far as \textit{local} physics is concerned (this
does not apply to global quantities, such the luminosity-distance--redshift
relation for example). So the
cheese is evolving as an FRW universe while the holes evolve differently.
In this way we can imagine putting in the cheese as many
holes as we want, even with different sizes and density profiles, and still have
an exact solution of the Einstein equations (as long as there is no
superposition among the holes and the correct matching is achieved).
The limiting picture of this procedure is
the Apollonian Gasket of Fig.\ \ref{apo}, where all the possible holes are
placed, and therefore the model has the strange property that it is FRW
nowhere, but it behaves as an FRW model on the average. This idea was first
proposed by Einstein and  Straus \cite{Einstein:1945id}.

\begin{figure}[htb]
\begin{center}
\includegraphics[width=11 cm]{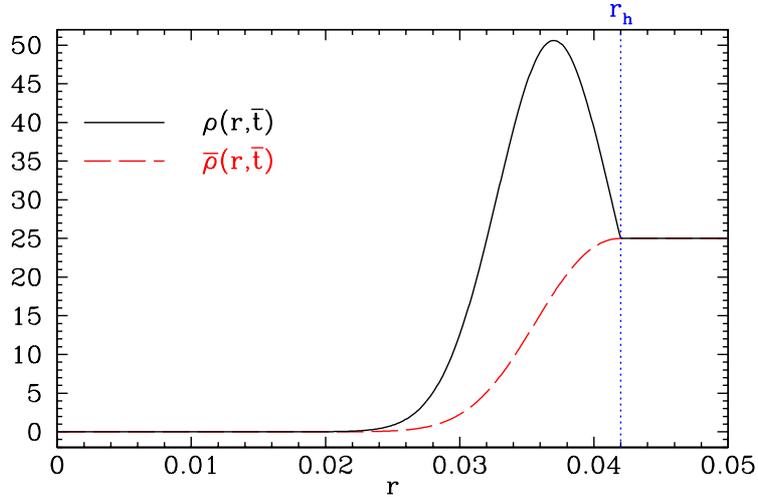}
\caption{\small \slshape The densities $\rho(r,\bar{t})$ (solid curve) and
$\bar{\rho}(r,\bar{t})$  (dashed curve). Here, $\bar{t}=-0.8$ (recall
$t_{BB}=-1$). The hole ends at $r_{h}=0.042$. The matching to the FRW solution
is achieved  as one can see from the plot of $\bar{\rho}(r,\bar{t})$.}
\label{rho0}
\end{center}
\end{figure}

\begin{figure}[htb]
\begin{center}
\includegraphics[width=10cm]{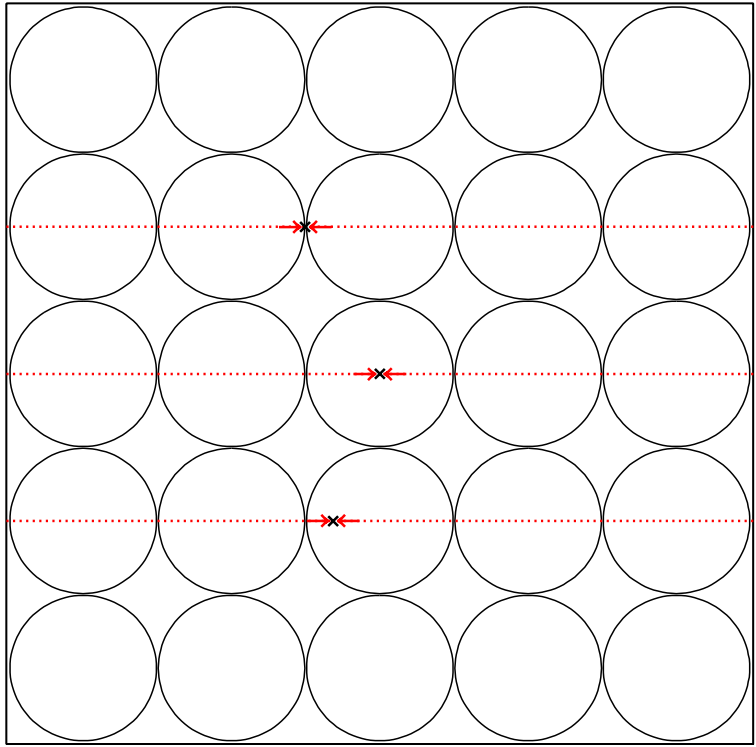}
\caption{\small \slshape Sketch of our Swiss-cheese model. An observer at rest with 
respect to a comoving cheese-FRW observer will see an isotropic CMB 
along the two directions of sight marked with dotted red lines. 
Three possible positions for an observer are shown.}
\label{imodel}
\end{center}
\end{figure}

To be specific, we choose $\rho(r,\bar{t})$
to be
\begin{equation}
\begin{array}{ll}
\rho(r,\bar{t}) = A \exp[-(r-r_M)^2/2\sigma^2] + \epsilon \quad & (r<r_h) \\
\rho(r,\bar{t}) = \rho_{FRW}(\bar{t}) & (r > r_h),
\end{array}
\end{equation}
where $\epsilon = 0.0025$, $r_h=0.42$, $\sigma=r_h/10$, $r_M=0.037$, $A=50.59$,
and $\rho_{FRW}(\bar{t})=25$.  In Fig.\ \ref{rho0} we plot this chosen Gaussian
density profile.  The hole ends at $r_{h}=0.042$ which is\footnote{To get this number from Table \ref{units} you need to multiply $r_{h}$ by $a(t_{0})\simeq 2.92$.} $350$ Mpc and roughly
$25$ times smaller than $r_{BB}$. Note that this is not a very big bubble. But
it is an almost empty region: in the interior the matter density is roughly
$10^4$ times smaller than in the cheese. Our model consists of a sequence of up
to five holes and the observer is looking through them. The idea, however, is
that the universe is completely filled with  these holes, which form a sort of
lattice as shown in Fig.\ \ref{imodel}. In this way an observer at rest  with
respect to a comoving cheese-FRW observer will see an isotropic  CMB along the
two directions of sight shown in Fig.\ \ref{imodel}.

\newpage

It is useful to consider the velocity of a shell relative to 
the FRW background.  
We define
\begin{equation}
\Delta v_{sh}(r,t)= \dot{a}_{LTB}(r,t)-\dot{a}_{FRW}(t),
\end{equation}
where $a_{LTB}(r,t)= Y(r,t)/ r$.
To have a realistic evolution, we demand that there are no initial peculiar
velocities at time $\bar{t}$, that is, to have an initial expansion $H$
independent of $r$: $\Delta v_{sh}(r,\bar{t})=0$. From Eq.\ (\ref{cucu})
this implies
\begin{equation} 
\label{Er}
E(r)=\frac{1}{2} H_{FRW}^2(\bar{t}) \, r^{2}-\frac{1}{6\pi}\frac{M(r)}{r}.
\end{equation}
The graph of $E(r)$ chosen in this way is shown in Fig.\ \ref{E}. As seen from
the figure, the curvature $E(r)$ is small compared with unity. Indeed, in many
formulae $W=(1+ 2 E)^{1/2}\simeq 1+E$ appears, therefore one should compare $E$
with $1$. In spite of its smallness, the curvature will play a crucial role to
allow a realistic evolution of structures, as we will see in the next section.

Also in Fig.\ \ref{E} we graph $k(r)=-2 E(r)/r^{2}$, which is the
generalization of the factor $k$ in the usual FRW models. (It is not normalized
to unity.) As one can see, $k(r)$ is very nearly constant  in the empty region
inside the hole. This is another way to see the reason for our choice of the
curvature function: we want to have in the center an empty bubble dominated by
negative curvature.

It is important to note that the dynamics of the hole is scale-independent:
small holes will evolve in the same way as big holes. To show this, we just
have to express Eq.\ (\ref{motion}) with respect to a generic variable
$\tilde{r}=r/g$ where $g$ fixes the scale. If we change $g$, \textit{i.e.,}
if we scale the density profile, we will find the same scaled shape for $k(r)$ and
the same time evolution. This property is again due to spherical symmetry
which frees the inner shells from the influence of the outer ones: We can think
of a shell as an infinitesimal FRW solution and its behavior is 
scale independent because it is a homogeneous and isotropic solution.

\begin{figure}[htb]
\begin{center}
\includegraphics[width=13 cm]{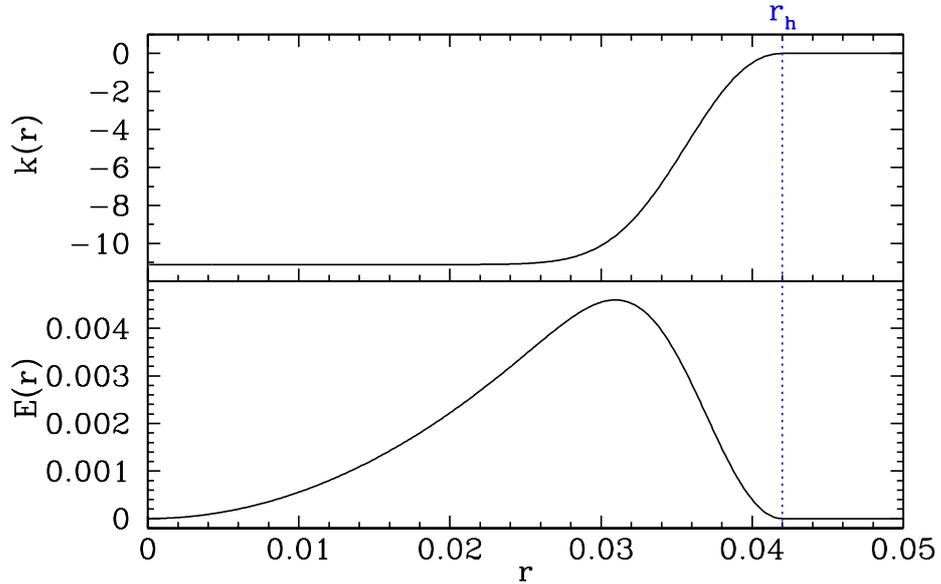}
\caption{\small \slshape Curvature $E(r)$ and $k(r)$ necessary for the initial conditions of no
peculiar velocities.}
\label{E}
\end{center}
\end{figure}

\clearpage
\section{The dynamics} \label{dynamics}

Now we explore the dynamics of this Swiss-cheese model.  As we have said, the
cheese evolves as in the standard FRW model. Of course, inside the holes the
evolution is different. This will become clear from the plots given below. 

We will discuss two illustrative cases: a flat case where $E(r)=0$, and a
curved case where $E(r)$ is given by Eq.\ (\ref{Er}). We are really interested
only in the second case because the first will turn out to be unrealistic.  But
the flat case is useful to understand the dynamics.

\subsection{The flat case}

\begin{figure}[p]
\begin{center}
\includegraphics[width= 15 cm]{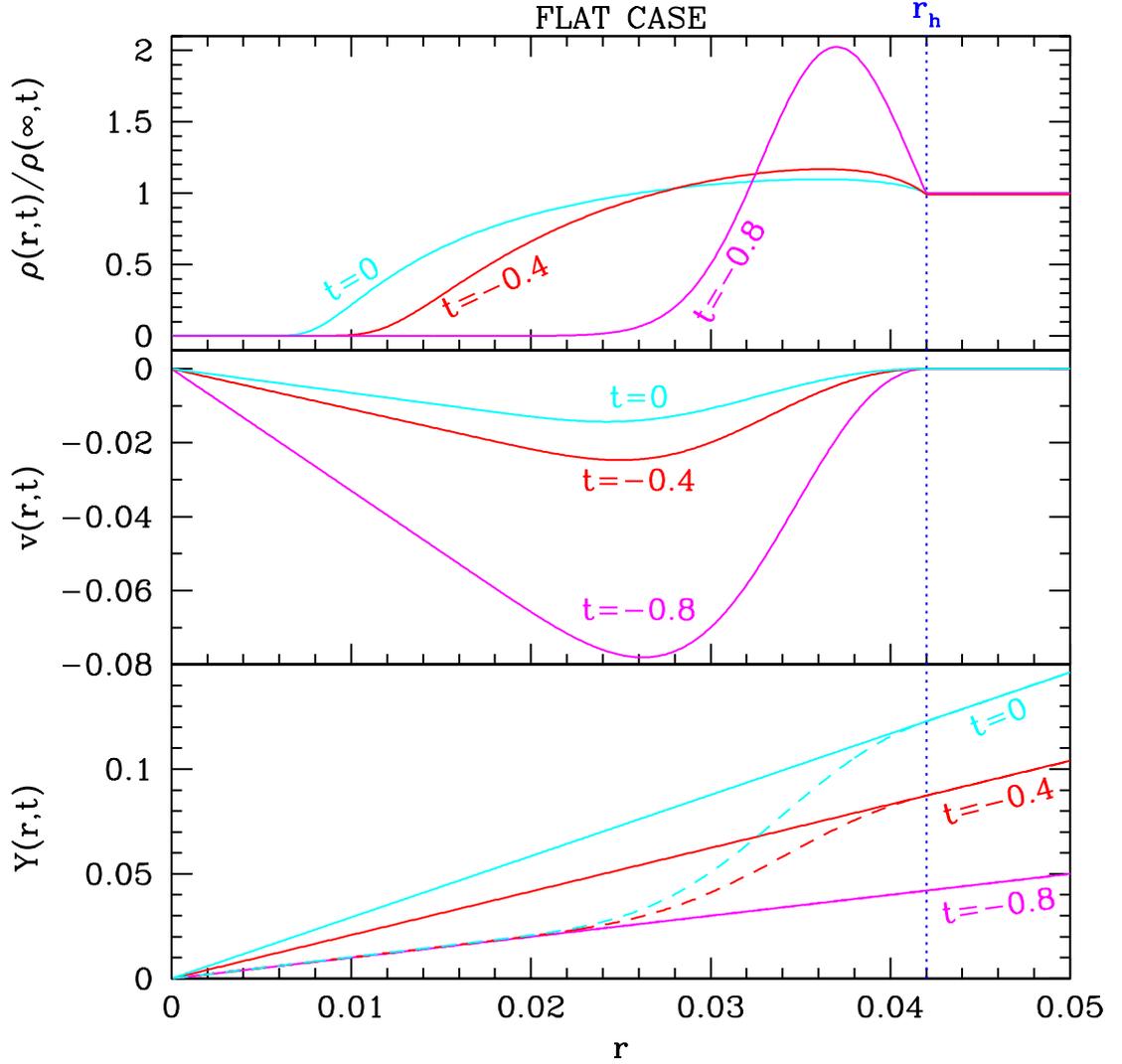}
\caption{\small \slshape Behavior of $Y(r,t)$ with respect to $r$, the peculiar velocities $v(r,t)$ with respect to $r$, and the
density profiles $\rho(r,t)$ with respect to $r_{FRW}=Y(r,t)/a(t)$,
for the flat case at times $t=\bar{t}=-0.8$, 
$t=-0.4$ and $t=t_0=0$.
The straight lines for $Y(r,t)$ are the FRW solutions
while the dashed lines are the LTB solutions.  For the peculiar velocities,
matter is  escaping from high density regions. The center has no peculiar
velocity  because of spherical symmetry,  and the maximum of negative peculiar
velocity is before the peak in density.  Finally, the values of $\rho(\infty,t)$ are $1,\ 2.8,$ 
and $25$, for $t=0,\ -0.4,\ -0.8$,  respectively.}
\label{flat}
\end{center}
\end{figure}

In Fig.\ \ref{flat} we show the evolution of $Y(r,t)$ for the flat case,
$E(r)=0$.  In the figure $Y(r,t)$ is plotted for three times: $t=\bar{t}=-0.8$
(recall $t_{BB}=-1$), $t = -0.4$, and $t=0$ (corresponding to today).

{}From Fig.\ \ref{flat} it is clear that outside the hole, \textit{i.e.,} for
$r \geq r_{h}$, $Y(r,t)$ evolves as a FRW solution, $Y(r,t)\propto r$. However,
deep inside the hole  where it is almost empty, there is no time evolution to
$Y(r,t)$:  it is Minkowski space.   Indeed, thanks to spherical symmetry, the
outer shells do not influence the interior. If we place additional matter
inside the empty space, it will start expanding as an FRW universe, but at a
lower rate because of the lower density. It is interesting to point out that a
photon passing the empty region will undergo no redshift: again, it is just
Minkowski space.

This counterintuitive behavior (empty regions expanding slowly) is due to the
fact that the spatial curvature vanishes.  This corresponds to an unrealistic
choice of initial peculiar velocities. To see this we plot the peculiar
velocity that an observer following a shell $r$ has with respect to an FRW
observer passing through that same spatial point. The result is also shown in
Fig.\ \ref{flat} where it is seen that matter is escaping from the high density
regions. This causes the evolution to be reversed as one can see in Fig.\
\ref{flat} from the density profile at different times: structures are not
forming, but spreading out.

Remember that $r$ is only a label for the shell
whose Euclidean position at time $t$ is $Y(r,t)$. 
In the plots of the energy density we have
normalized $Y(r,t)$ using $r_{FRW}=Y(r,t)/a(t)$.

\subsection{The curved case}

\begin{figure}[p]
\begin{center}
\includegraphics[width= 15 cm]{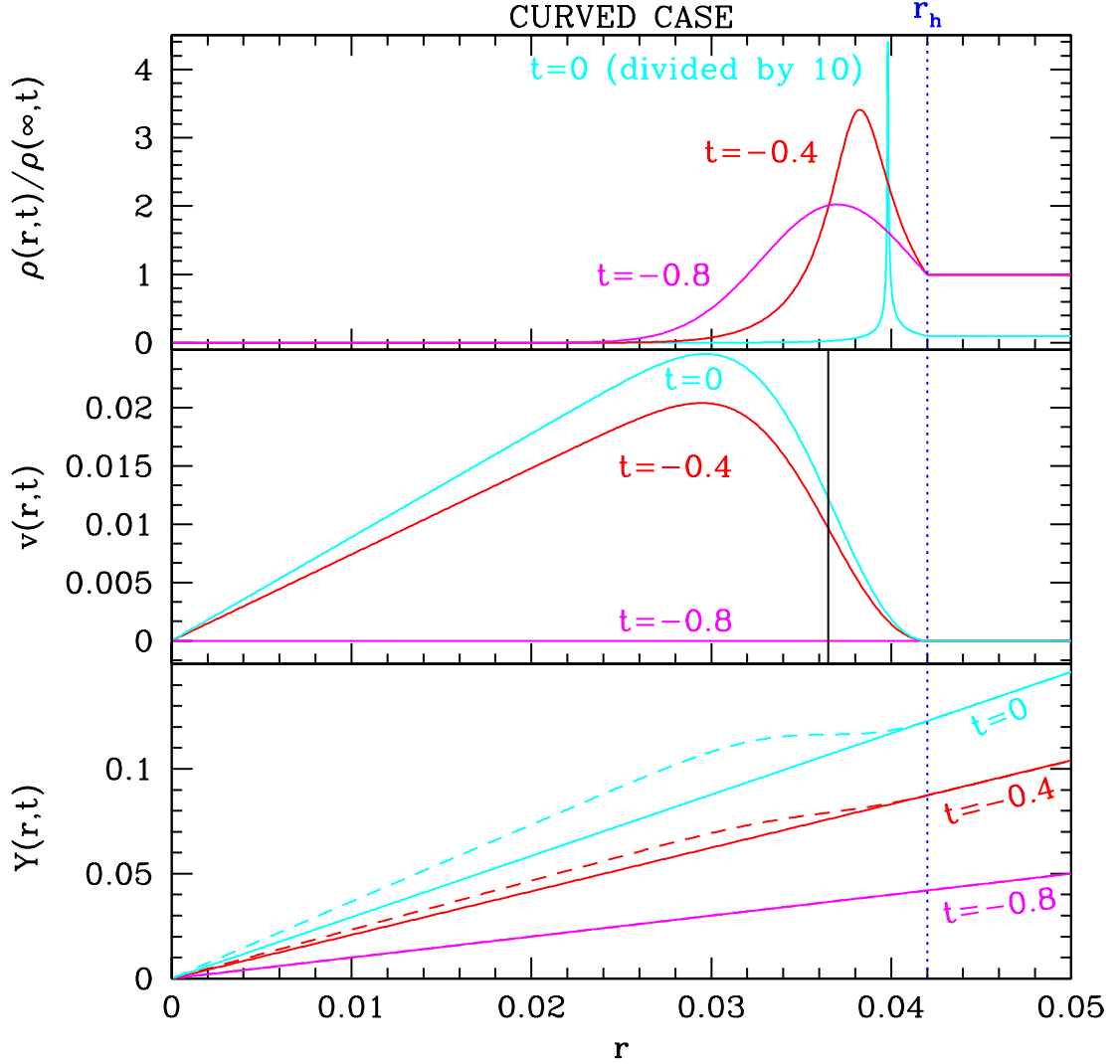}
\caption{\small \slshape Behavior of $Y(r,t)$ with respect to $r$, 
the peculiar velocities $v(r,t)$ with respect to $r$, and the
density profiles $\rho(r,t)$ with respect to $r_{FRW}=Y(r,t)/a(t)$,
for the curved case at times $t=\bar{t}=-0.8$, 
$t=-0.4$ and $t=t_0=0$. The straight lines for $Y(r,t)$ are the FRW solutions
while the dashed lines are the LTB solutions.  For the peculiar velocities,
the matter gradually starts to move toward high density regions. The solid
vertical line marks the position of the peak in the density with respect 
to $r$. For the
densities, note that the curve for $\rho(r,0)$ has been divided by $10$.  
Finally, the values of $\rho(\infty,t)$ are $1,\ 2.8,$ and $25$, for $t=0,\
-0.4,\ -0.8$,  respectively. }
\label{curved}
\end{center}
\end{figure}

Now we will move to a more interesting and relevant case. We are going to use the
$E(r)$ given by Eq.\ (\ref{Er}); the other parameters will stay the same.
Comparison with the flat case is useful to understand how the model behaves,
and in particular the role of the curvature.
In Fig.\ \ref{curved} the results for $Y(r,t)$ in the curved case are plotted.
Time goes from $t=\bar{t}=-0.8$ to $t=0$.

\begin{figure}[htb]
\begin{center}
\includegraphics[width= 13 cm]{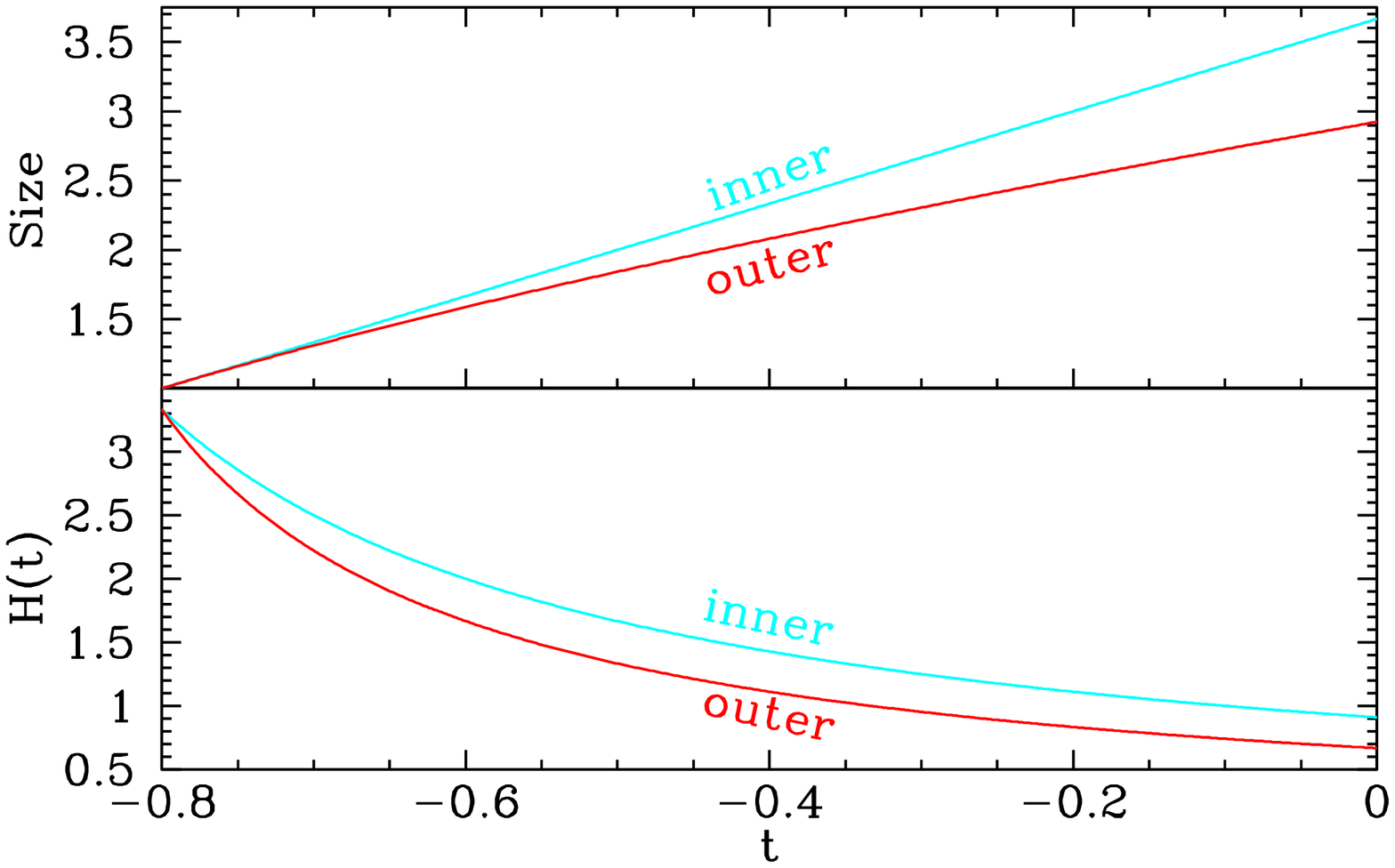}
\caption{\small \slshape Evolution of the expansion rate and the size for the inner 
and outer regions. Here ``inner'' refers to a point deep inside the hole, and
``outer'' refers to a point in the cheese.}
\label{outinc}
\end{center}
\end{figure}

As one can see, now the inner almost empty region is expanding faster than the
outer (cheese) region. This is shown clearly in Fig.\ \ref{outinc}, where also
the evolution of the inner and outer sizes is shown. Now the density ratio
between the cheese and the interior region of the hole increases by a factor of
$2$ between $t=\bar{t}$ and $t=0$. Initially the density ratio was $10^{4}$,
but the model is not sensitive to this number since the evolution in the
interior region is dominated by the curvature ($k(r)$ is much larger than the
matter density). We stress now the fact that the crucial ingredient is
to have a faster-than-cheese expanding void.

The peculiar velocities are now natural: as can be seen from Fig.\
\ref{curved}, matter is falling towards the peak in the density. The evolution
is now realistic, as one can see from Fig.\ \ref{curved}, which shows the
density profile at different times. Overdense regions start contracting and
they become thin shells (mimicking structures), while underdense regions become
larger (mimicking voids), and eventually they occupy most of the volume.

Let us explain why the high density shell forms and the nature of the shell
crossing. Because of the distribution of matter, the inner part of the hole is
expanding faster than the cheese; between these two regions  there is the
initial overdensity. It is because of this that there is less matter in the
interior part. (Remember that we matched the FRW density at the end of the
hole.) Now we clearly see what is happening: the overdense region is squeezed
by the interior and exterior regions which act as a clamp. Shell crossing
eventually happens when more shells -- each labeled by its own $r$ -- are so
squeezed that they occupy the same position $Y$, i.e.~when $Y'=0$.
Nothing happens to the photons other than passing through more shells at the
same time: this is the meaning of the $g_{r r}$ metric coefficient going to
zero.

A remark is in order here: In the inner part of the hole there is almost no
matter, it is empty. Therefore it has only negative curvature, which is largely
dominant over the matter: it is close to a Milne universe.

\clearpage
\section{Photons} \label{photons}

\begin{figure}[htb]
\begin{center}
\includegraphics[width= 15 cm]{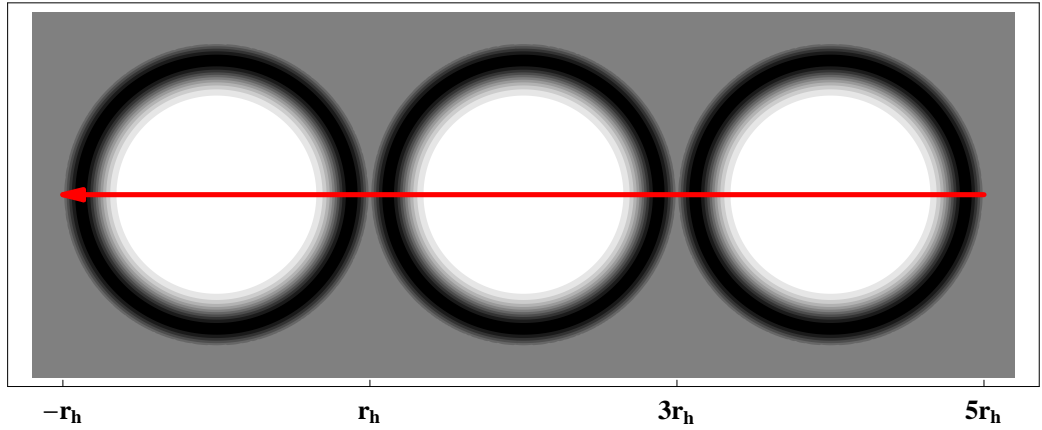}
\caption{\small \slshape Sketch of our model in comoving coordinates. The shading represents 
the initial density profile: darker shading implies larger densities. The uniform 
gray is the FRW cheese. The photons pass through the holes as shown by the 
arrow.}
\label{schizzo}
\end{center}
\end{figure}

We are mostly interested in observables associated with the  propagation of
photons in our Swiss-cheese model: indeed, our aim is to calculate the
luminosity-distance--redshift relation $d_{L}(z)$ in order to understand the
effects of inhomogeneities on observables. Our setup is illustrated in Fig.\
\ref{schizzo}, where there is a sketch of the model with only $3$ holes for the
sake of clarity. Notice that photons are propagating through the centers.

We will discuss two categories of cases: 1) when the observer is just outside
the last hole as in Fig.\ \ref{schizzo}, and 2) when the observer is inside the
hole. The observer in the hole will have two subcases: a) the observer located
on a high-density shell, and b) the observer in the center of the hole.  We are
mostly interested in the first case: the observer is still a usual FRW
observer, but looking through the holes in the Swiss cheese.

\clearpage
\subsection{Finding the photon path: an observer in the cheese} \label{ciccio}

We will discuss now the equations we will use to find the path of a photon
through the Swiss cheese. The geodesic equations can be reduced to a set of
four first-order differential equations (we are in the plane $\theta=\pi /2$):
\begin{equation}
\begin{array}{lll}
\Frac{dz}{d\lambda} & = -\Frac{\dot{Y}'}{Y'}\left((z+1)^2-
\Frac{c_\phi^2 }{Y^{2}}\right)
- c_{\phi}^{2} \Frac{\dot{Y}}{Y^{3}} & \qquad \qquad z(0)=0 \\
\Frac{dt}{d\lambda} & = z+1 & \qquad \qquad t(0)=t_{0}=0 \\
\Frac{dr}{d\lambda} & = \Frac{W}{Y'}\sqrt{(z+1)^2-\Frac{c_{\phi}^{2}}{Y^{2}}} 
 & \qquad \qquad r(0)=r_{h} \label{rgeo} \\
\Frac{d\phi}{d\lambda} & = \Frac{c_{\phi}}{Y^{2}} &  \qquad \qquad \phi(0)= \pi
\end{array}
\end{equation}
where $\lambda$ is an affine parameter that grows with time. The third
equation is actually the null condition for the geodesic. Thanks to the initial
conditions chosen we have $z(0)=0$. These equations describe the general 
path of a photon. To solve the equations we need to specify the constant 
$c_{\phi}$, a sort of angular momentum density. A first observation is 
that setting
$c_{\phi}=0$ allows us to recover the equations that describe a photon passing
radially trough the centers: $dt/dr=Y'/ W$. 

\begin{figure}[htb]
\begin{center}
\includegraphics[width=9 cm]{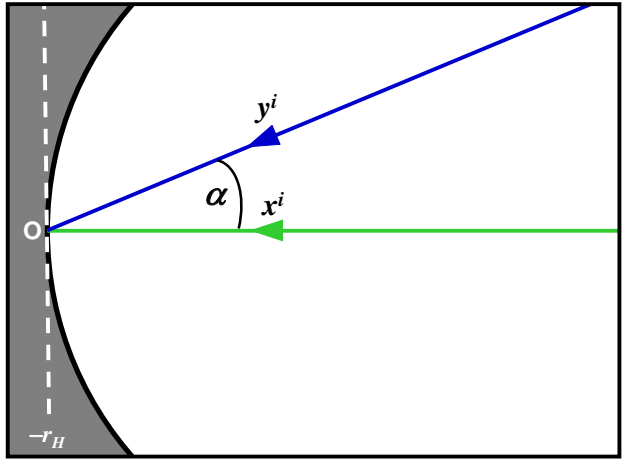}
\caption{\small \slshape A photon hitting the observer at an angle $\alpha$.}
\label{frecce}
\end{center}
\end{figure}

We are interested in photons that hit the observer at an angle $\alpha$ and are
passing trough all the holes as shown in Fig.\ \ref{schizzo}. To do this we
must compute the inner product of $x^{i}$ and $y^{i}$, which are the normalized
spatial vectors tangent to the radial axis and the geodesic as shown in Fig.\
\ref{frecce}. A similar approach was used in Ref.\ \cite{Alnes:2006pf}.

The inner product of $x^i$ and $y^i$ is expressed through
\begin{eqnarray}
x^{i} & = & -\frac{W}{Y'}\; (1,0,0)\vert_{\lambda=0} \\
y^{i} & = & \frac{1}{dt/d\lambda}\left. \left (\frac{d}{d\lambda},0,
\frac{d\phi}{d\lambda}\right) \right|_{\lambda=0} = \left. 
\left (\frac{dr}{d\lambda},0,\frac{d\phi}{d\lambda}\right) \right|_{\lambda=0} 
\\
x^{i}\, y^{i}\, g_{i \, j} & = & \left. \frac{Y'}{W} \; 
\frac{dr}{d\lambda}\right|_{\lambda=0}=\cos \alpha\\
c_{\phi} & = & \left. Y \sin \alpha \right| _{\lambda=0}  .
\end{eqnarray}
The vectors are anchored to the shell labeled by the value of the affine
parameter $\lambda=0$, that is, to the border of the hole. Therefore, they are
relative to the comoving observer located there. In the second equation we have
used the initial conditions given in the previous set of equations, while to
find the last equation we have used the null condition evaluated at
$\lambda=0$. 

The above calculations use coordinates relative to the center. However, the
angle $\alpha$ is a scalar in the hypersurface we have chosen: we are using the
synchronous and comoving gauge. Therefore, $\alpha$ is the same angle 
measured by a comoving observer of Fig.\ \ref{frecce} located on the shell 
$r=-r_{h}$: it is a coordinate transformation within the same hypersurface.

Given an angle $\alpha$ we can solve the equations. We have to change the sign
in Eq.\ (\ref{rgeo}) when the photon is approaching the center with respect to
the previous case where it is moving away. Also, we have to sew together the
solutions between one hole and another, giving not only the right initial 
conditions, but also the appropriate constants $c_{\phi}$  (see Appendix
\ref{sewing}).

Eventually we end up with the solution $t(\lambda)$, $r(\lambda)$,
$\phi(\lambda)$ and $z(\lambda)$ from which we can calculate the observables of
interest.

\subsection{Finding the photon path: an observer in the hole}

Finding the solution in this case is the same as in the previous case with the
only difference that in Eq.\ (\ref{rgeo}) the initial condition is now
$r(0)=r_{obs}$. But this observer has a peculiar velocity with respect to an
FRW observer passing by. This, for example, will make the observer see an
anisotropic cosmic microwave background as it is clear from Fig.\ \ref{imodel}.
This Doppler effect, however, is already corrected in the solution we are going
to find since we have chosen $z(0)=0$ as initial condition.

\begin{figure}[htb]
\begin{center}
\includegraphics[width=9 cm]{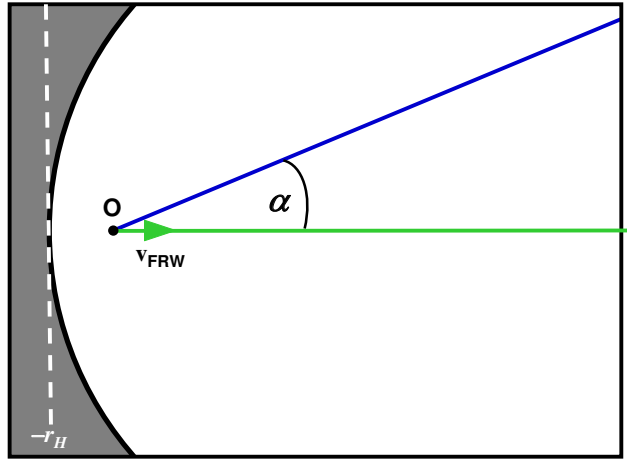}
\caption{\small \slshape A comoving observer and a FRW observer live in different frames, this
results in a relative velocity $v_{FRW}$ between observers.}
\label{doppler}
\end{center}
\end{figure}

There is however also the effect of light aberration which changes the angle
$\alpha$ seen by the comoving observer with respect to the angle
$\alpha_{\scriptscriptstyle FRW}$ seen by an FRW observer. The photon can be
thought as coming from a source very close to the comoving observer: therefore
there is no peculiar motion between them.  The FRW observer is instead moving
with respect to this reference frame as pictured in Fig.\ \ref{doppler}. The
relation between $\alpha$ and $\alpha_{\scriptscriptstyle FRW}$ is given by the
relativistic aberration formula:
\begin{equation}
\cos \alpha_{\scriptscriptstyle FRW}=\frac{\cos \alpha+ v/c}{1+v/c 
\; \cos \alpha} .
\end{equation}
The angle changes because the hypersurface has been changed. The velocity will
be taken from the calculation (see Fig.\ \ref{curved} for the magnitude of the
effect).

\subsection{Distances}

The angular diameter distance is defined as:
\begin{equation}
d_{A}=\frac{D}{\alpha_{\scriptscriptstyle FRW}} ,
\end{equation}
where $D$ is the proper diameter of the source and $\alpha$ is the angle at
which the source is seen by the observer. Using this definition to find $d_{A}$
we have
\begin{equation}
d_A = \frac{2 \, Y(r(\lambda),t(\lambda)) \; \sin \phi(\lambda)}{2 \,  
\alpha_{\scriptscriptstyle FRW}}  .
\end{equation}
The luminosity distance will then be:
\begin{equation}
d_{L}=(1+z)^{2} d_{A} .
\end{equation}
The formula we are going to use for $d_{A}$ is exact in the limit of zero
curvature. However in our model $E(r)$ is on average less than $0.3\%$  and
never more than $0.4\%$, as it can be seen from Fig.\ \ref{E}:  therefore the
approximation is good. Moreover, we are interested mainly in the case when the
source is out of the last hole as pictured in Fig.\ \ref{schizzo}, and in this
case the curvature is exactly zero and the result is exact.

We have checked that the computation of $d_{A}$ is independent of $\alpha$ for
small angles and that the result using the usual FRW equation coincides with
theoretical prediction for $d_{A}$. We also checked that $d_{A}$ reduces to
$Y(r,t)$ when the observer is in the center.

Finally we checked our procedure in comparison with the formula ($E.31$) of
Ref.\ \cite{Biswas:2006ub}: this is a rather different way to find the angular
distance and therefore this agreement serves as a consistency check. We placed
the observer in the same way and we found the same results provided that we use
the angle $\alpha$ uncorrected for the light-aberration effect.

\section{Results: observer in the cheese} \label{cheese}

Now we will look through the Swiss cheese comparing the results with respect to
a FRW-EdS universe and a $\Lambda$CDM case.

We will first analyze in detail the model with five holes, which is the one 
which we are most interested in. 
For comparison, we will study models with one big
hole and one small hole. In the model with one big hole, the hole will be
five-times bigger in size than in the model with five holes: \textit{i.e.,}
they will cover the same piece of the universe.

The observables on which we will focus are the changes in redshift
$z(\lambda)$, angular-diameter distance $d_{A}(z)$, luminosity distance
$d_{L}(z)$, and the corresponding distance modulus $\Delta m(z)$.

\subsection{Redshift histories} \label{histories}

\begin{figure}[htb]
\begin{center}
\includegraphics[width= 15 cm]{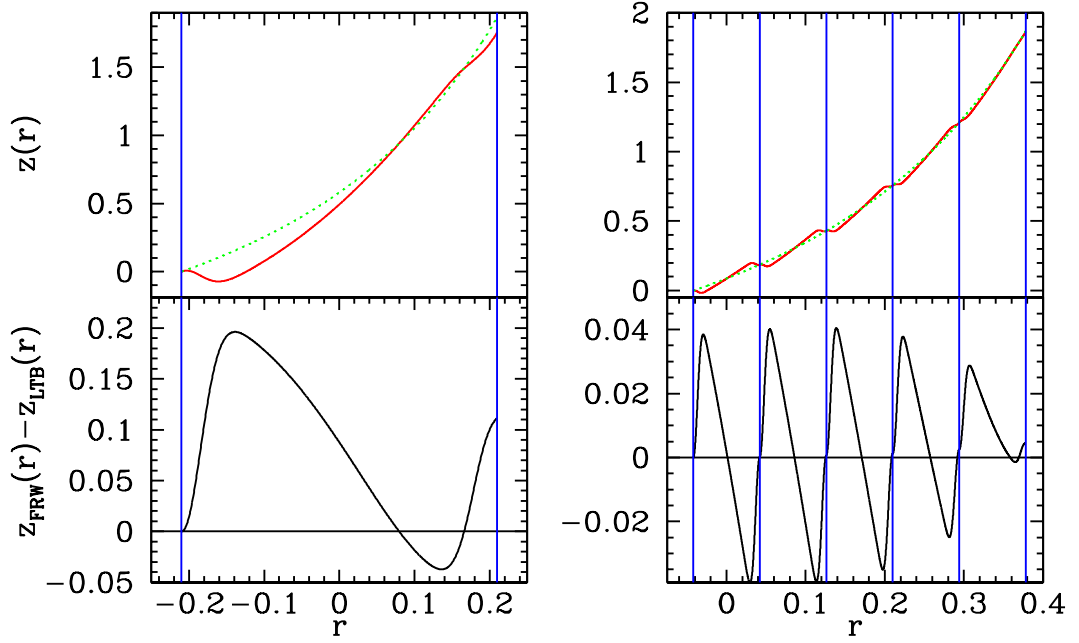}
\caption{\small \slshape Redshift histories for a photon that travels from one side of the 
one-hole chain (left) and five-hole chain (right) to the other where the 
observer will detect it at present time. The ``regular'' curve is for the FRW
model. The vertical lines mark the edges of the holes. The plots are with
respect to the coordinate radius $r$. Notice also that along the voids the
redshift is increasing  faster: indeed $z'(r)=H(z)$ and the voids are expanding
faster.}
\label{zorro}
\end{center}
\end{figure}

Now we will first compare the redshift undergone by photons that travel through
the model with either five holes or one hole to the FRW solution of the cheese.
In Fig.\ \ref{zorro} the results are shown for a photon passing through the
center with respect to the coordinate radius. As one can see, the effects 
of the inhomogeneities on the redshift are smaller in the five-hole case.

It is natural to expect a compensation, due to the spherical symmetry,
between the ingoing path and the outgoing one inside the same hole. This
compensation is evident in Fig.~\ref{zorro}.

However, there is a compensation already on the scale of half a hole as it is
clear from the plots. This mechanism is due to the density profile chosen, that
is one whose average matches the FRW density of the cheese: roughly speaking we
know that $z'=H \propto \rho = \rho_{\scriptscriptstyle FRW} + \delta \rho$. We
chose the density profile in order to have $\langle \delta \rho \rangle=0$, and
therefore in its journey from the center to the border of the hole the photon
will see a $\langle H\rangle \sim H_{\scriptscriptstyle FRW}$ and therefore
there will be compensation for $z'$.
It is somewhat similar to the screening among positive and negative 
charges.

Let us see this analytically. We are interested in computing a line average of
the expansion along the photon path in order to track what is going on.
Therefore, we shall not use the complete expansion scalar:
\begin{equation}
\theta=\Gamma_{0k}^{k}=2\frac{\dot{Y}}{Y}+\frac{\dot{Y}'}{Y'} ,
\end{equation}
but, instead, only the part of it pertinent to a radial line average:
\begin{equation} \label{dito}
\theta_r=\Gamma_{01}^{1}=\frac{\dot{Y}'}{Y'}\equiv H_{r} ,
\end{equation}
where $\Gamma_{0k}^{k}$ are the Christoffel symbols and $\theta$ is the trace
of the extrinsic curvature.

Using $H_r$, we obtain:
\begin{equation}
\langle H_r \rangle =
\frac{\int_{0}^{r_{h}}dr \; H_r \; Y' / W}
{\int_{0}^{r_{h}} dr \; Y' / W} \simeq \left. \frac{\dot{Y}}{Y}\right|_{r=r_{h}}
= H_{\scriptscriptstyle FRW} ,
\end{equation}
where the approximation comes from neglecting the (small) curvature and the
last equality holds thanks to the density profile chosen. This is exactly the
result we wanted to find. However, we have performed an average at constant time
and therefore we did not let the hole and its structures  evolve while the
photon is passing: this effect will partially break the compensation. This
sheds light on the fact that photon physics seems to be affected by the
evolution of inhomogeneities more than by inhomogeneities themselves.
We can argue that there should be perfect compensation if the hole will have a
static metric such as the Schwarzschild one. In the end, this is a limitation
of our assumption of spherical symmetry.

This compensation is almost perfect in the five-hole case, while it is not in
the one-hole case: in the latter case the evolution has more time to change the
hole while the photon is passing. Summarizing, the compensation is working on
the scale $r_{h}$ of half a hole. These results are in agreement  with Ref.\
\cite{Biswas:2007gi}.

\begin{figure}[htb]
\begin{center}
\includegraphics[width= 13 cm]{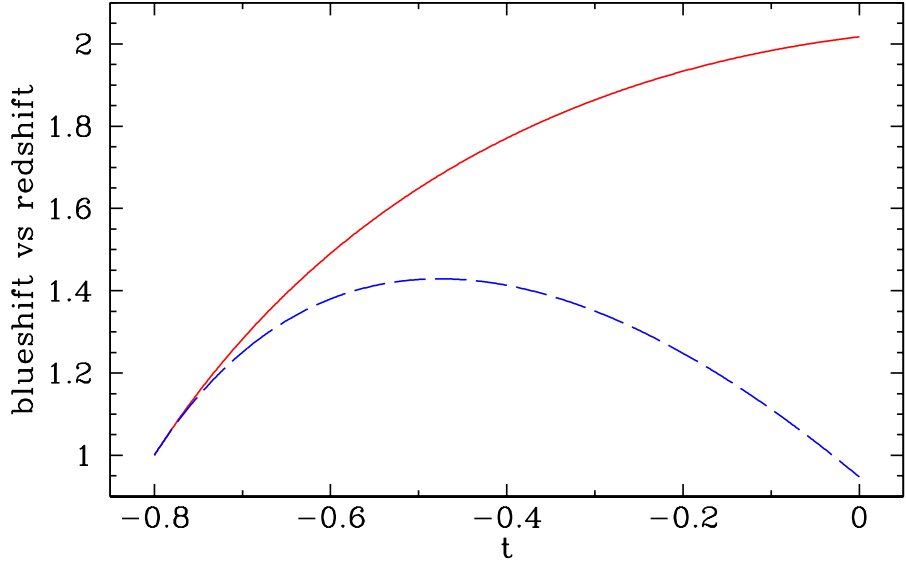}
\caption{\small \slshape Distance between the observer and two different shells. In the
solid curve $r=0.55 \, r_{h}$ will be redshifted, while in the dashed curve,
$r=0.8 \, r_{h}$ will be blueshifted. The latter indeed will start to collapse 
toward the observer. Time goes from $t=-0.8$ to $t=0$.
The observer is located just outside of the hole at $r=r_{h}$.}
\label{blue}
\end{center}
\end{figure}

From the plot of the redshift one can see that function $z(r)$ is not
monotonic. This happens at recent times when the high-density thin shell forms.
This blueshift is due to the peculiar movement of the matter that is forming
the shell. This feature is shown in Fig.\ \ref{blue} where the distance between
the observer located just out of the hole at $r=r_{h}$ and two different shells
is plotted. In the solid curve one can see the behavior with respect to a
normal redshifted shell, while in the dashed curve one can see the behavior with
respect to a shell that will be blueshifted: initially the distance increases
following the Hubble flow, but when the shell starts forming, the peculiar
motion prevails on the Hubble flow and the distance decreases during the
collapse.

It is finally interesting to interpret the redshift that a photon undergoes
passing the inner void. The small amount of matter is subdominant with respect
to the curvature which is governing the evolution, but still it is important to
define the space: in the limit of zero matter in the interior of the hole, we
recover a Milne universe, which is just (half of) Minkowski space in unusual
coordinates. Before this limit the redshift was conceptually due to the
expansion of the spacetime, after this limit it is instead due to the peculiar
motion of the shells which now carry no matter: it is a Doppler effect.

\subsection{Luminosity and Angular-Diameter Distances}

\subsubsection{The five-hole model} \label{5holes}

\begin{figure}[p]
\begin{center}
\includegraphics[width= 15 cm]{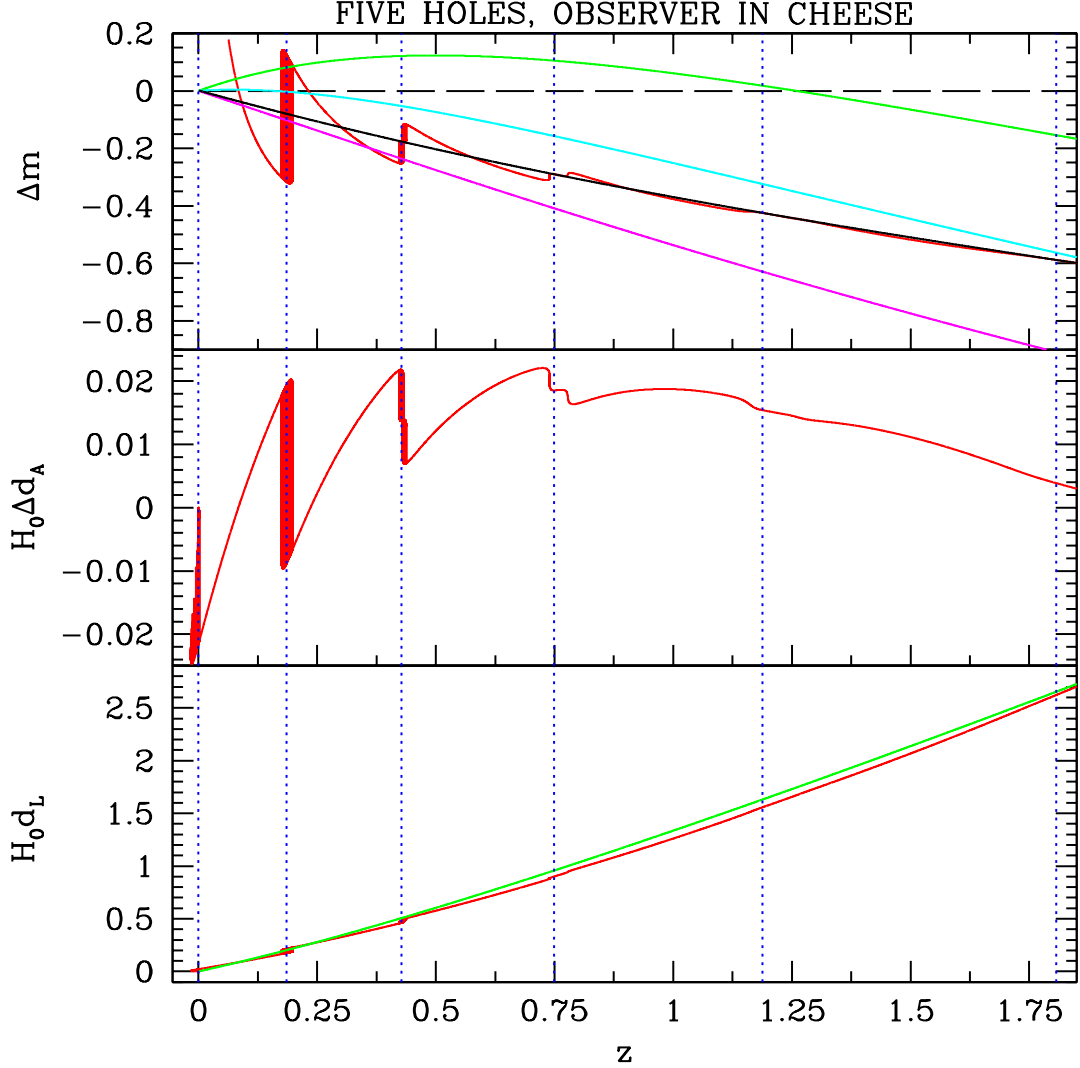}
\caption{\small \slshape On the bottom the luminosity distance $d_L(z)$ in the  five-hole model
(jagged curve) and the $\Lambda$CDM solution with $\Omega_{M}=0.6$ and
$\Omega_{DE}=0.4$ (regular curve) are shown.  In the middle is the change in
the angular diameter distance, $\Delta d_A(z)$, compared to a $\Lambda$CDM
model with  $\Omega_{M}=0.6$ and $\Omega_{DE}=0.4$. The top panel shows the
distance modulus in various cosmological models. The jagged line is for the
five-hole LTB model. The regular curves, from top to bottom, are a $\Lambda$CDM
model with  $\Omega_{M}=0.3$ and $\Omega_{DE}=0.7$, a $\Lambda$CDM model with  
$\Omega_{M}=0.6$ and $\Omega_{DE}=0.4$, the best smooth fit to the LTB model,
and the EdS model.  The vertical lines mark the edges of the five holes.}
\label{5incheese}
\end{center}
\end{figure}

In Fig.\ \ref{5incheese} the results for the luminosity distance and angular
distance are shown. The solution is compared to the one of the $\Lambda$CDM
model with $\Omega_{M}=0.6$ and $\Omega_{DE}=0.4$. Therefore, we have an
effective $q_{0}=\Omega_{M}/2-\Omega_{DE}=-0.1$. In all the plots we will
compare this $\Lambda$CDM solution to our Swiss-cheese solution. The strange
features which appear near the contact region of the holes at recent times are
due to the non-monotonic behavior of $z(r)$, which was explained in the
previous section.

The distance modulus is plotted in the top panel of Fig.\ \ref{5incheese}. The
solution shows an oscillating behavior which is due to the simplification of
this toy model in which all the voids are concentrated inside the holes and all
the structures are in thin spherical shells. For this reason a fitting curve
was plotted: it is passing through the points of the photon path that are in
the cheese between the holes. Indeed, they are points of average behavior and well
represent the coarse graining of this oscillating curve. The
simplification of this model also tells us that the most interesting part of
the plot is farthest from the observer, let us say at $z>1$. In this region we
can see the effect of the holes clearly: they move the curve from the EdS
solution (in purple) to the $\Lambda$CDM one with $\Omega_{M}=0.6$ and
$\Omega_{DE}=0.4$ (in blue). Of course, the model in not realistic enough to
reach the ``concordance'' solution.

Here we are discussing a comparison of our results with those of Ref.\
\cite{Biswas:2007gi}. In that paper they do not find the big difference from FRW
results that we do.  First of all, we can notice that we are able to reproduce their
results using our techniques.  The difference between their results and ours is
that our model has very strong nonlinear evolution, in particular, close to
shell crossing where we have to stop our calculations.  Also, the authors of Ref.\
\cite{Biswas:2007gi} used smaller holes with a different 
density/initial-velocity profile.  This demonstrated that a big change in
observables may require either non-spherical inhomogeneities, or evolution very
close to shell crossing.  (We remind the reader that caustics are certainly
expected to form in cold dark matter models.)

\begin{figure}[htb]
\begin{center}
\includegraphics[width= 13 cm]{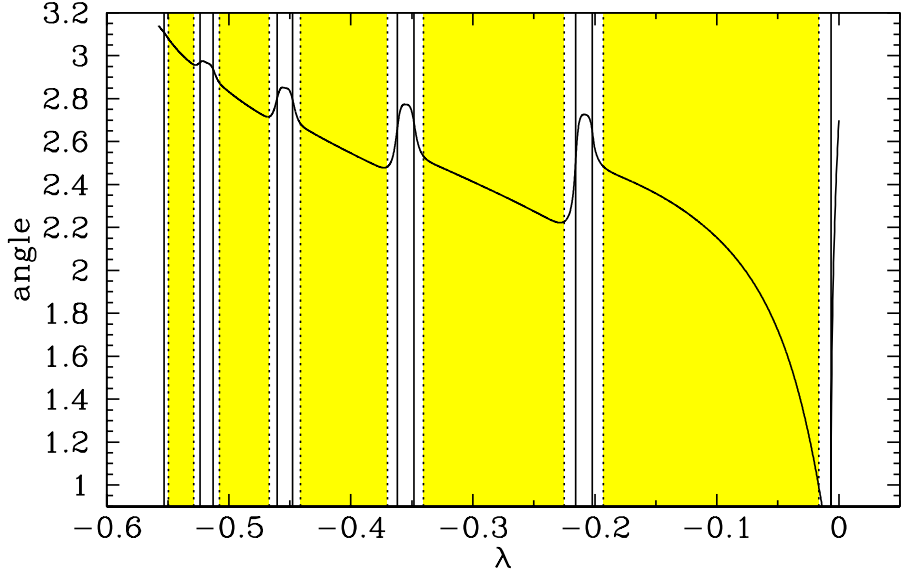}
\caption{\small \slshape The angle from the observer is plotted. The dashed vertical lines 
near the empty region mark the shell of maximum peculiar velocities of 
Fig.\ \ref{curved}. The shaded regions represent the inner FRW solution. The 
solid vertical lines mark the peak in density. The angle at which the photon 
hits the observer is $2.7\,^{\circ}$ on the left.}
\label{bend}
\end{center}
\end{figure}

Let us return now to the reason for our results. As we have seen previously,
due to spherical symmetry there are no significant redshift effects in the
five-hole case. Therefore, these effects must be due to changes in the
angular-diameter distance. Fig.\ \ref{bend} is useful to understand what is
going on: the angle from the observer is plotted. Through the inner void and
the cheese the photon is going straight: they are both $FRW$ solutions even if
with different parameters. This is shown in the plot by constancy of the slope.
The bending occurs near the peak in the density where the $g_{r \, r}$
coefficient of the metric goes toward zero. Indeed the coordinate velocity of
the photon can be split into an angular part: $v_{\phi}=d\phi/dt=1/\sqrt{g_{\phi
\, \phi}}$ and a radial part $v_{r}=dr/dt=1/\sqrt{g_{r \, r}}$. While
$v_{\phi}$ behaves well near the peak, $v_{r}$ goes to infinity in the limit
where shell crossing is reached: the photons are passing more and more
matter shells in a short interval of time as the evolution approaches the
shell-crossing point.  Although in our model we do not reach shell crossing,
this is the reason for the bending. We can, therefore, see that all the effects in
this model, redshift and angular effects, are due to the evolution of
inhomogeneities and this is primary due to the presence of a faster-than-cheese expanding void
which, we think, is a crucial ingredient.

\subsubsection{The one-hole model: the big hole case} 

\begin{figure}[p]
\begin{center}
\includegraphics[width= 15 cm]{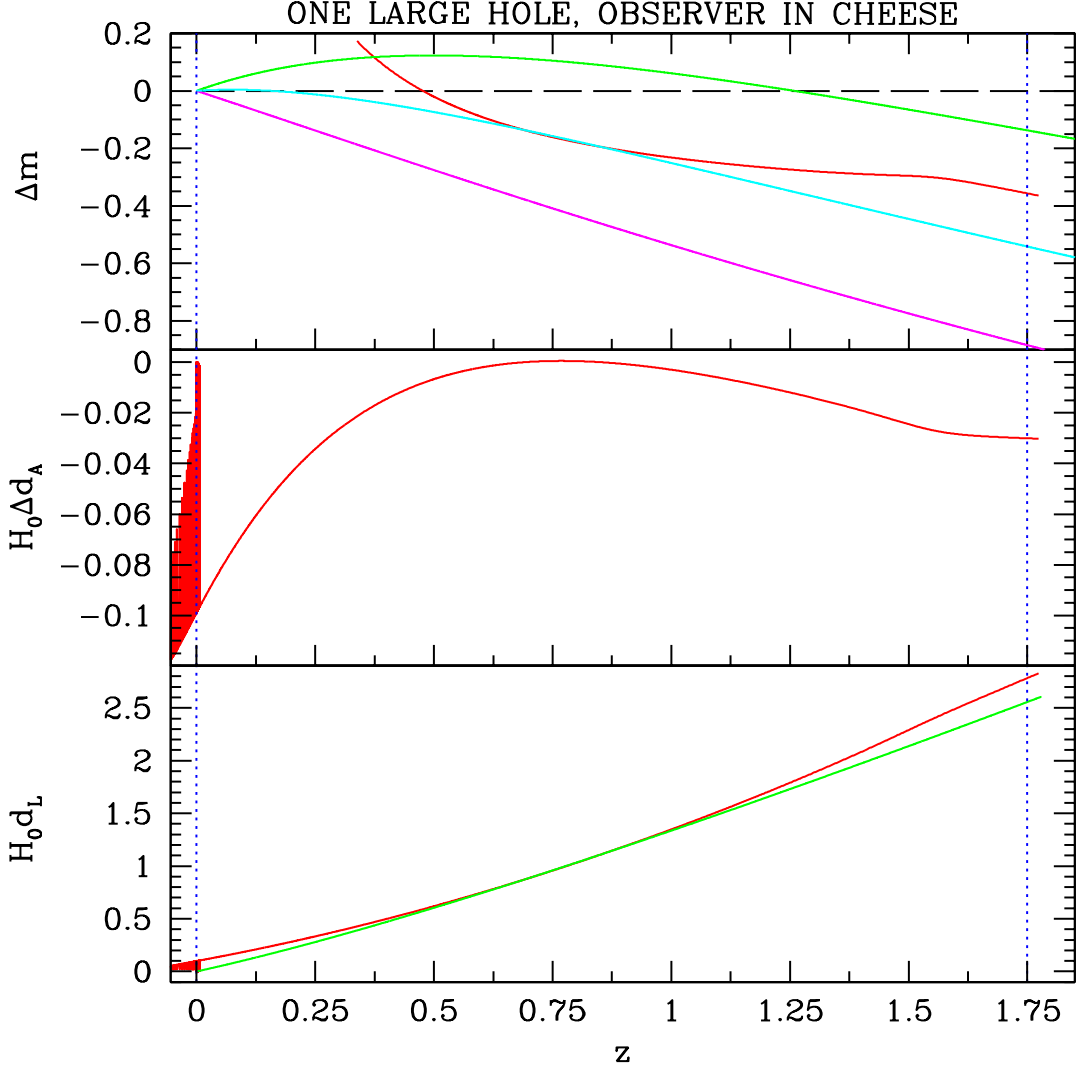}
\caption{\small \slshape On the bottom is shown the luminosity distance $d_L(z)$ in the one-hole
model (jagged curve) and the $\Lambda$CDM solution with $\Omega_{M}=0.6$ and
$\Omega_{DE}=0.4$ (regular curve).  In the middle is the change in the the
angular diameter distance, $\Delta d_A(z)$, compared to a $\Lambda$CDM model
with  $\Omega_{M}=0.6$ and $\Omega_{DE}=0.4$. On the top is shown the distance
modulus in various cosmological models. The jagged line is for the one-hole LTB
model.  The regular curves, from top to bottom are a $\Lambda$CDM model with 
$\Omega_{M}=0.3$ and $\Omega_{DE}=0.7$, a $\Lambda$CDM model with
$\Omega_{M}=0.6$ and $\Omega_{DE}=0.4$ and the EdS model.  The vertical lines
mark the edges of the hole.}
\label{1big}
\end{center}
\end{figure}

Let us see now how the results change if instead of the five-hole model we use
the one-hole model.  We have already shown the redshift results in the previous
section. As one can see from Fig.\ \ref{1big} the results are more dramatic:
for high redshifts the Swiss-cheese curve can be fit by a $\Lambda$CDM model
with less dark energy than $\Omega_{DE}=0.6$ as in the five-hole model.
Nonetheless, the results have not changed so much compared to the change in the
redshift effects discussed in the previous section. Indeed the compensation
scale for angular effects is $2 r_{h}$ while the one for redshift effects is
$r_{h}$.

\subsubsection{The one-hole model: the small hole case}

\begin{figure}[p]
\begin{center}
\includegraphics[width= 15 cm]{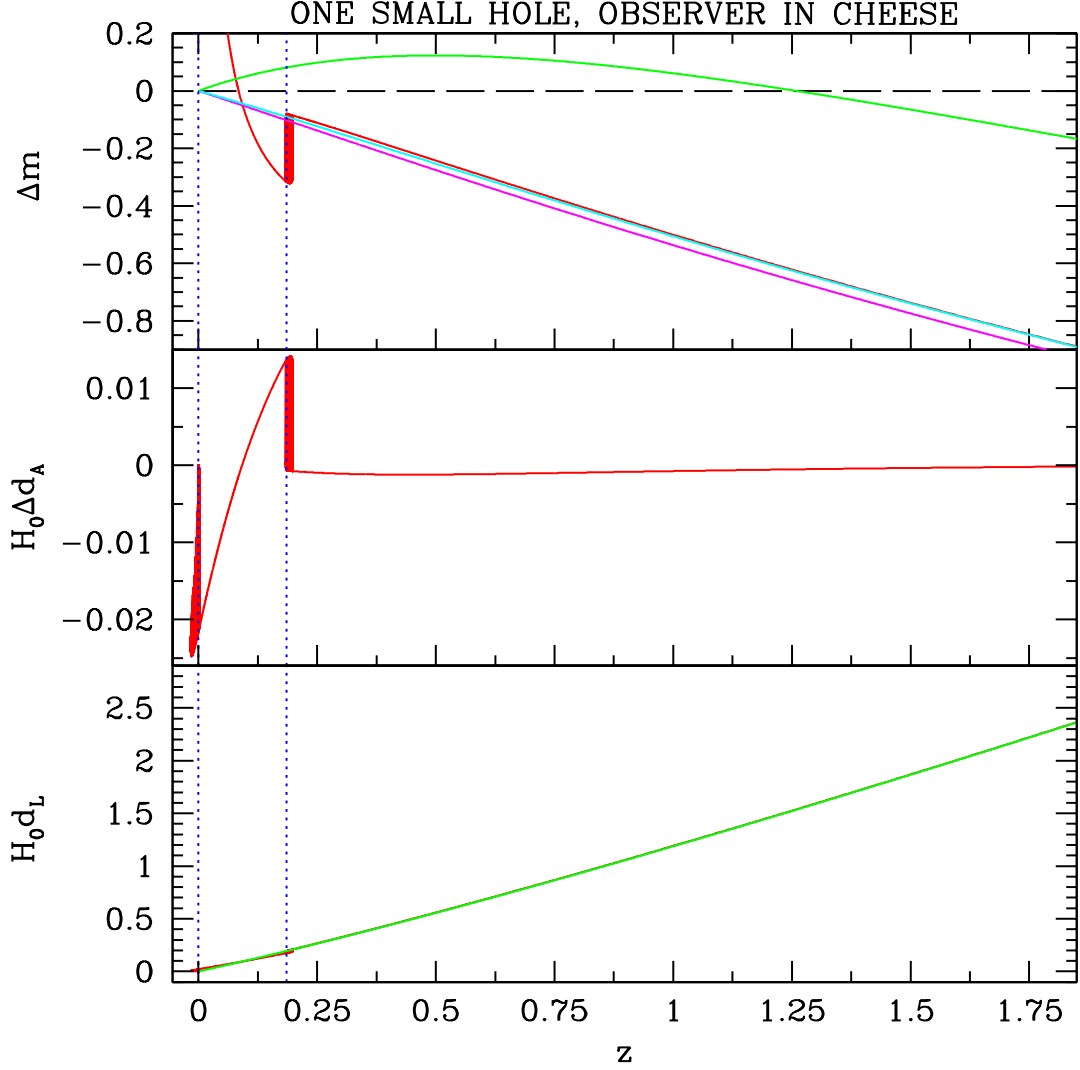}
\caption{\small \slshape On the bottom is shown the luminosity distance $d_L(z)$ in the 1-hole
model (jagged curve) and the $\Lambda$CDM solution with $\Omega_{M}=0.95$ and
$\Omega_{DE}=0.05$ (regular curve).  In the middle is the change in the the
angular diameter distance, $\Delta d_A(z)$, compared to a $\Lambda$CDM model
with  $\Omega_{M}=0.95$ and $\Omega_{DE}=0.05$. On the top is shown the distance
modulus in various cosmological models. The jagged line is for the one-hole LTB
model.  The regular curves, from top to bottom are a $\Lambda$CDM model with 
$\Omega_{M}=0.3$ and $\Omega_{DE}=0.7$, a $\Lambda$CDM model with
$\Omega_{M}=0.95$ and $\Omega_{DE}=0.05$ and the EdS model.  The vertical lines
mark the edges of the hole.}
\label{1small}
\end{center}
\end{figure}

Finally if we remove four holes from the five-hole model, we lose almost all
the effects. This is shown in Fig.\ \ref{1small}: now the model can be compared
to a $\Lambda$CDM model with $\Omega_{M}=0.95$ and $\Omega_{DE}=0.05$.

\clearpage
\section{Results: observer in the hole} \label{hole}

Now we will examine the case in which the observer is inside the last hole in
the five-hole model. We will first put the observer on the high-density shell
and then place the observer in the center.

\subsection{Observer on the high density shell}

\begin{figure}[htb]
\begin{center}
\includegraphics[width= 11 cm]{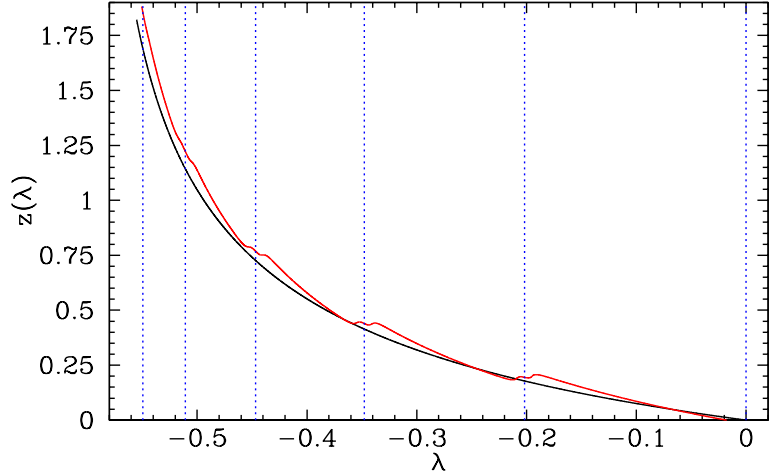}
\caption{\small \slshape Redshift histories for a photon that travels through the 
five-hole-chain to the observer placed on the high density shell.
The ``regular'' line is for the FRW model. $\lambda$ is the affine 
parameter and it grows with the time which go from the left to the right. 
The vertical lines mark the end and the beginning of the holes.}
\label{zpik}
\end{center}
\end{figure}

\begin{figure}[p]
\begin{center}
\includegraphics[width= 15 cm]{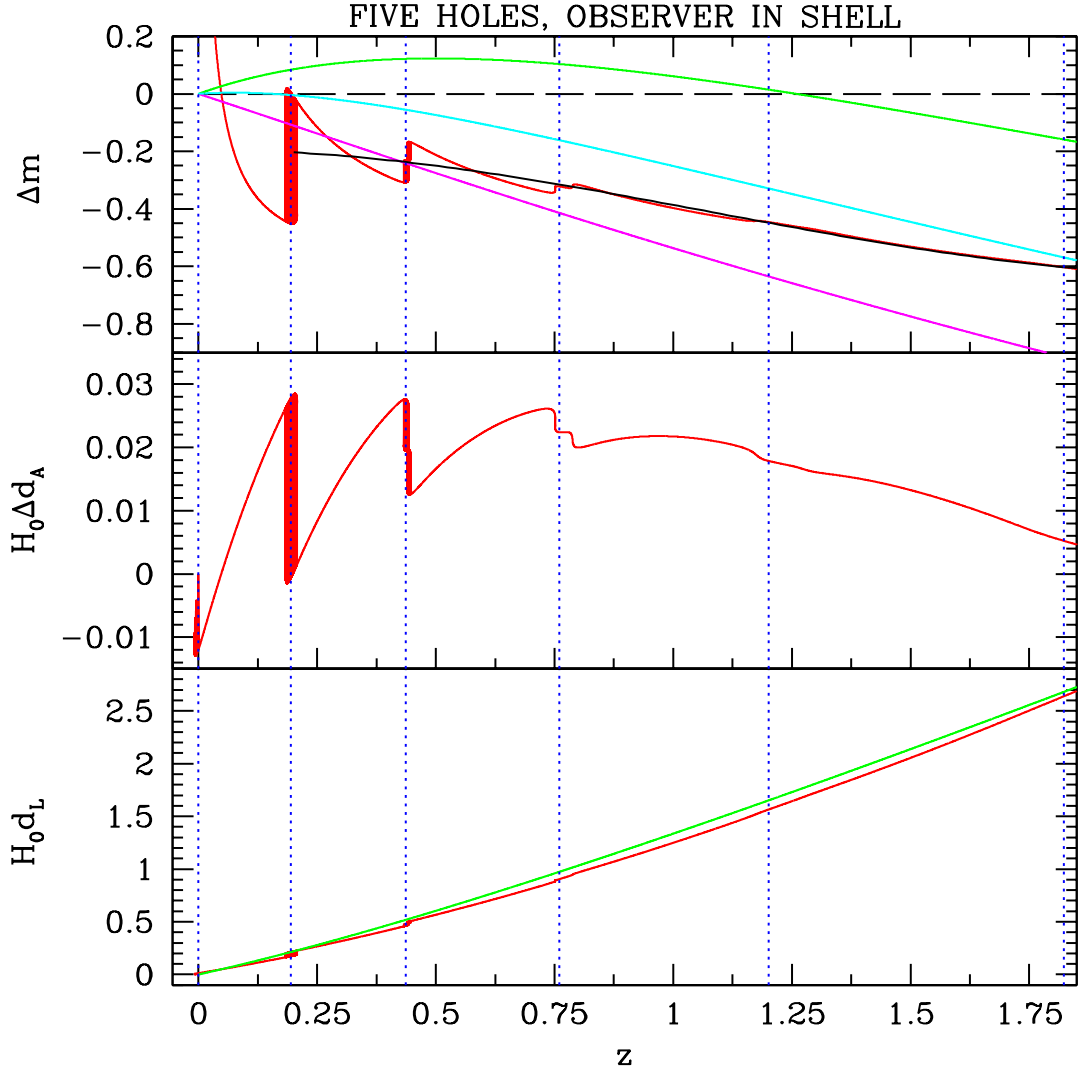}
\caption{\small \slshape On the bottom is shown the luminosity distance $d_L(z)$ in the
five-hole model (jagged curve) and the $\Lambda$CDM solution with
$\Omega_{M}=0.6$ and $\Omega_{DE}=0.4$ (regular curve). In the middle is the
change in the angular diameter distance, $\Delta d_A(z)$, compared to a
$\Lambda$CDM model with  $\Omega_{M}=0.6$ and $\Omega_{DE}=0.4$. On the top is
shown the distance modulus in various cosmological models. The jagged line is
for the five-hole LTB model.  The regular curves, from top to bottom are  a
$\Lambda$CDM model with  $\Omega_{M}=0.3$ and $\Omega_{DE}=0.7$, a $\Lambda$CDM
model with   $\Omega_{M}=0.6$ and $\Omega_{DE}=0.4$, the best smooth fit to the
LTB model, and the EdS model.  The vertical lines mark the edges of the five
holes.}
\label{shell}
\end{center}
\end{figure}

In the section we show the results for the observer on the high-density shell.
As one can see from Fig.\ \ref{zpik}, now the compensation in the redshift
effect is lost: the photon is not completing the entire last half of the last
hole. The results for the luminosity distance and the angular distance do not
change much as shown in Fig.\ \ref{shell}.

Remember that in this case the observer has a peculiar velocity compared to the
FRW observer passing through the same point. We correct the results taking into
account both the Doppler effect and the light aberration effect.

\clearpage
\subsection{Observer in the center}

\begin{figure}[htb]
\begin{center}
\includegraphics[width= 13 cm]{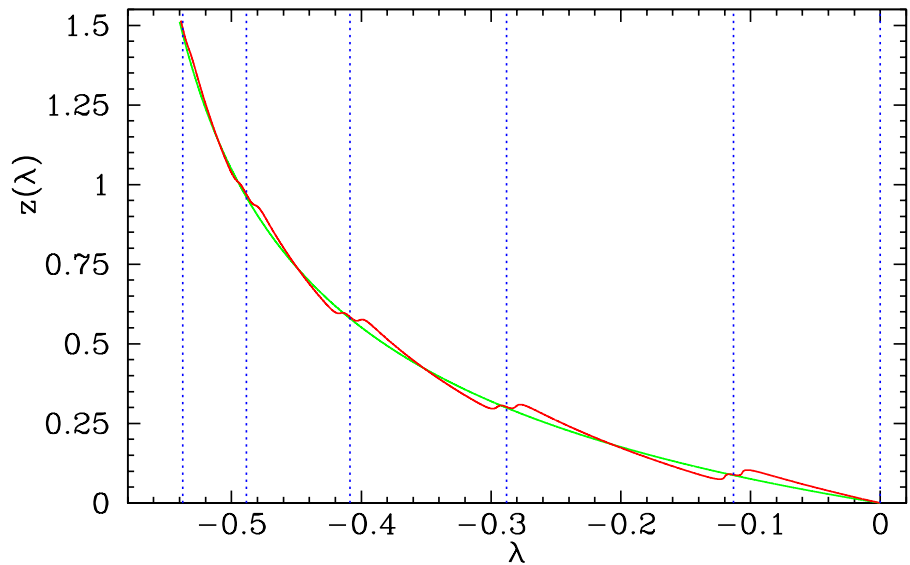}
\caption{\small \slshape Redshift histories for a photon that travels through the 
five-hole-chain to the observer placed in the center. The ``regular'' line is
for the FRW model. $\lambda$ is the affine parameter  and it grows with the
time which go from the left to the right. The vertical  lines mark the end and
the beginning of the holes.}
\label{zetacent}
\end{center}
\end{figure}

In this section we show the results for the observer in the center. As 
confirmed by Fig.\ \ref{zetacent}, the compensation in the redshift effect is
good: the photon is passing through an integer number of half holes.

\begin{figure}[p]
\begin{center}
\includegraphics[width= 15 cm]{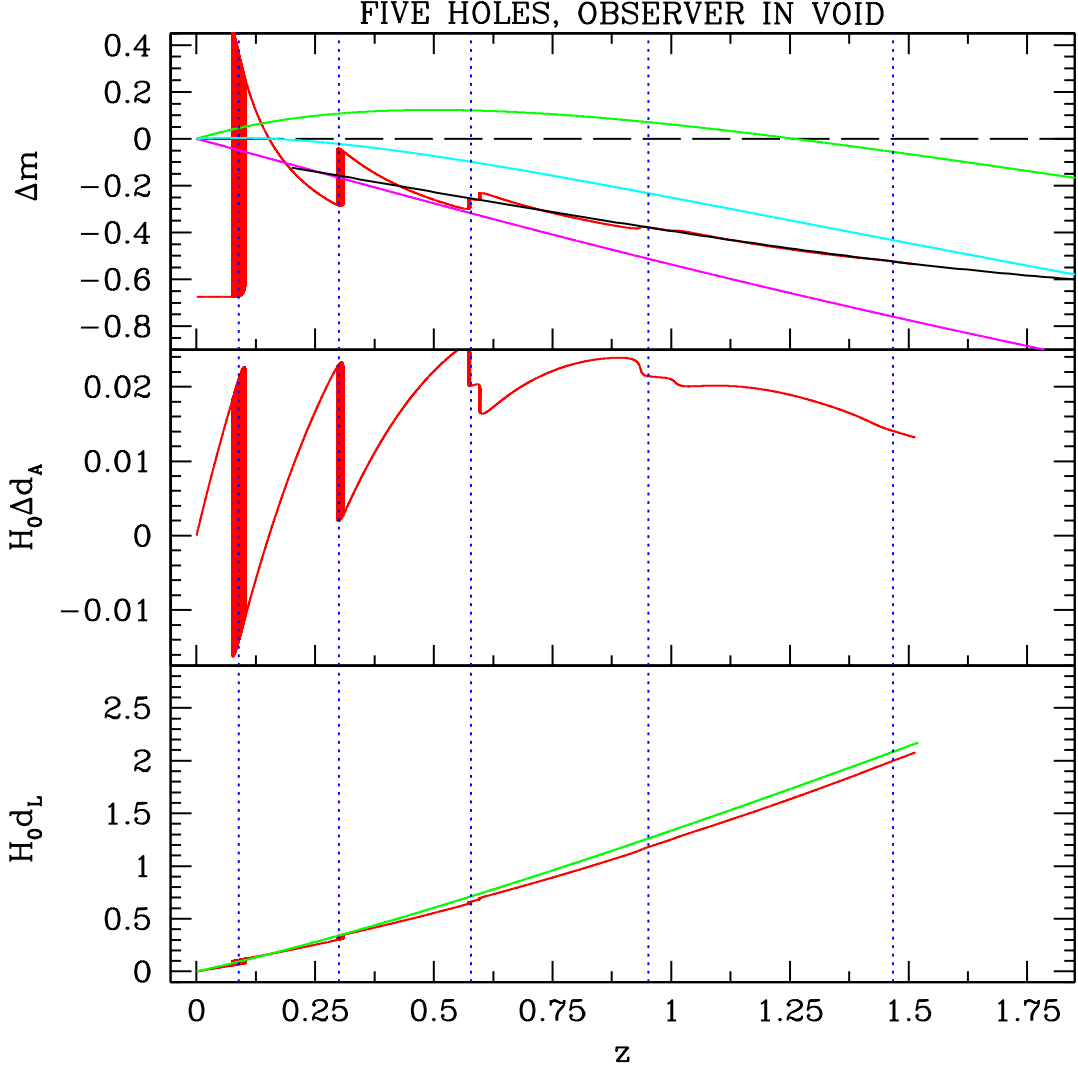}
\caption{\small \slshape The bottom panel shows the luminosity distance $d_L(z)$ in the
five-hole model (jagged curve) and the $\Lambda$CDM solution with
$\Omega_{M}=0.6$ and $\Omega_{DE}=0.4$ (regular curve). In the middle is the
change in the angular diameter distance, $\Delta d_A(z)$, compared to a
$\Lambda$CDM model with  $\Omega_{M}=0.6$ and $\Omega_{DE}=0.4$. On the top
panel the distance modulus in various cosmological models is shown.  The jagged
line is for the five-hole LTB model. The regular curves, from top to bottom are 
a $\Lambda$CDM model with  $\Omega_{M}=0.3$ and $\Omega_{DE}=0.7$, a
$\Lambda$CDM model with $\Omega_{M}=0.6$ and $\Omega_{DE}=0.4$, the best smooth
fit to the LTB model, and the EdS model.  The vertical lines mark the edges of
the five holes.}
\label{void}
\end{center}
\end{figure}

The results for the luminosity distance and the angular distance look worse as
shown in Fig.\ \ref{void}, but this is mainly due to the fact that now the
photon crosses half a hole less than in the previous cases and therefore it
undergoes less bending.

In this case the observer has no peculiar velocity compared to the FRW one:
this is a result of spherical symmetry.

\section{The fitting problem} \label{fitti}

Now that we have seen how the luminosity-distance--redshift relation is 
affected by inhomogeneities, we want to study the same model from the  point of
view of light-cone averaging to see if we can gain insights into how 
inhomogeneities renormalize the matter Swiss-cheese model and mimic a
dark-energy component.

We are going to work out in this section the ideas introduced in Section~\ref{fritto}.
A remark is, however, in order here: in the previous section we did not fit 
the $d_L(z)$ with an FRW solution. We have simply compared the shape of 
the $d_L(z)$ for the Swiss-cheese model with the one of a $\Lambda$CDM model.

We intend now to fit a phenomenological FRW model to our Swiss-cheese model. The
FRW model we  have in mind is a spatially flat model with a matter component
with present fraction of the energy density $\Omega_M=0.25$, and with a
phenomenological dark-energy component with present fraction of the energy
density $\Omega_\Lambda=0.75$.  We will assume that the dark-energy component
has an equation of state
\begin{equation}
\label{fame}
w(a)=w_{0}+ w_{a}\left(1-\frac{a}{a_0}\right)=w_0+w_a \; \frac{z}{1+z} .
\end{equation}
Thus, the total energy density in the phenomenological model evolves as
\begin{equation} 
\label{linda}
\frac{\rho^\textrm{\scriptsize FIT}}{\rho_{0}} = \Omega_M(1+z)^3 + \Omega_\Lambda
(1+z)^{3(1+w_{0}+w_{a})}\; \exp\left(-3w_{a}\frac{z}{1+z}\right) .
\end{equation}
We will refer to this model as the {\it phenomenological model}.

Our Swiss-cheese model is a lattice of holes as sketched in Fig.\
\ref{schizzo}:  the scale of inhomogeneities is therefore simply the size of
a hole.  We are interested in understanding how the equation of state of
``dark energy'' in the phenomenological model changes with respect to $r_{h}$,
and in  particular, why. Of course, in the limit $r_{h} \rightarrow 0$, we
expect to find  $w=0$, that is, the underlying EdS model out of which the cheese
is constructed.

\clearpage

\begin{figure}[htb]
\begin{center}
\includegraphics[width=10 cm]{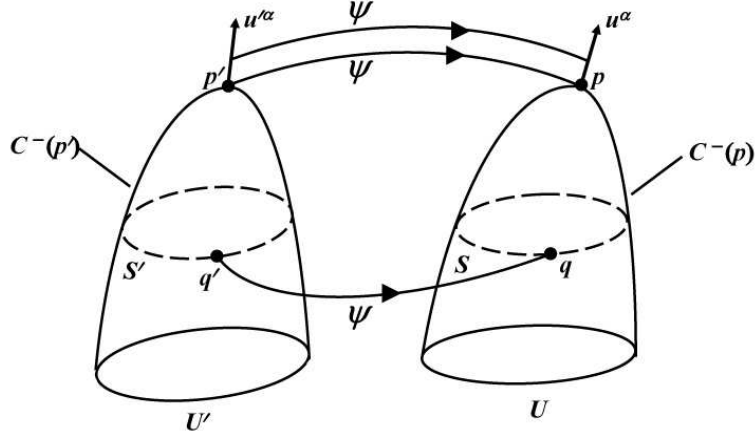}
\caption{\small \slshape In the null data best fitting, one successively chooses maps  from the
real cosmological model $U$ to the FRW model $U'$ of the null  cone vertex
$p'$, the matter 4-velocity at $p'$, a two-sphere $S'$ on the  null cone of
$p'$ and a point $q'$ on the 2-sphere. This establishes the  correspondence
$\psi$ of points on the past null cone of $p'$, $C^-(p^\prime)$, to the past 
null cone of $p$, $C^-(p^\prime)$, and then compares initial data at $q'$ and
at $q$. From  Figure 2 of \cite{ellis-1987}.}
\label{map}
\end{center}
\end{figure}

The procedure developed by Ref.\ \cite{ellis-1987} is summarized by Fig.\
\ref{map}.  We refer the reader to that reference for a more thorough analysis
and to Ref.\ \cite{Celerier:2007tp} and references therein for recent
developments. We will focus now in using our Swiss-cheese model as a cosmological (toy)
model.

\subsection{Choice of vertex points}

We start choosing the two observers to be compared. In the homogeneous FRW
model every observer is the same thanks to spatial  homogeneity. We choose an
observer in the cheese as the corresponding observer in our Swiss-cheese model,
in particular the one shown in Fig.\ \ref{schizzo}.

Our model allows us to choose also the time of observation, which, in general,
is a final product of the comparison. We now explain why.

The FRW model we will obtain from the fit will evolve differently from the 
Swiss cheese: the latter evolves as an EdS model, while the former will 
evolve as a quintessence-like model. They are really different models. 
They will agree only along the light cone, that is, on our observations.

Now, for consistency, when we make local measurements\footnote{Conceptually, 
it could not be possible with a realistic universe model to make 
local measurements that could be directly compared to the smoothed FRW 
model. We are allowed to do so thanks to our particular Swiss-cheese 
model in which the cheese well represents the average properties of the 
model.} the two models have to give us the same answer: local
measurements indeed can be seen as averaging measurements with a small enough 
scale of averaging, and the two models agree along the past light cone. 

Therefore, we choose the time in order that the two observers measure  the same
local density. This feature is already inherent in Eq.\ (\ref{linda}):  the
phenomenological model and the Swiss-cheese model evolve in order to have the
same local density, and therefore the same Hubble parameter, at the present
time.

\subsection{Fitting the 4-velocity}

The next step is to fit the four-velocities of the observers.  In the FRW model
we will choose a comoving observer, the only one who experiences an isotropic
CMB. In the Swiss-cheese model, we will choose, for the same reason, a 
cheese-comoving observer. Again, our Swiss-cheese model considerably simplifies
our work.

\subsection{Choice of comparison points on the null cones} \label{compa}

Now that the past null cones are uniquely determined, we have to choose a 
measure of distance to compare points along each null cone.

First, let us point out that instead of the entire two-sphere along the null
cone, we will examine, only a point on it. This is because of the simplified
set-up of our Swiss-cheese model in which  the observer is observing only in
two opposite directions, as  illustrated in Fig.\ \ref{schizzo}. This means
that we can skip the step consisting in averaging our observable quantities
over the surface of constant redshift,  which is generally necessary in order
to be able to compare an  inhomogeneous model with the FRW model
\cite{ellis-1987}.

Coming back to the main issue of this section,  we will use the observed
redshift  $z$ to compare points along the null cones. Generally, the
disadvantage of  using it is that it does not directly represent distances
along the null  cone. Rather, the observed value $z$ is related to the
cosmological redshift  $z_{C}$ by the relation:
\begin{equation}
1+z=(1+z_{O})(1+z_{C})(1+z_{S})
\end{equation}
where $z_{O}$ is the redshift due to the peculiar velocity of the observer 
$O$ and $z_{S}$ that due to the peculiar velocity of the source. The latter, 
in particular, is a problem because local observations cannot distinguish 
$z_{S}$ from $z_{C}$. 

However, our set up again simplifies this task. The chosen observers are, 
indeed, both comoving (in the Swiss-cheese model because the observer is in the
cheese, and in the phenomenological model by construction), and therefore
$z_{O}=0$.  Regarding the sources,  we know exactly their behavior because we
have a model to work with.

The sources are also comoving; however, there are structure-formation  effects
that should be disentangled from the average evolution. For this  reason we
will perform averages between points in the cheese (the  meaning of this will
be clear in the next section) in order to  smooth out these structure-formation
effects.

\clearpage
\subsection{Fitting the null data} \label{fitting}

\begin{figure}[htb]
\begin{center}
\includegraphics[width=16.2 cm]{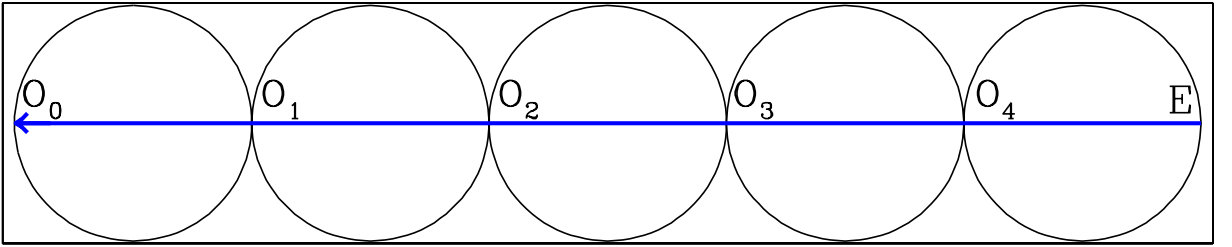}
\caption{\small \slshape An illustration of the points chosen for the averaging procedure.}
\label{shime}
\end{center}
\end{figure}

Now we are ready to set up the fitting of our Swiss-cheese model. Ref.\
\cite{Hellaby:1988zz} studied the approach based on volume averaging outlined
in Ref.\ \cite{ellis-1987}. This approach, however, is appropriate for studies 
concerning global dynamics, as in Refs.\ \cite{Buchert:2002ht, Buchert:2007ik}. As
stressed previously, here we are instead interested in averages  {\it directly}
related to observational quantities, and we constructed our  model following
this idea: it is a model that is exactly solvable and  ``realistic'' (even if
still a toy) at the price of no interesting volume-averaged dynamics.

Therefore, we will follow a slightly different approach from the ones outlined
in Ref.\ \cite{ellis-1987}:  we are going to fit averages along the light cone.
This method will be intermediate between the fitting approach and the
averaging approach.

We will focus on the expansion scalar and the density. We will see that these
two quantities behave differently under averaging. We denote by
$Q^\textrm{\scriptsize SC}(r, t)$ a quantity in the Swiss-cheese model we want to average.
We denote by $Q^\textrm{\scriptsize FIT}(t)$ the corresponding quantity we want to fit to
the average of $Q^\textrm{\scriptsize SC}(r, t)$. Note that $Q^\textrm{\scriptsize FIT}(t)$ does not
depend on $r$ because the phenomenological model we will employ to describe the
Swiss-cheese model is homogeneous.

Again, the fit model is a phenomenological homogeneous model (just refereed to
as the phenomenological model).  It need not be the model of the cheese.

The procedure is as follows. First we will average $Q^\textrm{\scriptsize SC}(r, t)$ for a
photon that starts from the emission point $E$ of the five-hole chain and
arrives at the locations of observers $O_{i}$ of Fig.\ \ref{shime}. We have
chosen those points because they well represent the average dynamics of the
model. Indeed, these points are not affected by structure evolution because they
are in the cheese.  Then, we will compare this result with the average of
$Q^\textrm{\scriptsize FIT}(t)$ for  the phenomenological and homogeneous source with
density given by Eq.\ (\ref{linda}) with an equation of state $w$ given by Eq.\
(\ref{fame}). 

The two quantities to be compared are therefore:
\begin{eqnarray}
\langle Q^\textrm{\scriptsize SC} \rangle_{\overline{\scriptscriptstyle EO}_{i}}
& = & \left[  \int_{E}^{O_{i}} dr \; Y' / W \right]^{-1} 
\int_{E}^{O_{i}} dr \ Q^\textrm{\scriptsize SC}(r,t(r)) \; Y'(r,t(r)) / W(r)  \nonumber \\
\langle Q^\textrm{\scriptsize FIT} \rangle_{\overline{\scriptscriptstyle EO}_{i}} 
\label{ququ1}
& = & \left[  \int_{E}^{O_{i}} dr \;  a_{\scriptscriptstyle FIT} \right]^{-1}
\int_{E}^{O_{i}} dr \ Q^\textrm{\scriptsize FIT}(t_{\scriptscriptstyle FIT}(r))\; 
a_{\scriptscriptstyle FIT}(t_{\scriptscriptstyle FIT}(r))  ,
\end{eqnarray}
where $t(r)$ and $t_{\scriptscriptstyle FIT}(r)$ are the photon  geodesics in
the Swiss-cheese model and in the phenomenological one, respectively. The
functions $t_{\scriptscriptstyle FIT}(r)$, $a_{\scriptscriptstyle FIT}$  and
other  quantities we will need are obtained solving the Friedman  equations
with a source described by Eq.\ (\ref{linda}) with no curvature. The points
$O_{i}$ in the Swiss-cheese model of Fig.\ \ref{shime}  are associated to
points in the phenomenological model with the same redshift, as discussed in
Sec.\ \ref{compa}.

We will then find the $w$ that gives the best fit between 
$\langle Q^\textrm{\scriptsize FIT}\rangle$ and $\langle Q^\textrm{\scriptsize SC}\rangle$, that 
is, the choice that minimizes:
\begin{equation} \label{ququ2}
\sum_{i} \left ( \langle Q^\textrm{\scriptsize FIT} 
\rangle_{\overline{\scriptscriptstyle EO}_{i}}-
\langle Q^\textrm{\scriptsize SC} 
\rangle_{\overline{\scriptscriptstyle EO}_{i}} \right )^{2}  .
\end{equation}
Of course, in the absence of inhomogeneities, this method would give $w=0$.


\ 

Let us summarize the approach:
\begin{itemize}

\item We choose a phenomenological quintessence-like model that, at the
present time, has the same density and Hubble parameter as the EdS-cheese
model. 

\item We make this phenomenological model and the Swiss-cheese model
correspond along the light cone via light-cone averages of $Q$.

\item We can substitute the Swiss-cheese model with the phenomenological 
model as far as the averaged quantity $Q$ is concerned.

\end{itemize} 

The ultimate question is if it is observationally meaningful to consider $Q$,
as opposed to the other choice of domain averaging at constant time, which is
not directly  related to observations. We will come back to this issue after
having obtained the results.

\clearpage
\subsubsection{Averaged expansion}

The first quantity in which we are interested is the expansion rate. To average
the expansion rate we will follow the formalism developed in Sec.\
\ref{histories}. We will therefore apply Eqs.\ (\ref{ququ1}-\ref{ququ2}) to 
$Q^\textrm{\scriptsize SC}=H_{r}\equiv\dot{Y}' / Y'$, where we remember that $H_r$ is the
radial expansion rate. The corresponding  quantity in the phenomenological
model is $Q^\textrm{\scriptsize FIT}=\dot{a}_{\scriptscriptstyle 
FIT}/a_{\scriptscriptstyle FIT}$.

\begin{figure}[htb]
\begin{center}
\includegraphics[width=13 cm]{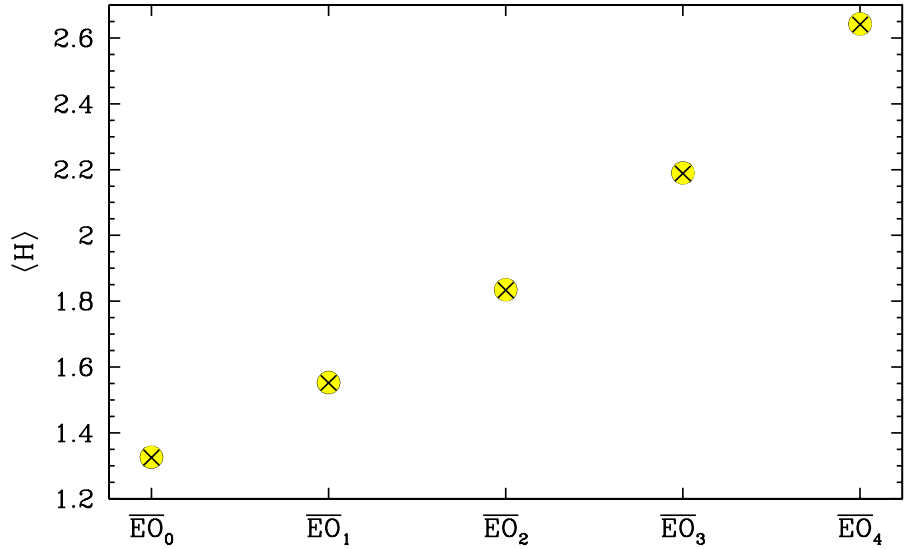}
\caption{\small \slshape Average expansion rate. The yellow points are $\langle H^\textrm{\scriptsize SC} 
\rangle_{\overline{\scriptscriptstyle EO}_{i}}$
 while the crosses are $\langle H^\textrm{\scriptsize FIT} 
\rangle_{\overline{\scriptscriptstyle EO}_{i}}$.
 $\overline{EO}_{i}$ means that the average was performed from $E$
and $O_{i}$ with respect to Fig.\ \ref{shime}. The best fit is 
found for
$w \simeq0$, that is, the phenomenological model is the cheese-FRW solution 
itself as far as the expansion rate is concerned.}
\label{hfit}
\end{center}
\end{figure}

For the same reason there is good compensation in redshift  effects (see Sec.\
\ref{histories}), we expect $\langle H_{r} \rangle$ to behave very similarly to
the FRW cheese solution. Indeed, as one can see in Fig.\ \ref{hfit}, the best
fit of the  Swiss-cheese model is given by a phenomenological source with
$w\simeq0$, that  is, the phenomenological model is the cheese-FRW solution
itself as far as the  expansion rate is concerned.

\clearpage
\subsubsection{Averaged density}

\begin{figure}[htb]
\begin{center}
\includegraphics[width=13 cm]{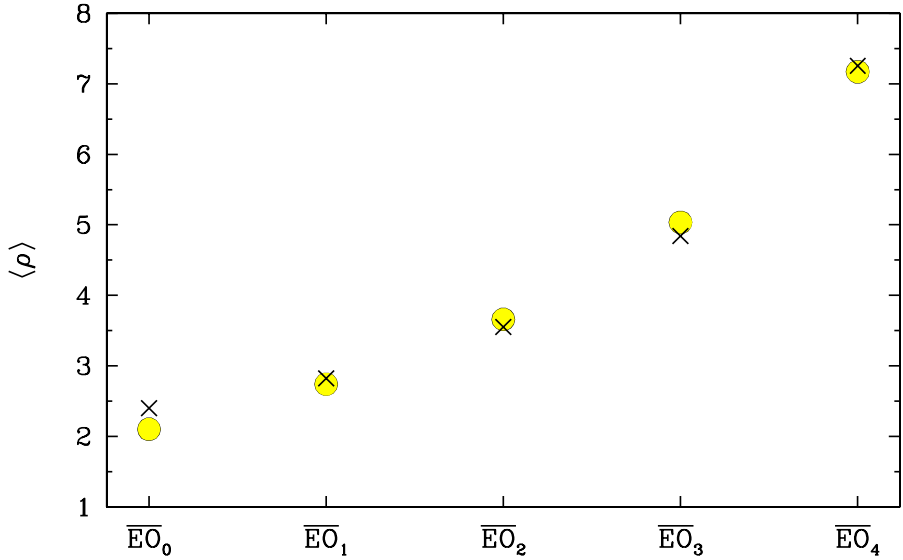}
\caption{\small \slshape Average density in $\rho_{C 0}$ units. The  yellow points
are  $\langle \rho^\textrm{\scriptsize SC} \rangle_{\overline{\scriptscriptstyle EO}_{i}}$
while the crosses are $\langle \rho^\textrm{\scriptsize FIT} 
\rangle_{\overline{\scriptscriptstyle EO}_{i}}$. $\overline{EO}_{i}$ means that
the average was performed from $E$ and $O_{i}$ with respect to Fig.\
\ref{shime}. The parametrization of  $\rho^\textrm{\scriptsize FIT}$ is from Eq.\
(\ref{linda}). The best fit is found for $w_{0}=-1.95$ and $w_{a}=4.28$.}
\label{qfit}
\end{center}
\end{figure}

The situation for the density is very different. The photon is spending more and
more time in the (large) voids than in the (thin) high density structures. We
apply Eqs.\ (\ref{ququ1}-\ref{ququ2}) to $Q^\textrm{\scriptsize SC}=\rho^\textrm{\scriptsize SC}$. The
corresponding quantity in the phenomenological model is 
$Q^\textrm{\scriptsize FIT}=\rho^\textrm{\scriptsize FIT}$ where $\rho^\textrm{\scriptsize FIT}$ is given by Eq.\
(\ref{linda}). The results are illustrated in Fig.\ \ref{qfit}: the best fit is
for  $w_{0}=-1.95$ and $w_{a}=4.28$.

As we will see in Sec.\ \ref{dressing}, we can achieve a better fit to the
concordance model with smaller holes than the ones of $350$ Mpc considered
here. We anticipate that for a holes of radius $r_{h}=250$ Mpc, we have
$w_{0}=-1.03$ and $w_{a}=2.19$.

We see, therefore, that this Swiss-cheese model could be interpreted, in the 
FRW hypothesis, as a homogeneous model that is initially dominated by matter
and subsequently by dark energy: this is what the concordance model suggests.
We stress that this holds only for the light-cone averages of the density.

\clearpage
\subsection{Explanation}

\begin{figure}[htb]
\begin{center}
\includegraphics[width=12 cm]{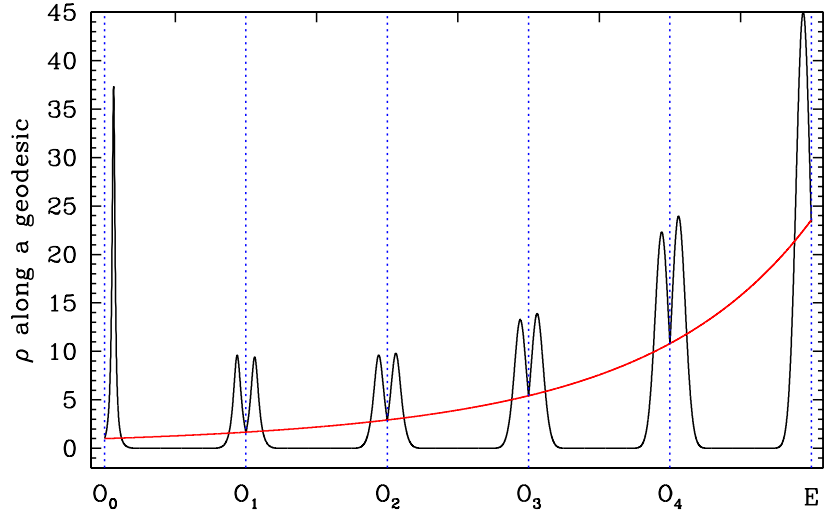}
\caption{\small \slshape Density along the light-cone for the Swiss-cheese model (the spiky
curve) and the EdS model of the cheese (the regular curve). The labeling of 
the $x$-axis is the same one of Fig.\ \ref{shime}.}
\label{photorho}
\end{center}
\end{figure}

Let us first explore the basis for what we found. In  Fig.\ \ref{photorho} we
show the density along the light cone for both the Swiss-cheese model and the
EdS model for the cheese. It is clear that the  photon is spending more and more
time in the (large) voids than in the (thin) high density structures.

\begin{figure}[htb]
\begin{center}
\includegraphics[width=10 cm]{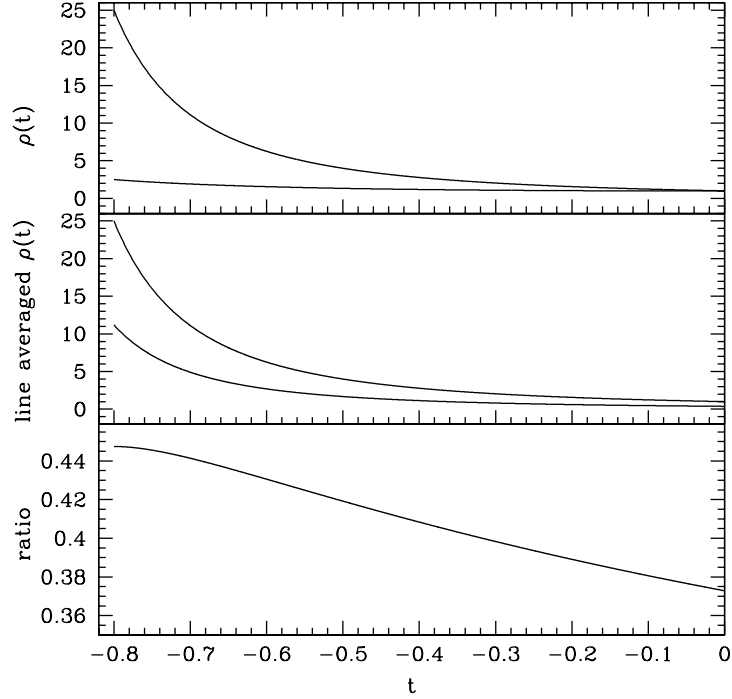}
\caption{\small \slshape At the top is the evolution of the energy density for the  Eds cheese
model (higher curve) and for the phenomenological model with  $w_{0}=-1.95$ and
$w_{a}=4.28$. In the middle is the constant-time line averaged density as a
function of  time for the Swiss-cheese model (lower curve) and the cheese-EdS
model  (higher curve).  At the bottom is their ratio of the last two quantities
as  a function of time.}
\label{opacity}
\end{center}
\end{figure}

To better show this, we plotted in Fig.\ \ref{opacity} the constant-time, 
line-averaged density as a function of time. The formula used for 
the Swiss-cheese model is
\begin{equation} \label{opacityf}
\int_{0}^{r_{h}} dr \ \rho(r,t) \; Y'(r,t) / W(r)  \left /  
\int_{0}^{r_{h}}dr \; Y' / W \right.  ,
\end{equation}
while for the cheese, because of homogeneity we can just use $\rho(t)$ of the
EdS model. As one can see, the photon is encountering less matter in the
Swiss-cheese model than in the EdS cheese model. Moreover, this becomes
increasingly true with the formation of high-density regions as illustrated in 
Fig.\ \ref{opacity} by the evolution  of the ratio of the previously calculated
average density: it  decreases by $17 \%$ from the starting to the ending time. 

The calculation of Eq.\ (\ref{opacityf}) is actually, except for some factors 
like the cross-section, the opacity of the Swiss-cheese model. Therefore, a
photon propagating through the Swiss-cheese model has a different average
absorption history; that is, the observer looking through the cheese will
measure  a different flux compared to the case with only cheese and no holes.
For the moment, in order to explore the physics, let us make the approximation
that during the entire evolution of the universe, the matter is transparent to
photons.

{}From the plots just shown we can now understand the reason for the best fit 
values of $w_{0}=-1.95$ and $w_{a}=4.28$ found in the case of holes of
$r_{h}=350$ Mpc. We are using a homogeneous phenomenological model, which has at
the present time the  density of the cheese (see Fig.\ \ref{opacity}). We want
to use it to fit the line-averaged density of the Swiss cheese, which is lower
than the (volume) averaged one. Therefore, going backwards from the present
time, the phenomenological model must keep its density low, that is, to have a
small $w$. At some point, however, the density has to start to increase,
otherwise it will not match the line-averaged value that keeps increasing:
therefore  $w$ has to increase toward $0$. It is very interesting that this
simple mechanism mimics the behavior of the concordance-model equation of state.
We stress that this simple mechanism works thanks to the set-up and fitting
procedure we have chosen; that is, the fact that we matched the cheese-EdS
solution at the border of the hole, the position of the observer, and the
observer looking through the holes. Moreover, we did not tune the model to
achieve a best matching with the concordance model. The results shown are indeed
quite natural.

\subsection{Beyond spherical symmetry} \label{cosca}

\begin{figure}[h!]
\begin{center}
\includegraphics[width=15 cm]{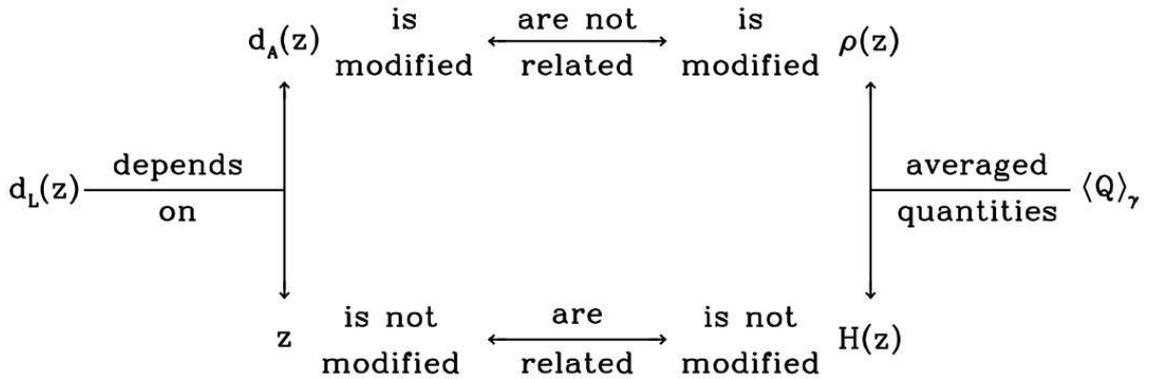}
\caption{\small \slshape Flow chart regarding relationships between the results obtained.}
\label{schema}
\end{center}
\end{figure}

We have summarized the relationships between the results obtained in this chapter till now 
in the flow chart of  Fig.\ \ref{schema}.

Regarding $d_{L}(z)$, we found no important effects from a change in the
redshift:  the effects on $d_L(z)$ all came from $d_{A}$ driven by the evolution
of the inhomogeneities.

Regarding light-cone averages, we found no important effects with respect 
to the expansion: this negative result is due to the compensation in 
redshift discussed in Sec.\ \ref{histories} and it is the same reason 
why we did not find redshift effects with $d_{L}(z)$. This is the main 
limitation of our model and it is due ultimately to the spherical
symmetry of the model as explained in Sec.\ \ref{histories}.

We found important effects with respect to the density: however this is 
not due to the effects driving the change in $d_{A}$. The latter is due to 
structure evolution while the former to the presence of voids, so the 
two causes are not directly connected. Indeed, it is possible to turn off 
the latter and not the former.

We can therefore make the point that the expansion is not affected by 
inhomogeneities because of the compensation due to the spherical symmetry.
Density, on the other hand, is not affected by spherical symmetry, so there
are no compensations, and the photon will systematically see more and more voids
than  structures. We can therefore argue that the average of density is more
relevant  than the average of expansion because it is less sensitive to the
assumption of spherical symmetry,  which is one of the limitations of this
model.

\begin{figure}[h!]
\begin{center}
\includegraphics[width=14 cm]{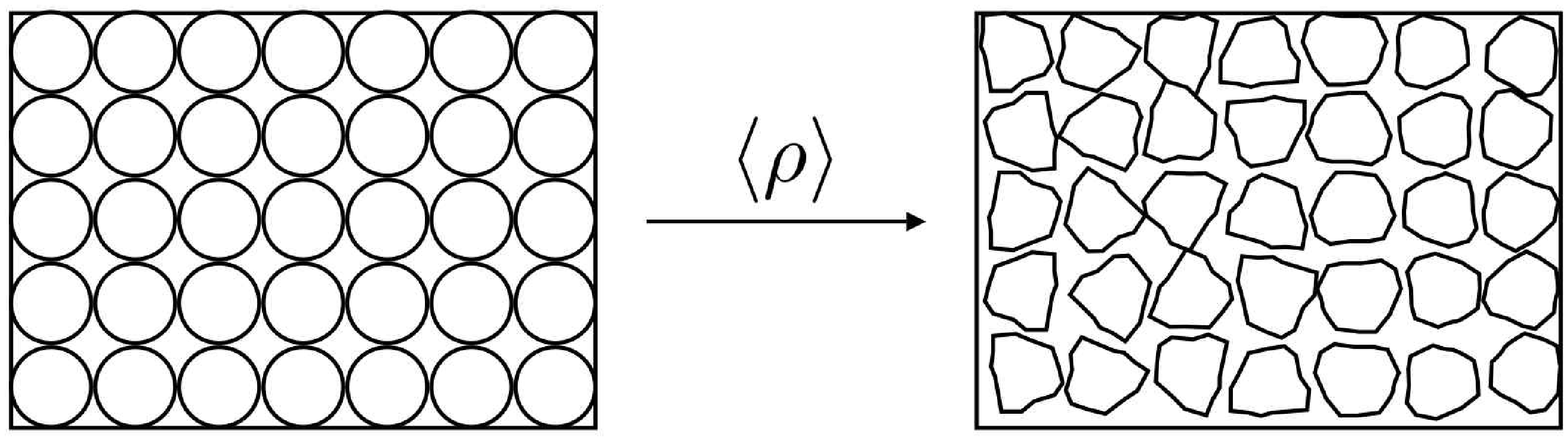}
\caption{\small \slshape Using $\langle \rho \rangle_{\gamma}$ as a tool to go from a Swiss-cheese model with 
spherically symmetric holes to a Swiss-cheese model with non-spherically symmetric holes.
The insensitivity of density to spherical symmetry makes us think that these two models would share the same light-cone averaged density.}
\label{scns}
\end{center}
\end{figure}

The next step is to define a Hubble parameter from this average density:  
$H^{2}\propto \langle \rho \rangle_{\gamma}$. In this way we are moving  from a
Swiss cheese made of spherically symmetric holes to a Swiss cheese  without
exact spherical symmetry. The correspondence is through the light-cone averaged
density which,  from this point of view, can be seen as a tool in performing
this step. See Fig.~\ref{scns} for a sketch.

\clearpage
Summarizing again:

\begin{itemize}

\item We started out from a Swiss-cheese model containing spherically
symmetric  holes only. A photon, during its journey through the Swiss cheese,
undergoes a  redshift that is not affected by inhomogeneities. However the
photon is  spending more and more time in the voids than in the structures. The
lack of an effect is due to spherical symmetry. We focused on this because a
photon spending most of its time in voids should have a different redshift
history than a photon propagating in a homogeneous background.

\item Since density is a quantity that is not particularly sensitive to
spherical symmetry, we tried to solve the mismatch  by focusing on 
density alone, getting from it expansion (and therefore the redshift 
history).

\item We ended up with a Swiss-cheese model with holes, which are  actually not
spherically symmetric. In this model there is an effect on the redshift history
of a photon due to the voids.

\item In practice this means that we will use the phenomenological best-fit
model  found, that is, we will use a model that behaves similarly to the
concordance  model.

\end{itemize}

\subsection{Motivations}

Let us go back to the discussion of Sec. \ref{fitting}, that is, if it is 
observationally meaningful to consider light-cone averages of $Q$  as the basis
for the correspondence. For example, domain averages at constant time are not
directly related  to observations.

Here, we are not claiming that light-cone averages are observationally 
relevant\footnote{However, a density light-cone average is an indicator 
of the opacity of the universe and, therefore, could be observationally 
relevant, as explained in the discussion around Fig.\ \ref{opacity}.}. 
Rather, we are using light-cone averages as tools to understand the model 
at hand. The approach has been explained in the previous section.

\clearpage
\section{Renormalization of the matter equation of state} \label{dressing}

\begin{figure}[htb]
\begin{center}
\includegraphics[width=14cm]{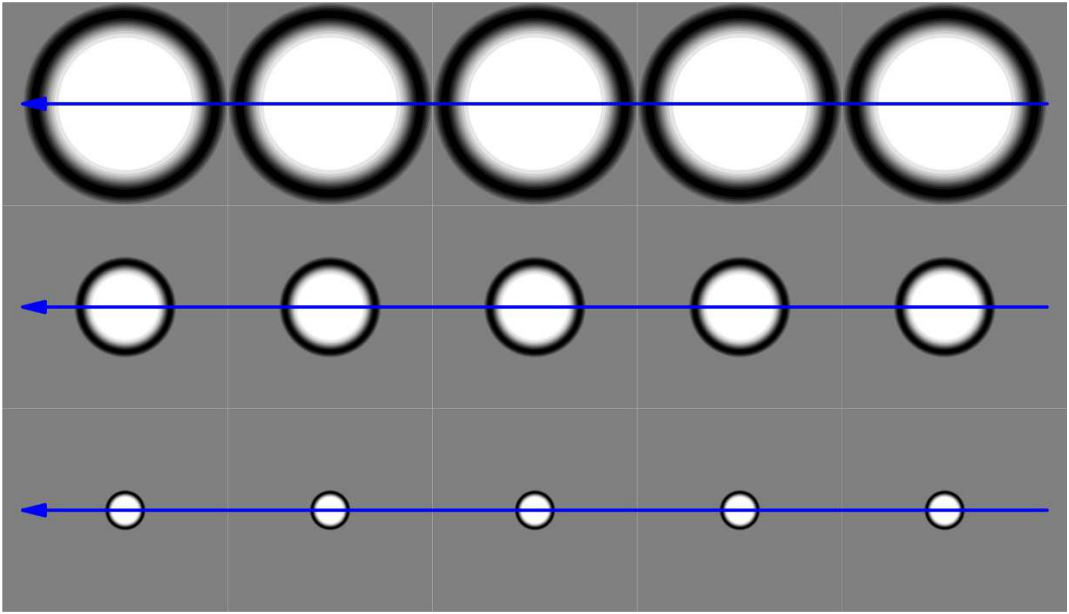}
\caption{\small \slshape Sketch of how the size of the inhomogeneity is changed in our  model.
The shading mimics  the initial density profile: darker shading implies larger
denser. The  uniform  gray is the FRW cheese. The photons pass through the holes
as shown by the  arrows and are revealed by the observer whose comoving position
in the  cheese does not change.  The size of the holes correspond to $n=0,$
$2$, $5$ of Eq.\ (\ref{scala}).}
\label{sdressing}
\end{center}
\end{figure}

In this section we will study how the parameters of the phenomenological  model
depend on the size of inhomogeneities, that is, on the size  of the hole. We
sketched in Fig.\ \ref{sdressing} our set-up: we keep the comoving position of
the centers of the holes fixed. The observer is located in  the same piece of
cheese.

We changed the radius of the hole according to
\begin{equation} \label{scala}
r_h(n) = \frac{r_h}{1.4^{n}} ,
\end{equation}
where $r_{h}$ is the radius we have been using till now, the one that results in
the holes touching. The choice of the $1.4$ in the scaling is only  for
convenience. We let $n$ run from $0$ to $7$.

In this analysis we will use instead of the energy density in Eq.\
(\ref{linda}), an  energy density in which only one effective source appears,
and the effective source evolves as
\begin{equation} 
\label{linda2}
\frac{\rho^\textrm{\scriptsize FIT}}{\rho_{0}} =
(1+z)^{3(1+w^{R}_{0}+w^{R}_{a})}\; \exp\left(-3w^{R}_{a}\frac{z}{1+z}\right)
\qquad \mbox{with} \qquad
w^{R}(z)=w^{R}_0+w^{R}_a \; \frac{z}{1+z} .
\end{equation}

We put $R$ as a superscript on the equation of state in order to differentiate
the parametrization of Eq.\ (\ref{linda2}), which we are now using to study
renormalization, from the parametrization of Eqs.\ (\ref{fame}-\ref{linda}),
which we used to compare the phenomenological model to the concordance model.
We are not disentangling different sources in Eq.\ (\ref{linda2}) because we
are interested in the renormalization of the matter equation of state of the
cheese,  that is, on the dependence of $w^{R}$ upon the size of the hole. To
this purpose we need only one source in order to keep track of the changes.

\begin{figure}[htb]
\begin{center}
\includegraphics[width=14.5 cm]{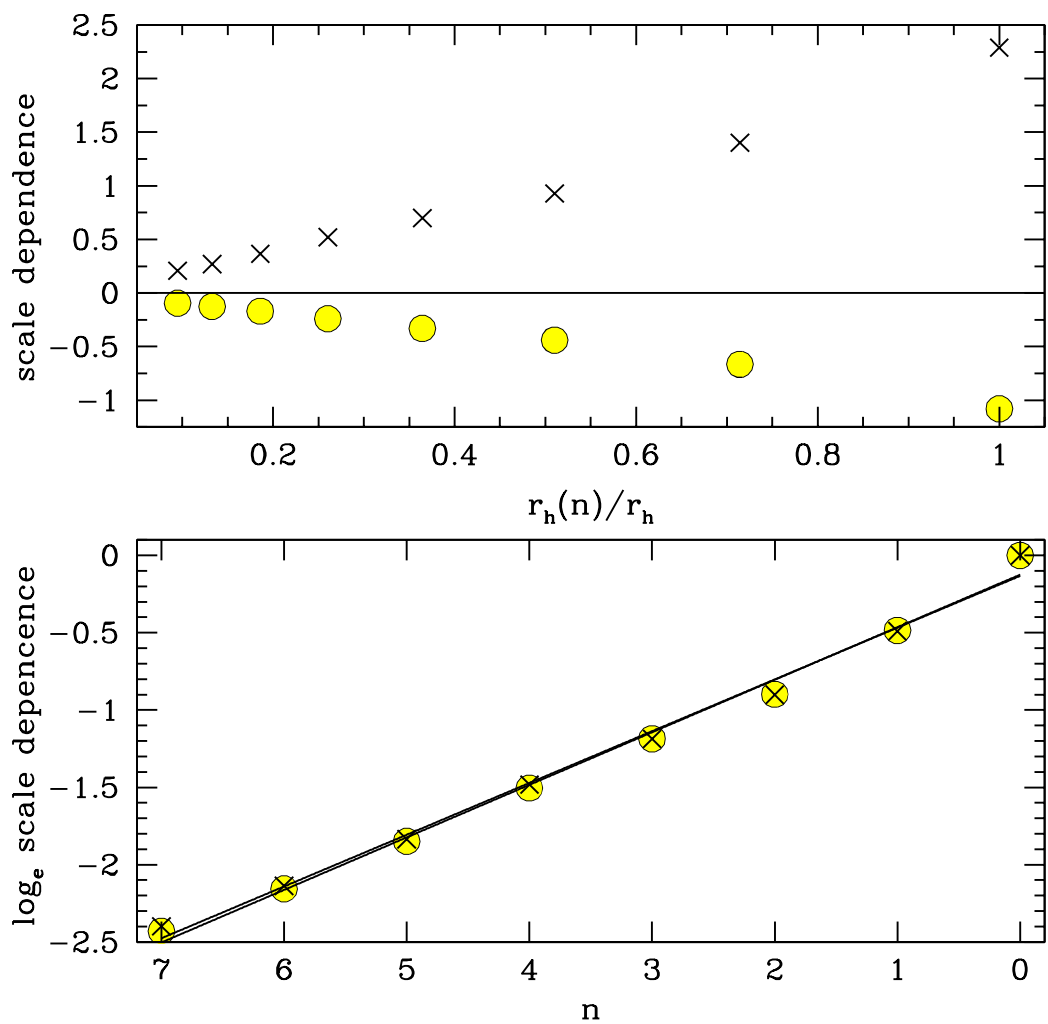}
\caption{\small \slshape At the top, dependence of $w^{R}_{0}$ (lower points denoted by circles)
and $w^{R}_{a}$  (upper points denoted by $\times$) with respect of the size of
the hole. At the bottom, fit as explained in the text. Recall that $r_{h}$ is
today $350$ Mpc.}
\label{dress}
\end{center}
\end{figure}

As one can see from Fig.\ \ref{dress},  we have verified that $w^{R}=0$ for
$r_{h} \rightarrow 0$, \textit{i.e.,} we recover the EdS model as the best-fit
phenomenological model.

We are interested to see if the equation of state exhibits a  power-law behavior
and, therefore, we use  the following functions to fit $w^{R}_{0}$ and
$w^{R}_{a}$:
\begin{eqnarray}
\frac{w^{R}_{0}(n)}{w^{R}_{0}(0)} & = & q_{0} \left ( \frac{r_{h}(n)}{r_{h}(0)} 
\right )^{p_{0}} \nonumber \\
\frac{w^{R}_{a}(n)}{w^{R}_{a}(0)} & = & q_{a} \left ( \frac{r_{h}(n)}{r_{h}(0)} 
\right )^{p_{a}} .
\end{eqnarray}
We performed a fit with respect to the logarithm of the above  quantities, the
result is shown in Fig.\ \ref{dress}. We found:
\begin{eqnarray}
p_{0} & = & p_{a} \simeq 1.00 \nonumber \\
q_{0} & = & q_{a} \simeq 0.88  .
\end{eqnarray}

\clearpage
Summarizing, we found three important facts.
\begin{itemize}

\item The parameters of the equation of state as a function of the 
size of the hole exhibit a power-law behavior.

\item The power-laws of $w^{R}_{0}$ and $w^{R}_{a}$ have the same scaling
exponent.  This is actually a check: once a physical quantity exhibits a
power-law  behavior, we expect that all its parameters share the same scaling
exponent.

\item The scale dependence is linear: the equation of state depends  linearly on
the length of holes the photon propagates through.  We stress that the
dependence we are talking about is not on the scale  of the universe, but on the
size of the holes.

\end{itemize}

We can finally ask which size of the holes will give us a phenomenological
model able to mimic the concordance model. We found that for $n=1$, that is for
a holes of radius $r_{h}=250$ Mpc, we have $w^R_0=-1.4$ and $w^R_a=0.67$,
which in terms of the energy density parametrization of Eq.\ (\ref{linda}),
corresponds to  $w_{0}=-1.03$  and $w_{a}=2.19$.

\cleardoublepage

\chapter*{Conclusions}
\addcontentsline{toc}{chapter}{Conclusions}
\chaptermark{Conclusions}

The aim of this thesis was to understand the role of large-scale non-linear 
cosmic inhomogeneities in the interpretation of {\it observational} data.
We are stressing the word observational because we think that it is the guiding idea
to properly set the back-reaction problem.

Back-reaction has indeed many faces and possible interpretations. We split up our efforts
into two different directions of research. The first is about building a realistic (even if still a toy)
and exact inhomogeneous universe model in order to study directly how inhomogeneities
affect observables like the luminosity-distance--redshift relation.
The other direction is about describing an inhomogeneous universe by means of an effective
homogeneous phenomenological model, that is, to {\it average} inhomogeneities.
We think that both approaches are worth studying.

About the latter, we explored two different paths, using two different models respectively.
The first averaging approach is based on Buchert's work.
We introduced it and discussed it in Chapter~\ref{brset} and we worked out a model to test it in
Chapter~\ref{pool}.
We think that this framework is very interesting and gives consistent approach to the averaging problem.
However, the back-reaction source it gives is not clearly related to the observed dark energy.
It features constant-time averages that do not match with an observational approach based on the light cone.
Anyway, this scenario gives precious hints toward the understanding of the back-reaction problem and allows us to check it quantitatively.\\
The model we built is an approximate ``homogeneous'' universe model in which the LTB free functions were chosen in order to not single out the center.
It was shown indeed that it is possible to fit the observed luminosity-distance--redshift relation by adjusting the LTB free functions. To achieve this result, however, it is necessary to place the observer at the center of a rather big underdensity. Even though we built nothing more than a toy model, we tried to avoid this fine-tuning.\\
We found a negligible back-reaction. We traced its cause back to the spherical symmetry of the model, which is the main limitation of the LTB metric. Indeed spherical symmetry introduces self-averaging in the equations, which erases any effect in the flat case where the back-reaction is exactly zero.
The curved case then does not depart significantly from the flat case.
We think that the lack of back-reaction effects comes from a inefficient use of LTB models and that both the Buchert framework and LTB models can give interesting insights.

In Chapter \ref{s-c}, therefore, we built another model, a Swiss-cheese model where the cheese consists of the usual Friedmann-Robertson-Walker (FRW) solution and the holes are constructed out of a LTB solution.
We focused on a Swiss-cheese model because, even if it is made of spherical symmetric holes, it is not a spherical symmetric model as a whole.
It is a first step to go beyond spherical symmetry which is the main limitation of LTB solutions.
In order not to fine-tune the position of the observer we placed the latter 
in the cheese and have the observer look through the Swiss-cheese holes.\\
We first focused on the effects of inhomogeneities on photons. The observables on which we focused are the change in redshift $\Delta z(\lambda)$, in angular-diameter distance $\Delta d_{A}(z)$,
in the luminosity distance-redshift relation $d_{L}(z)$, and in the distance modulus $\Delta m(z)$.
We found that redshift effects are suppressed when the hole is small because of
a compensation effect acting on the scale of a hole, due to spherical
symmetry, and on the scale of half a hole, due to the matching to the cheese metric.  The latter is somewhat similar to the screening among positive and negative charges.
However, we found interesting effects in the calculation of the angular
distance: the evolution of the inhomogeneities bends the photon path compared
to the FRW case. Therefore, inhomogeneities will be able (at least partly) to
mimic the effects of dark energy.
We stress that this non-trivial result depends on the presence of a faster-than-cheese expanding void which, we think, is the {\it crucial} ingredient in studying inhomogeneous cosmological models.\\
After having analyzed the model from an observational point of view,  we set up
the fitting problem -- our second approach to the averaging problem -- in order to better  understand how
inhomogeneities renormalize the matter Swiss-cheese model allowing us to eschew
a primary dark energy. We followed the scheme developed by Ellis and Stoeger,
but modified it in order to fit the phenomenological model to the Swiss-cheese
one. We chose a method that is intermediate between the fitting approach  and
the averaging one: we fitted with respect to light-cone averages.
In particular, we focused on the expansion and the density. While the 
expansion behaved as in the FRW case because of the compensation  effect
mentioned above, we found that the density behaved differently, thanks to its
insensitiveness to that compensation effect: a photon is spending more and more
time in the (big) voids than in the (thin) high density  structures. This
effect is not directly linked to the one giving us an  interesting $d_{A}$.\\
The insensitivity to the compensation effect made us think that a Swiss cheese made
of spherical-symmetric holes and a Swiss cheese without an exact spherical
symmetry  would share the same light-cone averaged density.  Knowing the density
behavior we will, therefore, be able to know the behavior of the Hubble
parameter, which, in turn, will be the one of the FRW solution with a phenomenological
source characterized by the fit equation of state. In this way we can think to
go beyond the main limitation of this model, that is, the assumption of
spherical symmetry. From this point of view, light-cone averaged density
can be seen as a  tool in performing this step.\\
Then we studied how the equation of state of a phenomenological model
with only one effective source depends on the size of inhomogeneity.
We found that $w^{R}_{0}$ and $w^{R}_{a}$ follow a power-law dependence with the same scaling 
exponent which is equal to unity. That is, the equation of state depends 
linearly on the distance the photon travels through voids.\\
We finally dealt with the size of the holes in order to mimic the concordance model.  We found out that the best fit is for
holes of radius $r_{h}=250$ Mpc and an equation of state with $w_{0}=-1.03$ and $w_{a}=2.19$.

\ 

We reached the conclusion that it is possible to set the back-reaction problem in a physically meaningful way and that it is by no means worth examining the physics of the back-reaction in depth.

\appendix
\chapter[Cosmological evolution of Alpha]{Cosmological evolution of Alpha driven by a general coupling with Quintessence}
\label{alphaevo}
We have here presented (see \cite{Marra:2005yt}) a general model for the cosmological evolution of the fine structure constant $\a$ driven by a typical Quintessence scenario.  
We have considered a coupling, between the Quintessence scalar $\phi$ and  the electromagnetic kinetic term $F_{\mu\nu}F^{\mu\nu}$, given by a general function $B_F(\phi)$. 
We have studied the dependence of the cosmological $\D\a(t)$ upon the functional form of $\BF$ and discussed the constraints imposed by the data.
We have found that different cosmological histories for $\D\a(t)$ are possible within the avaliable constraints. We have also found that Quasar absorption spectra evidence for a time variation of $\a$, if confirmed, is not incompatible with Oklo and meteorites limits.

We have referred to \cite{Marra:tesi} for a general analysis about the fundamental constants and their variation induced by a cosmological scalar.

\section{Introduction}      
Over the last few years there has been an increasing interest in the possibility of varying the fundamental constants over cosmological time-scales. This has a twofold motivation. 
On one side, several observations point towards the existence of a smooth dark energy component in the universe, which could be modeled via a dynamical scalar field called Quintessence (for recent reviews see \cite{Carroll:2000fy, Peebles:2002gy, Padmanabhan:2002ji}). In general, we expect such a cosmological scalar to couple with some, if not all, the terms in the matter-radiation Lagrangian, thus inducing a time variation of physical masses and couplings.
On the other side, recent improved measurements on possible variations of the fundamental constants are opening up the possibility of testing the theoretical models to a good degree of precision over a wide range of cosmological epochs. It should also be mentioned that, although controversial, some evidence of time variation of the fine structure constant $\a$ in Quasar absorption spectra was recently reported \cite{Webb}.
The cosmological variation of fundamental constants induced by couplings with the Quintessence scalar is then worth studying  in order to see if such a field could be responsible for a measurable effect.

Among all the possibilities, the time-variation of the fine-structure constant is the simplest to study both from the theoretical and experimental points of view. In this work we have restricted ourselves to this issue.
The theoretical study of a time-varying fine structure constant dates back to 1982 when Beckenstein \cite{Bekenstein:1982eu} first considered the possibility of introducing a linear coupling between a scalar field and the electromagnetic field. More recently the Beckenstein model has been revived, generalized and confronted with updated experimental limits 
\cite{Carroll:1998zi,Olive:2001vz,Damour:2002nv}.
The concrete case of the Quintessence scalar  has been considered too 
\cite{Copeland:2003cv,Lee:2004vm,Dvali:2001dd,Wetterich:2003jt}. However, as we have seen, most authors restrict their studies to the simplest case of a linear  or quadratic coupling. The possibility of reconstructing the dark energy equation of state from a measure of $\a$-variation has also been proposed in the literature \cite{Nunes:2003ff,Parkinson:2003kf}.

In this work we have discussed a general model for the variation of the fine structure constant $\a$ driven by a typical Quintessence scenario. After briefly reviewing the most recent observational and experimental constraints on the variation of $\a$, we have constructed the theoretical framework. 
In particular we have considered the case of a general coupling $\BF$ (see Eq.~(\ref{BF}) below) which includes several classes of possible functions. In this way we have been able to study the dependence of the cosmological variation of alpha, $\D\a(t)$, upon the functional form of $\BF$ and discuss the constraints imposed by present data.
We have found that different cosmological histories for $\D\a(t)$ are possible within the avaliable constraints. We have also found that, Webb et al.~data \cite{Webb}, if confirmed, are not incompatible with Oklo and meteorites constraints \cite{Olive:2002tz,Damour:1996zw}.

\section{Overview of the constraints}
\begin{figure}[!htb]
\begin{flushright}
\includegraphics[width= 15cm]{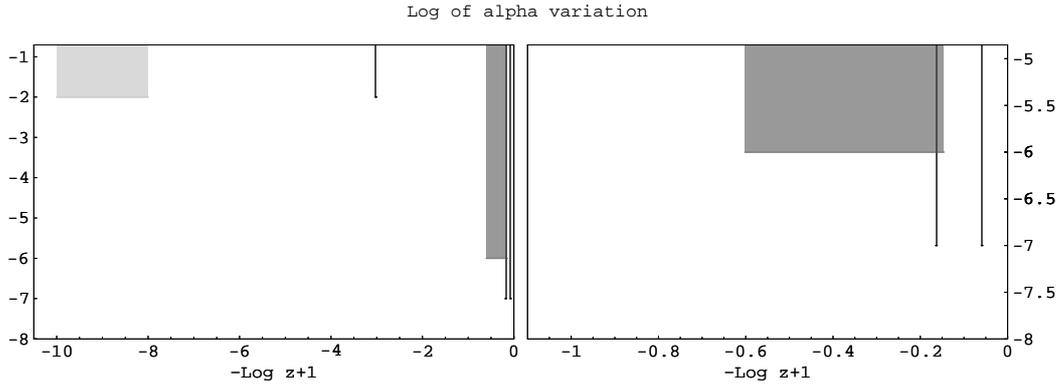}
\caption{\small \slshape The experimental constraints (1)-(6) discussed above are summarized in the picture: $\log | \Delta \alpha / \alpha |$ is plotted as a function of the redshift $z$. 
On the right-hand side we zoom on $z\lta 10$. The grey areas are those  excluded by present data.}
\label{limiti}
\end{flushright}
\end{figure}

Comprehensive reviews about the theoretical and experimental issues connected to the time variation of fundamental constants can be found in Refs.~\cite{Uzan:2002vq}, \cite{Olive:2002tz} and \cite{Martins:2004ni}. In the following we are summarizing the avaliable constraints on  the time variation of $\alpha$, expressed as functions of the redshift $z$ (see also Fig.~\ref{limiti}):
\begin{equation}
{\Delta \alpha(z) \over \alpha}\equiv {\alpha(z) - \alpha_{0}\over \alpha_{0}}
\end{equation}
where $\alpha_{0}=\alpha(0)$ is the value measured today. 

(1) The most ancient data come from Big Bang Nuclesynthesis (BBN) and give \cite{bbn,bbncmb}:
\begin{equation}
\left | {\Delta \alpha \over \alpha}\right | \lta 10^{-2}
\qquad \qquad z=10^{10}-10^{8}
\; \;  .
\end{equation}

(2) More recently we have the limit coming from the power spectrum of anisotropies in the Cosmic Microwave Background (CMB)
 \cite{bbncmb}:
\begin{equation}
\left | {\Delta \alpha \over \alpha}\right | < 10^{-2}
\qquad \qquad z=10^{3}
\;\; .
\end{equation}

(3) From absorption spectra of distant Quasars there are  more controversial data.
Webb and Murphy's groups combined data \cite{Webb} report  a $4 \, \sigma$ evidence for $\a$ variation: $\Delta \alpha / \alpha=(-0.543 \pm 0.116) \cdot 10^{-5}$ on a cosmological time span between $z=0.2$ and $z=3.7$. This result has not been confirmed by other groups. For example, Chand et al.~\cite{Chand:2004et,Levshakov:2004bg} give: ${\Delta \alpha / \alpha} =(-0.06 \pm 0.06) \cdot 10^{-5}$ for $z=2.3-0.4$ and
${\Delta \alpha / \alpha} =(0.15 \pm 0.43) \cdot 10^{-5}$ for $z=2.92-1.59$. 
We have chosen to be conservative, considering a limit based on the last two results, which is consistent with zero variation:
\begin{equation} \label{quasa}
\left |{\Delta \alpha \over \alpha} \right | \lta 10^{-6}
\qquad \qquad z=3-0.4
\;\;  .
\end{equation}

(4) From the analysis of the ratio Re/Os in meteorites dating around 4.56 billion years ago it is possible to compute $^{187}$Re half-life, which gives \cite{Olive:2002tz}:
\begin{equation}
\left | {\Delta \alpha \over \alpha}\right | \lta 10^{-7}
\qquad \qquad z=0.45
\;\;  .
\end{equation}

(5) From the Oklo natural nuclear reactor that operated 2 billion years ago in Gabon, we also have \cite{Olive:2002tz,Damour:1996zw}:
\begin{equation}
\left | {\Delta \alpha \over \alpha}\right | \lta 10^{-7}
\qquad \qquad z=0.14
\;\; .
\end{equation}

(6) We then have limits coming from laboratory measurments which constrain the present rate of change of $\a$.
Comparing atomic clocks, which use different transitions and atoms, what is obtained is \cite{Marion:2002iw}:
\begin{equation} \label{labo}
\left | {\dot{\alpha} \over \alpha}\right | \lta 10^{-15} \mbox{ yr}^{-1}
 \qquad \;\; z=0
\end{equation}
where the dot represents differentiation w.r.t. cosmic time.

%
%

(7) In addition to the limits discussed above, there is a constraint coming from indirect violation of the Weak Equivalence Principle (WEP), parametrized by the  E\"otv\"os ratio
\begin{equation}
\eta=2{|a_{1}-a_{2}| \over |a_{1}+a_{2}|}
\end{equation}
where $a_{1}$ e $a_{2}$ are the accelerations of two different test bodies in the Earth gravitational field. 
These constraints come from the fact that nucleon masses get electromagnetic corrections from quark-quark interactions. As extensively discussed in \cite{Gasser:1982ap}, the leading term of the electromagnetic contribution comes from the electrostatic energy of the quark distribution, which is proportional to $\alpha$.
The corrected masses, to leading order in $\alpha$,  can then be written as \cite{Dvali:2001dd,Gasser:1982ap}:
\begin{eqnarray} \label{gama}
m_{p}&=&m+\alpha \: B_{p}
\nonumber\\
m_{n}&=&m+\alpha \: B_{n}
\end{eqnarray}
where $p,n$ stand for proton and neutron and $B_p \equiv 0.63 \mbox{MeV}/\a_0$,  $B_n \equiv -0.13 \mbox{MeV}/\a_0$.
If we suppose that $\a=\a(\phi)$, then we will induce a $\phi$-dependence on the nucleon masses:
\begin{equation} \label{me}
\delta m_{n}=B_n   \delta \alpha    \qquad  ;  \qquad  \delta m_{p}=B_p  \delta \alpha \;\;  .
\end{equation}
If we define
\beq
g_{i} =
{\partial m_{i} \over \partial \phi} =
{\partial \alpha \over \partial \phi} B_i
\label{gi}
\eeq
we get an indirect violation of the equivalence principle induced by the `fifth-force' mediated by the scalar field
\begin{equation} \label{eta}
\eta \simeq {M_{Pl}^{2} \over 4 \pi \bar{m}^{2}} \left ( R_{n}^{E} g_{n}+R_{p}^{E} g_{p} \right ) \left ( \Delta R_{n} g_{n}+\Delta R_{p} g_{p} \right )
\end{equation}
where:
\begin{equation}
R_{i}^{E}\equiv {n^{E}_{i} \over n^{E}_{n}+n^{E}_{p}}\simeq 0.5
\qquad
\Delta R_{i}\equiv {|n_{i,\: 1}-n_{i,\: 2}| \over n_{n}+n_{p}}\simeq 0.06-0.1 \,\,\, ,
\end{equation} 
and $\bar{m} \simeq 931~ \mbox{MeV}$ is the atomic mass unit. The suffix $E$ refers to the Earth, while $1$ and $2$ refer to two test bodies having equal mass but different composition.
From Eqs.~(\ref{gi})-(\ref{eta}) we see that any model of $\a$-variation will induce a characteristic $g_{p,n} \not = 0$ and hence WEP violation:  while the first two factors in Eq.~(\ref{eta}) are universal and depend on the Earth composition, the third term is not zero if and only if $g_{p,n}\not=0$ and the test bodies have different composition in neutrons and protons. The current limits on WEP violations impose \cite{Baessler:1999iv}:

\begin{equation} \label{eot}
\eta < 10^{-13}  \,\,\, .
\end{equation}

\section{The theoretical framework}
Following Olive et al.~\cite{Olive:2001vz}, the most generic action involving a scalar field, the Standard Model fields and an hypothetical Dark Matter particle $\chi$, can be written as
\beqra  \label{action}
S & = & \frac{1}{16 \pi G} \int d^4x \sqrt{-g}~R + 
\int d^4x \sqrt{-g} \left[   \frac{1}{2} \partial^{\mu}\phi\partial_{\mu}\phi - V(\phi) \right]
\nonumber \\
& - & {1 \over 4} \int d^4x \sqrt{-g} \ B_{F}(\phi)F_{\mu \nu}F^{\mu \nu}  
-  {1 \over 4} \int d^4x \sqrt{-g} \  B_{F_{i}}(\phi)F^{(i)}_{\mu \nu}F^{(i)\mu \nu} 
\nonumber  \\
& + & \int d^4x \sqrt{-g} \  \sum_{j} \left[ \bar{\psi}_{j} D\!\!\!\!/ \psi_{j}+i B_{j}(\phi)m_{j}\bar{\psi}_{j}\psi_{j} \right] 
\nonumber  \\
& + &  \int d^4x \sqrt{-g} \  \left[ \bar{\chi} \partial\!\!\!/ \chi - B_\chi (\phi) m_{\chi} \chi^T \chi  \right]
\eeqra
where $D\!\!\!\!/=\gamma_{\mu}D^{\mu}$ and for the electromagnetic term, for example, $D^{\mu}=\partial_{\mu}-i e_{0} A_{\mu}$. The index $i=1,2,3$ refers to the $SU(3)$ gauge group of the Standard Model and  $j$ runs over the various matter fields.

The form of the action (\ref{action}) follows from supplying $\phi$-dependent factors to all mass and kinetic terms to the standard Lagrangian (which would have all $B_i=1$).
In general we would expect that all of the $B_i(\phi)$'s are switched on, if not forbidden by any symmetry principle. However, the theoretical treatment of the full Lagrangian is very cumbersome and so the coupling functions $B_i(\phi)$ are usually switched on one at a time. In this way one can also disentangle the effects due to each single term. 
Since our focus is on the fine-structure constant $\a$, we will keep only $B_F(\phi)\not = 1$ and set all the other functions equal to 1. 

The relevant part of the action for the effect we have studied is then
\beqra  \label{action2}
S  = \frac{1}{16 \pi G} \int d^4x \sqrt{-g}~R + 
\int d^4x \sqrt{-g} \left[   \frac{1}{2} \partial^{\mu}\phi\partial_{\mu}\phi - V(\phi) \right]
\nonumber \\
 -  {1 \over 4} \int d^4x \sqrt{-g} \ B_{F}(\phi)F_{\mu \nu}F^{\mu \nu}  
\eeqra
which allows to define an ``effective'' fine structure constant
\begin{equation} \label{alpha}
\alpha(t)={\alpha_{0}\over B_F(\phi(t))}
\end{equation}
where $\alpha_{0}$ is the value measured today. From (\ref{alpha}) we obtain the relative variation relevant for each cosmological epoch
\begin{equation} \label{Dalpha}
{\Delta \alpha \over \alpha}\equiv {\alpha(t) - \alpha_{0}\over \alpha_{0}}={1-B_F(\phi(t))\over B_F(\phi(t))}
\end{equation}
It can immediately be seen that, depending on the cosmological evolution of $\phi(t)$ and on the functional form of $B_F(\phi)$, the fine structure constant $\a$ could in principle have had many possible histories during the life-time of the universe. 
What possibilities are allowed by a general coupling $\BF$ within the avaliable observational constraints is then worth studying. 

%
%
The relevant equations governing the cosmological evolution in a flat universe are the following
\beqra
\frac{\ddot{a}}{a} = - \frac{4 \pi}{3 M_p^2} ~\sum_i (1+3 w_i)\rho_i
\label{ddota} \\
H^2 \equiv \left( \frac{\dot{a}}{a} \right)^2 = \frac{8\pi}{3M_p^2}~ \sum \rho_i 
\label{H2}\\
\ddot{\phi} + 3H\dot{\phi} + \frac{dV}{d\phi} = 0
\label{phi-eq}
\eeqra
where $i=m,~r,~\phi$ runs over the matter (including dark matter), radiation and scalar components. The relevant equations of state are $w_m=0$ for matter, $w_r=1/3$ for radiation and $w_\phi$ as defined in Eq.~(\ref{wphi}).
It is important to note that the evolution equation of the Quintessence scalar (\ref{phi-eq}) does not depend on $B_F$ or its derivatives. This is due to the fact that the statistical average of the term $F^{\mu\nu}F_{\mu\nu}$ over a current state of the universe is zero. So the only term that drives $\phi$ during the cosmological evolution is the potential $V(\phi)$.

Since we are working under the hypothesis that the scalar field $\phi$ in Eq.~(\ref{phi-eq}) is the Quintessence scalar, we should also impose the additional constraints coming from Quintessence phenomenology.
In particular we choose a runaway potential which goes to zero as far as the field $\phi$ rolls to infinity, in accordance with the observational data. It is also required that the scalar dynamics gives the correct value for the equation of state 
\begin{equation}
\label{wphi}
w_\phi = \frac{\dot{\phi}^2/2 - V(\phi)}{\dot{\phi}^2/2 + V(\phi)} \;\;\; \left( \lta -0.7 \; \mbox{today} \right)
\end{equation}
The most general form for the Quintessence potential involves a combination of a power-law and exponential terms \cite{Ng:2001hs}. For the purpose of this work, however, we will consider the simplest case of an inverse power-law potential $V(\phi)=M^{4+n}\phi^{-n}$, which gives a late-time attractor equation of state $w_\phi=-2/(n+2)$ during matter domination \cite{Steinhardt:1999nw}. The potential should also be normalized  in  order to give the correct energy density today
($\rho_\phi^0 \simeq V(\phi) \simeq 2/3~ \rho_c^0$): this sets the mass scale $M$. In what follows we have chosen $n=1$ in the potential in order to have the correct attractor equation of state, and so obtain $M\simeq\sqrt[5]{2/3~\rho_c^0 ~M_p}$.

We have checked that choosing different Quintessence potentials gives a subdominant effect on the cosmological variation of $\a$, with respect to changing the coupling function $\BF$.
In what follows we will then fix $V(\phi)=M^5/\phi$ and study the effect of different $\BF$'s.
An interesting study, which is complementary to what is done here, is that of Ref.~\cite{Copeland:2003cv} where the effect induced by different Quintessence models on the cosmological $\D\a$ is examined in detail, while keeping the function $\BF$ fixed.

%
%

In order to be as general as possible we will consider a function $\BF$ which is a combination of different possible behaviors and characterized by a set of four parameters that are allowed to vary freely:
\beq
\label{BF}
B_F(\phi) = \left(\frac{\phi}{\phi_{0}}\right)^{\epsilon}  
\left[1-\zeta {(\phi-\phi_{0})}^{q} \right] \ e^{\tau(\phi-\phi_0)}   \,\,\,\, .
\eeq
This choice is not motivated by a specific theoretical model, but it is rather a working tool for  obtaining different functional forms of $\BF$ and thus cosmological histories of $\a$, according to Eq.~(\ref{Dalpha}).
We have chosen a combination of functions (power-law, polynomial, exponential) that can be switched on and off at will (depending on the values of the parameters $\epsilon$, $\zeta$, $\tau$ and $q$), thus giving rise to a variety of possibile $\BF$'s. In this way we can carry on a unified discussion of a number of different models of $\a$ variation.

%
\section{Cosmic evolution of $\alpha$}
\begin{figure}[htbp]
\begin{flushright}
\includegraphics[width= 15.7cm]{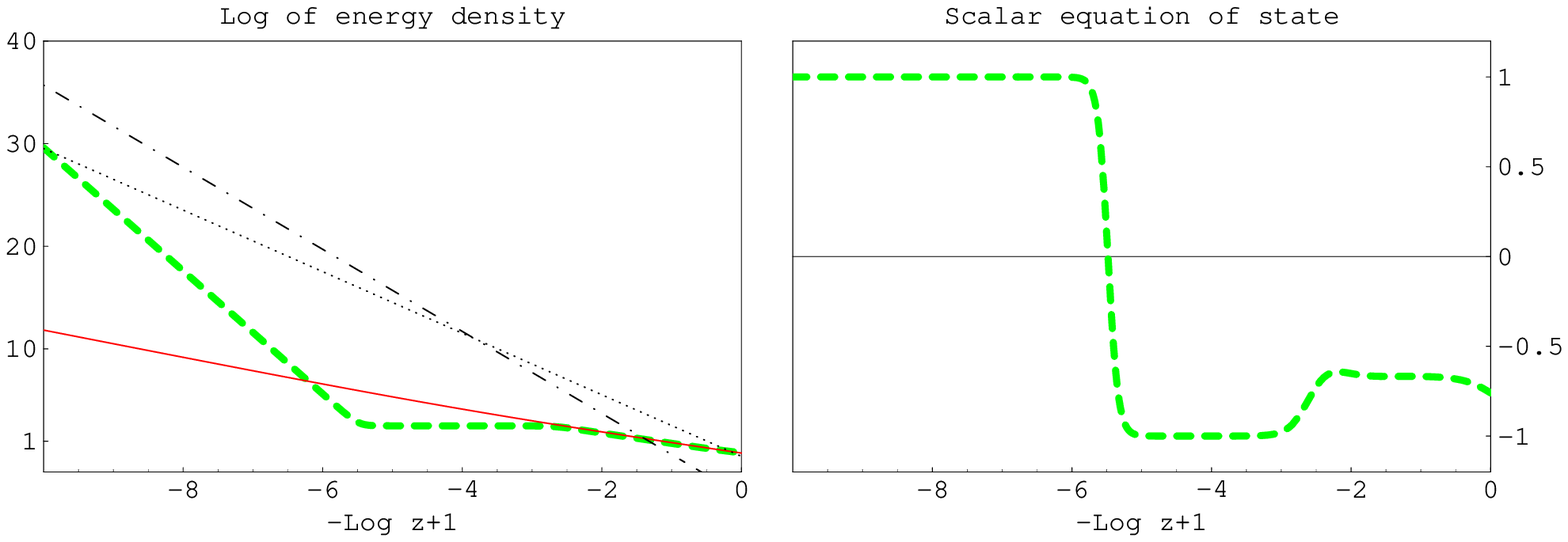}
\caption{\small \slshape Evolution of the energy densities (left) and scalar equation of state (right) for a  quintessence model with potential $V=1/\phi$ and  initial conditions $\rho_{\phi}^{in}/\rho_{c}^{0}=10^{30}$ at $z=10^{10}$.
The dot-dashed line represents the energy density of radiation, the dotted line the energy density of matter, the green dashed line the energy density of quintessence and the red solid line the attractor. All of the energy densities are expressed in units of the present critical energy density  $\rho_{c}^{0}$.}
\label{quinti}
\end{flushright}
\end{figure}

We have numerically solved the cosmological equations (\ref{ddota})-(\ref{phi-eq}) and then plotted the resulting cosmological history of $\D\a$ for various classes of functional forms of $\BF$, according to Eq.~(\ref{Dalpha}).
As already mentioned, for illustrative purposes we have chosen a scalar potential $V=1/\phi$ and initial conditions $\rho_{\phi}^{in}/\rho_{c}^{0}=10^{30}$ at $z=10^{10}$.
Fig.~\ref{quinti} shows the corresponding evolution of the energy densities and of the scalar equation of state parameter.

\subsection*{Linear coupling}

The simplest case is given by the choice $\epsilon=\tau=0$ and $q=1$ for the parameters in Eq.~(\ref{BF}):
\begin{equation} \label{beki}
B_{F}(\phi) = 1-\zeta  (\phi-\phi_{0})
\end{equation}
This case corresponds to the original Beckenstein proposal \cite{Bekenstein:1982eu}, which however did not supply a potential to the scalar field\footnote{To be precise,  Beckenstein actually invoked an exponential coupling, which however is practically equivalent to  eq.(\ref{beki}) due to the smallness of $\zeta\phi$.}. 
Copeland et al.~\cite{Copeland:2003cv} give a comprehensive discussion on various Quintessence models linearly coupled to the electromagnetic field, but assuming  Webb et al.~data \cite{Webb} to be correct and imposing on $\D\a(t)$ to agree with that measure.

\begin{figure}[!htbp]
\begin{flushright}
\includegraphics[width= 15cm]{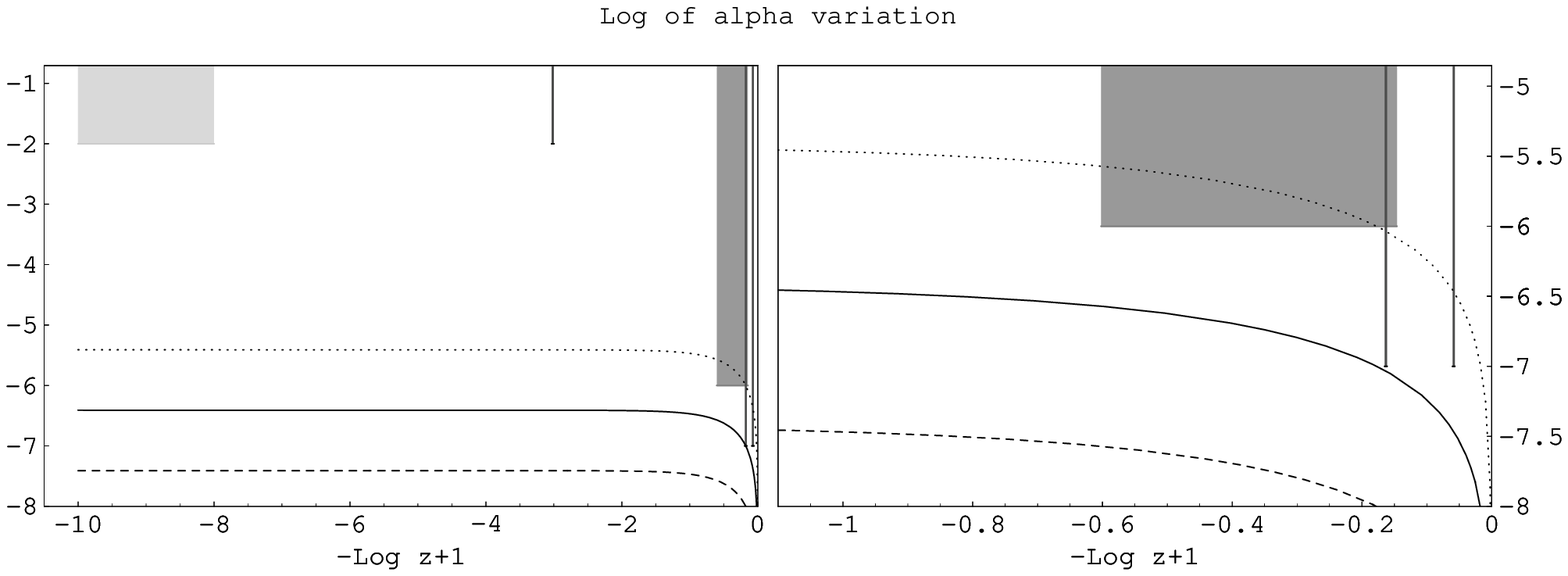}
\caption{\small \slshape The logarithm of $| \Delta \alpha / \alpha |$ is plotted as a function of Log$(z+1)$ for $B_{F}(\phi) = 1-\zeta  (\phi-\phi_{0})$ with $\zeta=0.6 \cdot 10^{-5}$ (dotted line), $\zeta=0.6 \cdot 10^{-6}$ (solid line) and $\zeta=0.6 \cdot 10^{-7}$ (dashed line).  
On the right-hand side we zoom on $z\lta 10$. 
Only the curves not overlapping the grey areas are phenomenologically viable.}
\label{linear}
\end{flushright}
\end{figure}

In our case, the resulting $\Delta \alpha$, as defined in Eq.~(\ref{Dalpha}), is plotted in Fig.~\ref{linear}. We tried a number of different values for $\zeta$, in order to verify in which cases all the available experimental constraints were simultaneously satisfied. We found that they are all respected for $\zeta \leq0.6 \cdot 10^{-6}$.
With this choice, the constraints on the violation of equivalence principle and the constraints derived from atomic clocks are automatically satisfied:
\begin{equation}
\eta \simeq 4 \cdot 10^{-21} \ll 10^{-13}
\qquad \qquad
\left | {\dot{\alpha} \over \alpha_{0}}\right |
=4 \cdot 10^{-17}
\ll 10^{-15} \mbox{ yr}^{-1}
\end{equation}
%

%
%

\subsection*{Polynomial coupling}

A slightly more complicated case is given by the choice  $\epsilon=\tau=0$, allowing the exponent $q$ to be $>1$:
\begin{equation} \label{beki2}
B_{F}(\phi)= 1-\zeta  (\phi-\phi_{0})^{q}
\end{equation}
The case  of a quadratic coupling ($q=2$) was considered in Ref.~\cite{Lee:2004vm}, but with the additional assumption of a proportionality relation between $\BF$ and $V(\phi)$.

\begin{figure}[!thbp]
\begin{flushright}
\includegraphics[width= 15cm]{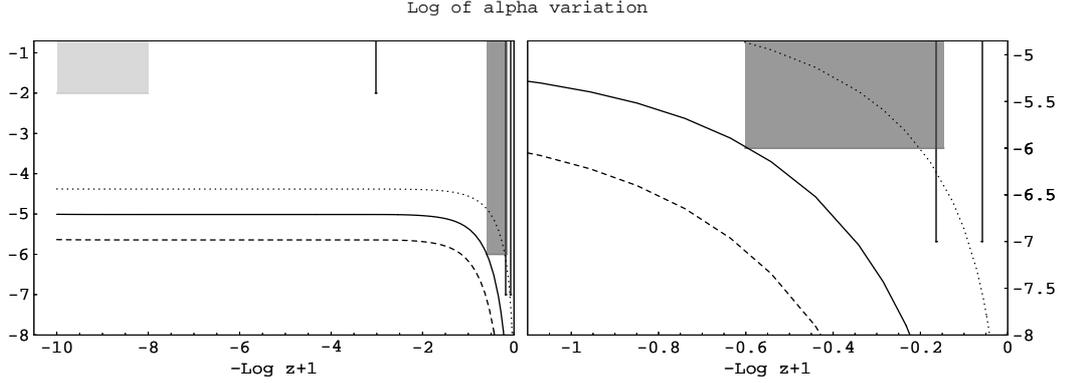}
\caption{\small \slshape The logarithm of $| \Delta \alpha / \alpha |$ is plotted as a function of Log$(z+1)$ for $B_{F}(\phi) = 1-\zeta  (\phi-\phi_{0})^{q}$ with $\zeta = 10^{-4}$ and $q=3$ (dotted line), $q=6$ (solid line) and $q=9$ (dashed line). 
On the right-hand side we zoom on $z\lta 10$.  
Only the curves not overlapping the grey areas are phenomenologically viable.}
\label{polinomial}
\end{flushright}
\end{figure}

We have found  that the data do not impose any upper limit on the exponent $q$ and that increasing $q$ makes it possible to reduce the fine-tuning in $\zeta$.
For example, choosing $\zeta = 10^{-4}$ the experimental limits are respected for $q=6$ and  the constraints on the violation of equivalence principle and the constraints derived from atomic clocks satisfied by many orders of magnitude.
In Fig.~\ref{polinomial} we plot Log~$|\Delta \alpha / \alpha |$ for $\zeta = 10^{-4}$ with $q=3, \, 6 \mbox{ and } 9$, as function of red-shift.
%
%

\begin{figure}[!thbp]
\begin{flushright}
\includegraphics[width= 15cm]{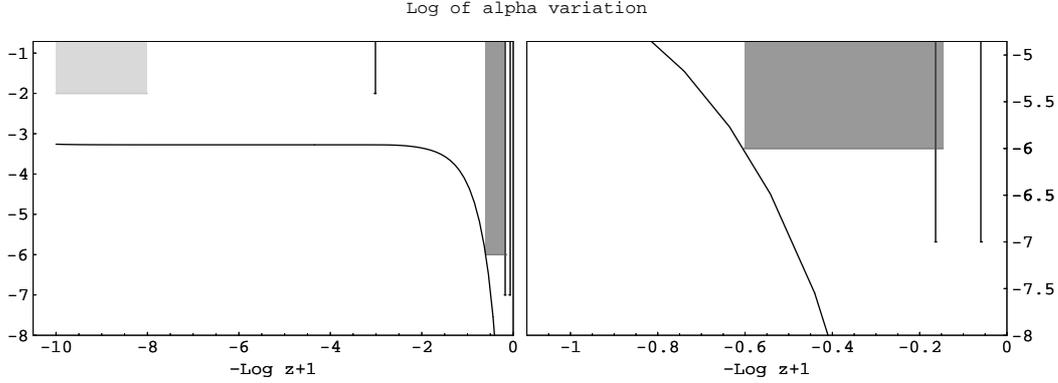}
\caption{\small \slshape The logarithm of $| \Delta \alpha / \alpha |$ is plotted as a function of Log$(z+1)$ for $B_{F}(\phi) = 1-\zeta  (\phi-\phi_{0})^{q}$ with $\zeta = 1$ and $q=17$  (solid line). 
On the right-hand side we zoom on $z\lta 10$.  Note that all the experimental limits  are satisfied without any fine--tuning in the parameters of the function $\BF$.}
\label{polyb}
\end{flushright}
\end{figure}

As already mentioned, by increasing the exponent $q$ we can do even better.
For example, with $q=17$ the experimental constraints are satisfied even for $\zeta=1$, as illustrated in Fig.~\ref{polyb}. It should be emphasized that among all the possibilities considered in this work, this choice of the parameters appears to be the most natural of all. A notable feature is also the fact that the value of $\D\a$ is enhanced in the past, with respect to the other cases, becoming closer to the observational limits, while falling off very steeply in recent times.
%
%

\subsection*{Power--law coupling}
\begin{figure}[htbp]
\begin{flushright}
\includegraphics[width= 15cm]{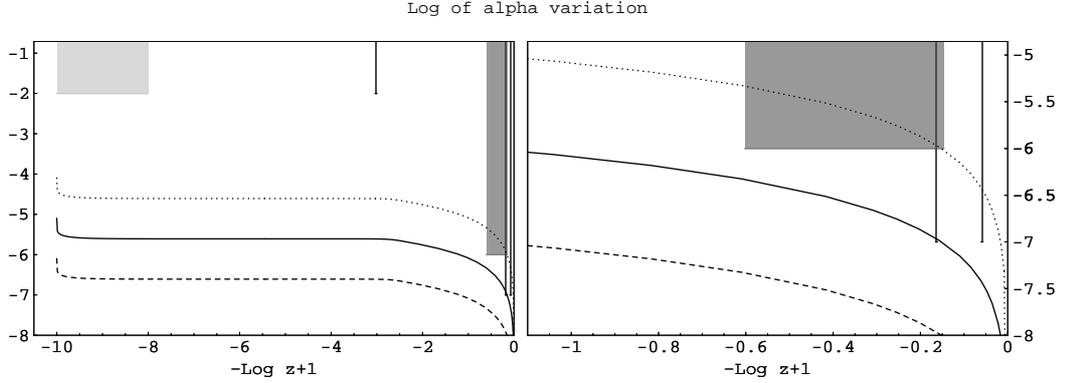}
\caption{\small \slshape The logarithm of $| \Delta \alpha / \alpha |$ is plotted as a function of Log$(z+1)$  for $B_{F}(\phi)=\left({\phi \over \phi_{0}}\right)^{\epsilon}$ with $\epsilon=4 \cdot 10^{-6}$ (dotted line), $\epsilon=4 \cdot 10^{-7}$ (solid line) and $\epsilon=4 \cdot 10^{-8}$ (dashed line). 
On the right-hand side we zoom on $z\lta 10$. 
Only the curves not overlapping the grey areas are phenomenologically viable.}
\label{power}
\end{flushright}
\end{figure}

With the choice $\zeta=\tau=0$ we obtain the following coupling function:
\begin{equation} \label{beki5}
B_{F}(\phi)=\left({\phi \over \phi_{0}}\right)^{\epsilon}  \,\, .
\end{equation}
In this case, it is necessary to fine-tune the exponent $\epsilon$ in order to satisfy the data, due to the smallness of $\phi$ in the early universe. Keeping $\epsilon$ of order one would violate even the constraints from BBN.
We found that the experimental limits are respected for  $|\epsilon| \leq 4 \cdot 10^{-7}$.
In Fig.~\ref{power} we plot Log~$|\Delta \alpha / \alpha |$  as a function of red-shift for different choices of $\epsilon$. Note that the sign of $\Delta \alpha$ depends on the sign of $\epsilon$.
With the choice $\epsilon=4 \cdot 10^{-7}$, the constraints on the violation of equivalence principle and the constraints derived from atomic clocks are automatically satisfied:
\begin{equation}
\eta \simeq 4 \cdot 10^{-21} \ll 10^{-13}
\qquad \qquad
\left | {\dot{\alpha} \over \alpha_{0}}\right |
=4 \cdot 10^{-17}
\ll 10^{-15} \mbox{ yr}^{-1}
\end{equation}
Such a small exponent might look quite unnatural, however Eq.~(\ref{beki5}) for $\epsilon \ll 1$ is equivalent to:
\begin{equation} \label{bk5}
B_{F}(\phi)=1+\epsilon \ln \left({\phi \over \phi_{0}}\right)    \,\, .
\end{equation}
In this way the fine tuning is moved from the exponent to the coefficient.
%
%

\subsection*{Exponential coupling}

The choice $\epsilon=\zeta=0$  of the parameters in (\ref{BF}) gives:
\begin{equation}
B_{F}(\phi) = e^{ -\tau (\phi-\phi_{0})}   \,\, .
\end{equation}
In this case, if $ \tau \gta 1$ it is not possible to satisfy all the constraints at the same time.  Depending on the sign of $\tau$, the resulting $| \Delta \alpha / \alpha |$ becomes too large in the early or late universe.
For $\tau \ll 1$, instead, the coupling function becomes equivalent to the linear case: 
$\BF = e^{ -\tau (\phi-\phi_{0})} \simeq 1 -\tau (\phi-\phi_{0})$.

\subsection*{Linear and power--law coupling combined}
\begin{figure}[!htbp]
\begin{flushright}
\includegraphics[width= 15cm]{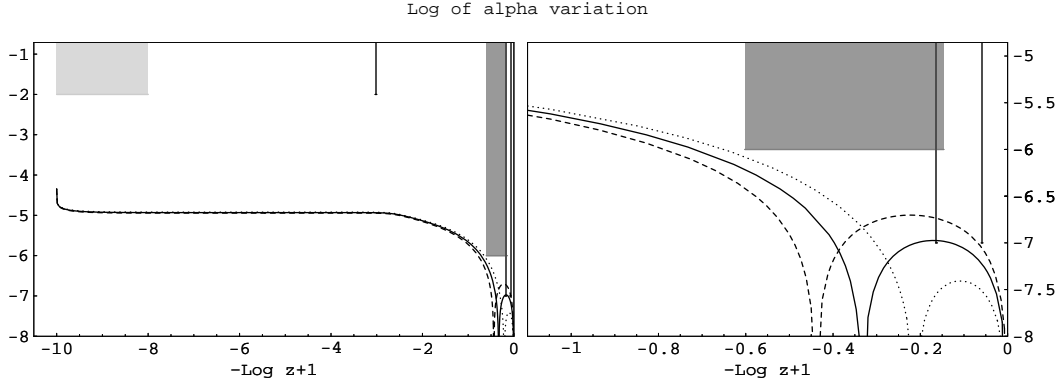}
\caption{\small \slshape The logarithm of $| \Delta \alpha / \alpha |$ is plotted as a function of Log$(z+1)$ for $B_{F}(\phi)= \left({\phi \over \phi_{0}}\right)^{\epsilon}(1-\gamma \, \epsilon \, (\phi-\phi_{0}))$ with $\epsilon = 2.4 \cdot 10^{-6}$,
$\gamma=2.2$ (solid line) and $\gamma=2.2 \pm 10\%$ (dashed and dotted respectively). 
On the right-hand side we zoom on $z\lta 10$. 
Only the curves not overlapping the grey areas are phenomenologically viable.}
\label{linear-power}
\end{flushright}
\end{figure}

Now let's consider two factors in (\ref{BF}) with $q=1$ and $\tau=0$. 
\begin{equation}
B_{F}(\phi)= \left({\phi \over \phi_{0}}\right)^{\epsilon} (1-\zeta \, (\phi-\phi_{0}))   \,\, .
\end{equation}
For an arbitrary choice of $\zeta$ and $\epsilon$, the resulting $\D \a$ is similar to the linear coupling or power--law coupling case, depending on which factor dominates.
It is instead interesting to consider the case $\zeta=\gamma \epsilon$ in which the two factors are of the same order of magnitude.
If $\gamma>0$, the two factors can contribute in an opposite way and it is easy to obtain $\Delta \alpha \simeq 0$  also at some time in the past. For example, with the choice $\epsilon=2.4 \cdot 10^{-6}$ and $\gamma=2.2$ we obtained the behavior plotted in Fig.~\ref{linear-power}, in which we have varied $\gamma$ of 10\%.
For the choice  $\epsilon = 2.4 \cdot 10^{-6}$, $\gamma=2.2$  the constraints on the violation of equivalence principle and the constraints derived from atomic clocks are automatically satisfied:
\begin{equation}
\eta \simeq 2 \cdot 10^{-20} \ll 10^{-13}
\qquad \qquad
\left | {\dot{\alpha} \over \alpha_{0}}\right |
=9 \cdot 10^{-17}
\ll 10^{-15} \mbox{ yr}^{-1}
\end{equation}
%
%
%

%
\subsection*{Power--law and exponential coupling combined}
Since, as already discussed, the exponential  coupling function case is equivalent to the linear one, this possibility falls within the previous example.

\subsection*{Polynomial and exponential combined}
\begin{figure}[htbp]
\begin{flushright}
\includegraphics[width= 15cm]{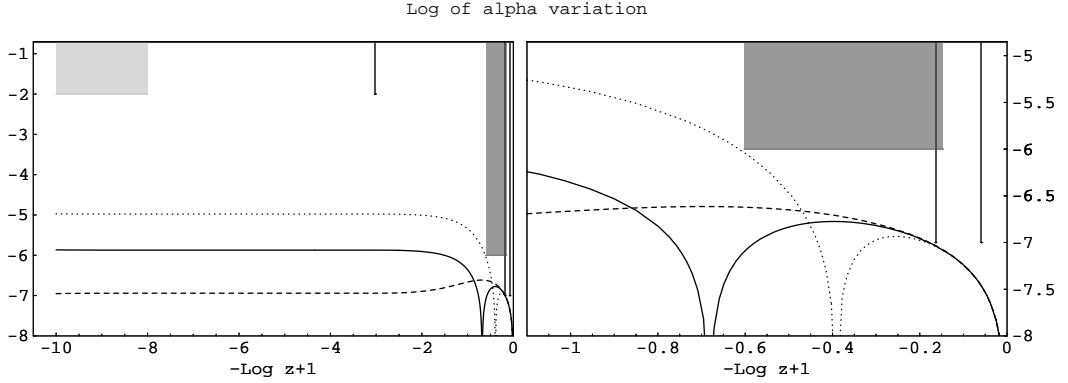}
\caption{\small \slshape The logarithm of $| \Delta \alpha / \alpha |$ is plotted as a function of Log$(z+1)$ for $B_{F}(\phi)=  (1-\zeta \, (\phi-\phi_{0})^6)\; e^{-\tau  (\phi-\phi_{0})}$ with $\tau=0.6 \cdot 10^{-6}$ and $\zeta=2 \cdot 10^{-4}$ (dotted line), $\zeta=3.2 \cdot 10^{-5}$ (solid line) and $\zeta=5 \cdot 10^{-6}$ (dashed line). 
On the right-hand side we zoom on $z\lta 10$. 
Only the curves not overlapping the grey areas are phenomenologically viable.}
\label{polinomial-exp}
\end{flushright}
\end{figure}

Now let's consider $\epsilon=0$ and $q\not = 1$. The case $q=1$ is not interesting since the two factors would be almost equivalent and the behavior  corresponding to the linear case.   Let's choose then, for example, $q=6$:
\begin{equation}
B_{F}(\phi)=  (1-\zeta \, (\phi-\phi_{0})^6)\; e^{-\tau  (\phi-\phi_{0})}
\end{equation}
For recent times ($z<1$) the exponential coupling $e^{-\tau  (\phi-\phi_{0})} \simeq 1- \tau  (\phi-\phi_{0})$ dominates, while in the past the two terms can be of the same order and, due to $q$ being even,  cancel at some time. This is shown in  Fig.~\ref{polinomial-exp}.
The constraints on the violation of equivalence principle and the constraints derived from atomic clocks are satisfied.

\section{Summary and Conclusions}
%
In this work we have carried out a comprehensive study of the cosmological variation of the fine-structure constant $\a$ induced by the coupling of the electromagnetic field with a typical Quintessence scalar.
We have considered a variety of functional forms for the coupling function $\BF$, obtainable from a general expression (see Eq.~(\ref{BF})) depending on four parameters.

We have found that very different cosmological histories for $\D \a$ are possible, depending on which parameters are switched on. For example, we can produce a $\D \a$ which is well below the observational constraints in the early universe and just within the experimental limits in recent times (linear coupling case). But also the converse is possible, if we choose a polynomial coupling.
In particular, the behavior at small redshift can be qualitatively very different depending on the model we choose: sharply decreasing in the polynomial coupling case or mildly decreasing with the power--law coupling.

By combining different functional forms, a notable feature emerged. In some cases it is possible that the scalar dynamics drives $\D\a$ to a zero at some time in the past, thus inverting the slope of its cosmological evolution. This happens for all the combined cases discussed here.

It is also worth remarking that in our parameter space span we have found solutions with extremely reduced fine-tuning, which are still compatible with the available constraints. This is the case of the polynomial coupling with exponent $q \geq 15$ lifting the fine-tuning of the coefficient $\zeta$ (usually constrained to be $\leq 10^{-6}$) to order~1.

It should also be emphasized that, while in the literature  the result by Webb et al. \cite{Webb} is usually said to be incompatible with the Oklo limit \cite{Damour:1996zw}, we have found that this is not always the case. For example, in the polynomial coupling case it is possible to obtain  several examples with a $\D \a$ which matches  the Quasars data and at the same time respects the Oklo bound.
\chapter{LTB models}
\label{ltbm}

We will introduce here the Lema\^{i}tre-Tolman-Bondi (LTB) solution, which will be used in Chapters \ref{pool} and \ref{s-c}.

First, in Table \ref{units} we list the units we will use for mass
density, time, the radial coordinate, the expansion rate, and two quantities,
$Y(r,t)$ and $W(r)$, that will appear in the metric.
\begin{table}[h!]
\caption{\label{units} \small \slshape  Units for various quantities.  We use 
geometrical units, $c=G=1$. Here, the present critical density is
$\rho_{C0}=3H^{2}_{0,\, Obs}/8 \pi$, with $H_{0,\, Obs}=70 \textrm{ km
s}^{-1}\textrm{ Mpc}^{-1}$.  In order to have the proper distance today 
we have to multiply the comoving distance by $a(t_{0})\simeq 2.92$.}
\begin{tabular}{lccr}
\hline
\hline
Quantity          & Notation    & Unit            & Value            \\ 
\hline
mass density & $\rho(r,t)$, $\bar{\rho}(r,t)$ & $\rho_{C 0}$ 
& $9.2\times10^{-30}\textrm{ g cm}^{-3}$             \\
time              & $t$, $T$, $\bar{t}$, $t_{BB}$, $T_0$ 
& $(6 \pi \rho_{C 0})^{-1/2}$  & $9.3\textrm{ Gyr}$ \\
comoving radial coordinate & $r$         & $(6 \pi \rho_{C 0})^{-1/2}$  
& $2857 \textrm{ Mpc}$ \\
metric quantity   & $Y(r,t)$    & $(6 \pi \rho_{C 0})^{-1/2}$  
& $2857 \textrm{ Mpc}$ \\
expansion rate    & $H(r,t)$    & $(6 \pi \rho_{C 0})^{1/2} $ 
& $\frac{3}{2}H_{0,\, Obs}$ \\
spatial curvature term    & $W(r)$      & $1$     &        ---             \\
\hline
\hline
\end{tabular}
\end{table}

The time $t$ appearing in Table \ref{units} is not the usual time in FRW
models.  Rather, $t=T-T_0$, where $T$ \textit{is} the usual cosmological time
and $T_0=2H_0^{-1}/3$ is, for an EdS model, the present age of the universe.  Thus, $t=0$ is the
present time and $t=t_{BB}=-T_0$ is the time of the big bang.  Finally, the
initial time of the LTB evolution is defined as $\bar{t}$. 

Both the FRW and the LTB metrics, after solving Einstein's equations, can be written in the form:
\begin{equation}
ds^2 = -dt^{2}+\frac{Y'^2(r,t)}{W^2(r)}dr^2+Y^2(r,t) \, d\Omega^2
\end{equation}
where the ``prime'' superscript will denote $d/dr$ and the ``dot'' superscript $d/dt$.
$Y(r,t)$ is the position of the shell $r$ at the time $t$. In order not to have shell crossing we will demand $Y'>0$.\\
It is clear that the Robertson--Walker metric is recovered with the substitution $Y(r,t)=a(t)r$ and $W^2(r)=1-kr^2$.
The above metric is expressed in the synchronous and comoving gauge.

\section{The EdS solution}

To connect the LTB solution to the FRW one, let's see how a flat matter-dominated universe (the
EdS model) is described within the LTB notation.  For the EdS solution there is no $r$ dependence to $\rho$ or $H$.
Furthermore, $Y(r,t)$ factors into a function of $t$ multiplying $r$ ($Y(r,t) =
a(t)r$), and in the EdS model $W(r)=1$.  In this model $\Omega_{M}=1$, so the value of $\rho$ today, denoted as $\rho_0$, is unity in the
units of Table \ref{units}. In order to connect with the LTB solution, we can
express the line element in the form:
\begin{equation}
ds^2=-dt^2+Y'^2(r,t) dr^2 + Y^2(r,t) \, d\Omega^2
\end{equation}
The Friedman equation and its solution are (recall $t = 0$
corresponds to the present time):
\begin{eqnarray}
H^{2}(t) & = & \frac{4}{9} \; \rho(t)=\frac{4}{9}(t+1)^{-2} \\
Y(r,t)& = & r \, a(t)=r \, \frac{(t+1)^{2/3}}{(\bar{t}+1)^{2/3}}
\end{eqnarray}
where the scale factor is normalized so that at the beginning of the LTB
evolution it is $a(\bar{t})=1$.

For the EdS model, $T_0=1$.  We also note that the comoving distance traveled
by a photon since the big bang is $r_{BB}=3/ a_{0}$.

\section{The LTB solution}

The LTB model \cite{Lemaitre:1933gd, Tolman:1934za, Bondi:1947av} is based on the assumptions that the system is spherically symmetric with purely radial motion and the motion is geodesic without shell crossing (otherwise we could not neglect the pressure).

It is useful to define a ``Euclidean'' mass $M(r)$ and an ``average'' mass
density $\bar\rho(r,t)$, defined as:
\begin{equation}
M(r) = 4\pi \int_0^r \rho(r,t) \: Y^2 Y' \: dr
= \frac{4 \pi}{3} Y^{3}(r,t) \: \bar{\rho}(r,t)
\end{equation}
In spherically symmetric models, in general there are two expansion rates: an
angular expansion rate, $H_\perp\equiv \dot{Y}(r,t)/Y(r,t)$, and a radial
expansion rate, $H_r\equiv \dot{Y}'(r,t)/Y'(r,t)$.  (Of course in the FRW model
$H_r=H_\perp$.)
The angular expansion rate is given by:
\begin{equation}
H^2_\perp(r,t) = \frac{4}{9} \; \bar{\rho}(r,t) +\frac{W^2(r)-1}{Y^{2}(r,t)}
\label{motion}
\end{equation}
Unless specified otherwise, we will identify $H_\perp=H$. The integral of (\ref{motion}) is:
\begin{equation} \label{integra}
t=\bar{t}(r)+\int_{\bar{Y}}^{Y}\left(W^{2}-1+{2G \over y}\right)^{-1/2}dy
\end{equation}
From the equation of motion (\ref{motion}) we see that the spherical shells of matter further away from the origin in relation to a particle P, do not affect the motion of P at all. This can be traced back to the Birkhoff's theorem.

Differentiating (\ref{motion}) gives the acceleration equation:
\begin{equation}
\label{eacce}
{\ddot{Y} \over Y}=-{2 \over 9}  \; \bar{\rho}(r,t)
\end{equation}
Equations (\ref{motion}) and (\ref{eacce}) are similar to the homogeneous ones with the crucial difference that there is an extra dependence on $r$.

Here we find the first potential problem of this metric as far as averages are concerned: in the equations (\ref{motion}-\ref{eacce}) the density is already averaged. This is the effect of the spherical symmetry. Moreover this ``automatic'' mean is taken from the center of symmetry which remains singled out in spite of our choices of the arbitrary functions. This mean that it could be inappropriate to average {\it again} the LTB model.

To specify the model we have to specify initial conditions, \textit{i.e.,} the
position $Y(r,\bar{t})$, the velocity $\dot{Y}(r,\bar{t})$ and the density
$\rho(\bar{t})$ of each shell $r$ at time $\bar{t}$. In the absence of 
shell crossing it is possible 
to give the initial conditions at different times for
different shells $r$: let us call this time $\bar{t}(r)$. The initial
conditions fix the arbitrary curvature function $W(r)$:
\begin{equation}
\label{cucu}
W^2(r)-1 \equiv 2 E(r)= \left. \left(\dot{Y}^2-  
\frac{1}{3 \pi}\frac{M}{Y}\right)\right|_{r,\bar{t}} \ ,
\end{equation}
where we can choose $Y(r,\bar{t})=r$ so that $M(r) = 4 \pi
\int_{0}^{r}\rho(\bar{r},\bar{t}) \: \bar{r}^{2} \: d\bar{r}$.

In a general LTB model there are therefore three arbitrary functions:
$\rho(r,\bar{t})$, $W(r)$ and $\bar{t}(r)$.
In Sect.~\ref{arbifunc} we provide a discussion about the number of 
independent arbitrary functions in a LTB model.

Finally here is the expression of the full Ricci scalar $R$:
\begin{equation} \label{ricci}
R=
\underbrace{2{\dot{Y}^{2} \over Y^{2}}+4{\ddot{Y} \over Y}}_{FLRW}
+ \underbrace{2{\ddot{Y}' \over Y'}+4{\dot{Y} \dot{Y}' \over Y Y'}}_{LTB}
-\underbrace{2{W^{2}-1 \over Y^{2}}-\overbrace{4{W W' \over Y Y'}}^{W \neq \textrm{const}}}_{W \neq 1}
\end{equation}
and of the spatial Ricci scalar ${\cal R}$:
\begin{equation} \label{riccis}
{\cal R}=-2{W^{2}-1 \over Y^{2}}-4{W W' \over Y Y'}=-4{E \over Y^{2}}-4{E' \over Y Y'}
\end{equation}
and of the expansion rate:
\begin{equation}
\theta=\Gamma_{0k}^{k}=2{\dot{Y} \over Y}+{\dot{Y}' \over Y'}
\end{equation}
and of the square shear:
\begin{equation}
\sigma^{2}=\half \sum_{k}\left(\Gamma_{0k}^{k} \right)^{2}- {1 \over 6} \left(\sum_{k} \Gamma_{0k}^{k} \right)^{2}=
{1 \over 3} \left({\dot{Y} \over Y}-{\dot{Y}' \over Y'}\right)^{2}
\end{equation}

\clearpage
\section{About the arbitrary functions in a LTB model} \label{arbifunc}

Here we illustrate, by means of an example, the choice of the arbitrary
functions in LTB models.  We are going to analyze the flat case.  Indeed we
have an analytical solution for it and this will help in understanding the
issues.

We said previously that there are three arbitrary functions in the LTB model: 
$\rho(r)$, $W(r)$ and $\bar{t}(r)$. They specify the position and velocities 
of the shells at a chosen time. In general, $\bar{t}$ depends on $r$; because
of the  absence of shell crossing it is possible to give the initial conditions
at different times for different shells labeled by $r$.

We start, therefore, by choosing the curvature $E(r)=(W^2(r)-1)/2$ to vanish,
which can be  thought as a choice of initial velocities $\dot{Y}$ at the time
$\bar{t}(r)$:
\begin{equation}
2 \, E(r)= \left. \dot{Y}^2-  
\frac{1}{3 \pi}\frac{M}{Y} \right|_{r, \, \bar{t}(r)} .
\end{equation}
For $E(r)=0$, the model becomes
\begin{equation}
ds^{2}=-dt^{2}+dY^{2}+Y^{2} d\Omega^{2} \;, 
\end{equation}
with solution
\begin{eqnarray} 
\label{flatso}
Y(r,t) & = & \left(\frac{3 \, M(r)}{4 \pi}\right)^{1/3}[t-\hat{t}(r)]^{2/3}
\nonumber \\
\bar{\rho}(r,t) & = & [t - \hat{t}(r)]^{-2} \;, 
\end{eqnarray}
where
\begin{eqnarray} 
\label{uffa}
\hat{t}(r) & \equiv & \bar{t}(r) - \bar{\rho}^{-1/2}(r, \bar{t}(r) )
\nonumber \\
\bar{\rho}(r, \bar{t}(r) ) & = & \left. \frac{3 \, M(r)}{4 \pi} 
\frac{1}{Y^{3}} \right|_{r, \, \bar{t}(r)} \;.
\end{eqnarray}
The next step is to choose the position of the shells, that is, to choose 
the density profile. As far as $\bar{t}(r)$ is concerned, only the 
combination $\hat{t}(r)$ matters. This, however, is not true for $M(r)$, 
which appears also by itself in Eq.\ (\ref{flatso}).

Looking at Eq.\ (\ref{uffa}) we see that to achieve an inhomogeneous profile 
we can either assign a homogeneous profile at an inhomogeneous initial time, or
an inhomogeneous density profile at a homogeneous initial time, or both.
Moreover, if we assign the function $M(r)$, then we can use our freedom to 
relabel $r$ in order to obtain all the possible $\bar{t}(r)$.  So we can see that
one of the three arbitrary functions expresses the gauge freedom.

In this thesis we fixed this freedom by choosing $\bar{t}(r)=\bar{t}$ and 
$Y(r, \bar{t})=r$ in order to have a better intuitive understanding 
of the initial conditions.

\chapter{Sewing the photon path} \label{sewing}

\begin{figure}[htb]
\begin{center}
\includegraphics[width=13 cm]{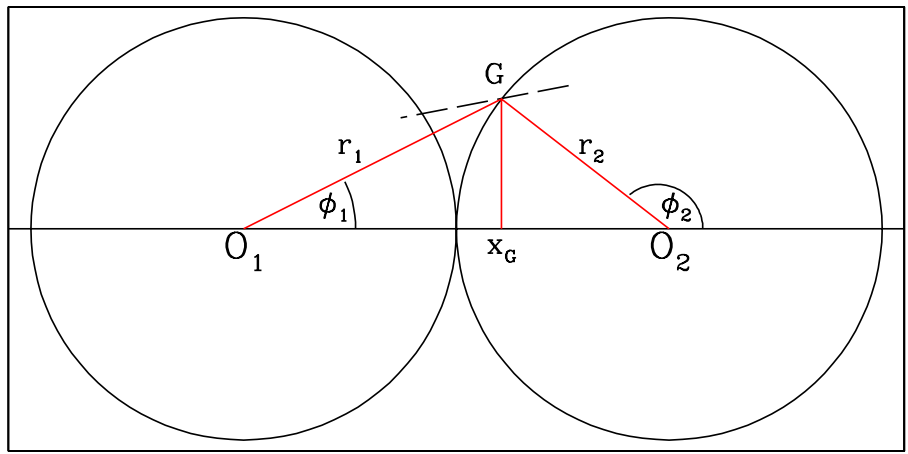}
\caption{Illustration of the procedure to calculate the transition between two
holes. The dashed line is a segment of the geodesic. $O_{1}$ and $O_{2}$
represent the two coordinate systems.}
\label{sew}
\end{center}
\end{figure}

In this Appendix we will show how to sew together the photon path between
two holes. We will always use center-of-symmetry coordinates, and therefore we
will move from the coordinates of $O_{1}$ to the ones of $O_{2}$ illustrated in
Fig.\ \ref{sew}. The geodesic near the contact point $G$ is represented by the
dashed line segment in Fig.\ \ref{sew}.

First, we want to find out the value $\bar{\lambda}$ of the affine parameter for
which the photon is at $G$. This is found by solving
\begin{eqnarray}
G_{2} & = & \left(r_{1}(\lambda) \cos \phi_{1}(\lambda)-2 r_{h},\; 
r_{1}(\lambda) \sin \phi_{1}(\lambda)\right) \nonumber \\
r_h^2 & = & x_{2 \, G}^{2}+y_{2 \, G}^{2}  .
\end{eqnarray}
These equations imply
\begin{equation}
r_{1}^{2}(\lambda)+3 r_{h}^{2}-4 r_{1}(\lambda)r_{h} \cos \phi_{1}(\lambda)=0 .
\end{equation}
Then we can give the initial conditions for the second hole:
\begin{eqnarray}
q_{2}(\bar{\lambda}) & = & q_{1}(\bar{\lambda}) \nonumber \\
t_{2}(\bar{\lambda}) & = & t_{1}(\bar{\lambda}) \nonumber \\
r_{2}(\bar{\lambda}) & = & r_{h} \nonumber \\
\phi_{2}(\bar{\lambda}) & = & \arccos(x_{2\, G}/r_{h}) .
\end{eqnarray}
Finally, we will need the constant $c_{\phi}$, a sort of constant angular momentum
density. Repeating the procedure of Sect.\ \ref{ciccio} for the first hole we will
find
\begin{equation}
c_{2\, \phi}= \sin \alpha_{2} \;  q_{1}(\bar{\lambda}) \; 
\left. Y_{2}\right|_{\bar{\lambda}} .
\end{equation}
Only $\alpha_{2}$ is missing. One way to find it is to calculate the inner
product in $O_{1}$ coordinates of the geodesic with the normalized spatial
vector parallel to $\overline{O_{2}G}$ (see Fig.\ \ref{sew}).

\chapter*{Acknowledgments}
\addcontentsline{toc}{chapter}{Acknowledgments}
\chaptermark{Acknowledgments}

\begin{quotation}


\ 

I would like, in the first place, to sincerely thank my advisors who guided me through my studies.
Professor Sabino Matarrese gave me the opportunity to work on the subject of this thesis, which is really fascinating to me. I will always be thankful for his support and for the opportunity he has given me. He also made my studies at the University of Chicago with Professor Rocky Kolb, possible. Working under their joint guidance has been and will be, I hope, very fruitful: their point of view and continual assistance has helped me to focus my aim and improve my analysis all along the development of this exciting and demanding research. Without their guide this work would not have been completed as it is now. I am really grateful to them for their knowledgeable advice and warm support.\\
I would also like to thank Professor Antonio Masiero and Doctor Francesca Rosati who helped me along the first steps of my research. I am very grateful for their patience, advice and support throughout my thesis of Laurea and for the interest they have shown in my work after graduation.

\ 

The strongest hug is for my dear family: my father, my mother, Lisa, Philippe, Leo, Caroline, Cinzia, Tommi, Ettore (due 04/08!), zio Filiberto, nonna Marisa, nonno Dano, zia Antonella, zio Paolo, Susi e Niki. Also for unforgettable nonna Elena and nonna Almerina, their recipes are still in good health!\\
My \textit{parentado} could not be warmer, it is always with me!

\newpage

Now my friends... I will try to group them, but this does not really make sense, we have done so many things together: mixing, changing, moving, returning.\\
Chronologically.. a big thank to 5C, in particular to Marco, D'0n, Ciux, Gale, Foffi, Punk, Qy, Just, Max, Lash. Then the GnazioQeA friends Pier, Teo, M, Benny, Gloria, Ignazio, Ilaria, Alessia, Ela, Enrico, Ele, A\&A, Cri, Mauro. A special thank to Gloria's warm \textit{taverna} as well as Qy's gdr-\textit{taverna} (pg included) for the many nice moments. \\ 
A word also for the VNP.\\
A big thank to all the p.a.p.a.~(there's a big book about..) and the ``365'', that is, Daniele, Sandro and aNdrea. A ``worm'' thank to Giovanni as well!\\
A thank also to Marcello for the many funny hours crashing airplanes and to the Astrofili Marcon, in particular Pagi, Eros, Alessia, Gigi, Andrea, Beppe (e i panzerotti della mamma!), Attilio.\\
Then a big thank to my lunch-dinner group in Chicago: Fabi, Yuki, Peggy, Jim, Pedro, Yuna, Mary, Andreas, Dominique, Hanna, Valentina, Sarah, Burcak.\\
A thank also to my foosball mates in the AAC basement, Carlos, Anibal, Vasilis, Felipe. To Valentin as well.

\ 

There are many things I could have said about each one of you, really many years of life.. I am sure you are thinking right now of something we have done together, a situation, a moment or many..\\
What I would like to say better now is that I hope to have more such moments and chances for each other!

\ 

\ 

...and.. a special kiss to Fabiana, my love, and many more..

\end{quotation}

\backmatter

\addcontentsline{toc}{chapter}{Bibliography}
\bibliographystyle{plain}
\bibliography{biblio}


\end{document}